\documentclass[12pt]{article}
\usepackage{graphicx} 
\usepackage{authblk} 
\usepackage{setspace} 
\usepackage{natbib} 
\usepackage{amsmath}
\usepackage{verbatim} 
\usepackage[a4paper, total={6in, 8in}]{geometry} 
\usepackage{float}  
\usepackage{hyperref} 
\usepackage{amssymb} 
\usepackage{amsfonts}
\usepackage{dutchcal} 
\usepackage{booktabs} 
\usepackage{algorithm} 
\usepackage{algorithmic}
\usepackage{multirow} 
\usepackage{makecell}
\usepackage{subcaption} 
\usepackage{orcidlink}
\usepackage{rotating}
\usepackage{lineno}

\usepackage{geometry}
\author{Anjana Wijayawardhana\thanks{Corresponding author: anjuwije99@gmail.com} \orcidlink{0000-0003-3847-2671}}
\author{David Gunawan}
\author{Thomas Suesse}
\affil{School of Mathematics and Applied Statistics, University of Wollongong, Wollongong, NSW, Australia}
\date{}

\title{Bayesian Inference for Non-Gaussian Simultaneous Autoregressive Models with Missing Data} 

\begin{document}
\doublespacing
\maketitle
\begin{abstract}

Standard simultaneous autoregressive (SAR) models typically assume normally distributed errors, an assumption often violated in real-world datasets that frequently exhibit non-normal, skewed, or heavy-tailed characteristics. New SAR models are proposed to capture these non-Gaussian features. The spatial error model (SEM), a widely used SAR-type model, is considered. 
Three novel SEMs are introduced, extending the standard Gaussian SEM. These extensions incorporate Student’s $t$-distributed errors to accommodate heavy-tailed behaviour, one-to-one transformations of the response variable to address skewness, or a combination of both.
Variational Bayes (VB) estimation methods are developed for these models, and the framework is further extended to handle missing response data under the missing not at random (MNAR) mechanism. Standard VB methods perform well with complete datasets; however, handling missing data requires a hybrid VB (HVB) approach, which integrates a Markov chain Monte Carlo (MCMC) sampler to generate missing values. The proposed VB methods are evaluated using both simulated and real-world datasets, demonstrating their robustness and effectiveness in dealing with non-Gaussian data and missing data in spatial models. Although the method is demonstrated using SAR models, the proposed model specifications and estimation approaches are widely applicable to various types of models for handling non-Gaussian data with missing values.


Keywords: Spatial error models; Student's $t$ errors; Yeo and Johnson (YJ) transformation; Missing not at random; Variational Bayes

\end{abstract}

\section{Introduction}



Simultaneous autoregressive (SAR) models are a broad class of spatial statistical models commonly used for analysing spatially correlated lattice data, with applications spanning ecology, social sciences, and finance~\citep{cressie1993statistics, chi2008spatial, calabrese2017measuring, ver2018spatial}. Examples of SAR models include spatial error model (SEM), spatial autoregressive models (SAM), and spatial Durbin model (SDM), among others~\citep{anselin1988spatial, cressie1993statistics}. Similar to other conventional spatial models, SAR models typically assume Gaussian-distributed data; see ~\citet{Besag1974, anselin1988spatial, cressie1993statistics}. However, this normality assumption is frequently violated in practice, as real-world datasets often exhibit non-normal characteristics such as skewness and heavy-tailed distributions. 

Several studies have examined the estimation of SAR models when the underlying error distributions are non-Gaussian, leading to methods that do not rely on strong distributional assumptions. For example, quasi-maximum likelihood estimators (QMLE) have been proposed for the SAM~\citep{lee2004asymptotic}, while various Generalised Method of Moments (GMM) approaches have been developed for different types of SAR models, which are robust to unknown heteroskedasticity and deviations from Gaussianity~\citep{kelejian1998generalized, lee2010efficient, Breitung03042018}. However, these approaches require the response variable to be fully observed and therefore cannot accommodate missing data.

Missing values frequently arise in spatial datasets. When estimating SAR models, ignoring missing response values can lead to inconsistency and bias~\citep{wang2013estimation, benedetti2020spatial}. Various estimation methods for SAR models with randomly missing data have been well developed (see \citet{lesage2004models, wang2013estimation, suesse2017computational, suesse2018marginal, math12233870}). However, research on estimating SAR models when data are missing not at random (MNAR) remains relatively scarce. \citet{flores2012estimation} proposed a GMM estimator for this setting, while \citet{Dougan2018bayesian} and \citet{Seya2021parameter} employed the Metropolis–Hastings (MH) algorithm.

Bayesian estimation of SAR models using Markov chain Monte Carlo (MCMC) methods can be computationally demanding, particularly when the data contain missing values~\citep{wijayawardhana2025variational}. Variational Bayes (VB) provides a computationally efficient alternative, yet it remains underutilised in the context of SAR models. For instance, \citet{wu2018fast} and \citet{bansal2021fast} applied VB to fully observed SAR models, while \citet{wijayawardhana2025variational} proposed hybrid VB (HVB) approaches for SAR models with missing data. Moreover, Bayesian inference methods, including both VB and MCMC, offer full posterior uncertainty quantification, in contrast to GMM, which yields only point estimates with asymptotic standard errors.

Our article makes several key contributions aimed at addressing limitations in the current literature on SAR models, particularly in handling non-Gaussian responses, missing data, and the computational challenges arising from these complex scenarios. First, we introduce novel SAR models designed to capture non-Gaussian characteristics of the response variable, such as skewness and heavy tails. We extend conventional Gaussian SAR models in two ways: by incorporating Student's $t$-distributed errors to address heavy-tailedness, and by applying the Yeo-Johnson (YJ) transformation~\citep{yeo2000new} to the response variable to handle skewness (see Section~\ref{sec:SEM_YJ}). Unlike the traditional two-step approach, where skewness is first reduced using a transformation such as YJ or Box-Cox and then a model is fitted, our integrated framework combines these steps within the SAR structure. These innovations yield three novel SAR models that jointly capture skewness and heavy-tailedness, making them well-suited for complex real-world data. While we demonstrate these models using SEM, yielding three SEM variants, the framework is broadly applicable to other SAR models.


Second, we propose efficient VB methods for estimating these new SEMs, both with and without missing values in the response variable. For SEMs with missing values, we use an extended version of the HVB algorithm of \citet{wijayawardhana2025variational} to handle non-Gaussian response variables and missing data under the MNAR mechanism. While the proposed HVB algorithm is an approximation compared to the MCMC methods \citep{Dougan2018bayesian, Seya2021parameter} and Hamiltonian Monte Carlo (HMC)~\citep{neal2011HMCchapter}, it offers substantial computational advantages. Our simulation study in Section~\ref{sec:simulationstudy-1} shows that the HVB algorithm is more than ten times faster than the HMC implementation in \texttt{RStan}~\citep{stanpkg} for a dataset of $n=625$ observations with approximately $50\%$ missing responses. Furthermore, Sections~\ref{sec:simulationstudy-2} and \ref{sec:real} show that the proposed HVB approach scales efficiently to much larger datasets, for which HMC becomes computationally impractical.

The rest of the paper is organised as follows. Section~\ref{sec:models} introduces the new SEMs, and Section~\ref{sec:SEMmissing} discusses SEMs with missing responses. In Section~\ref{sec:methods}, we present the VB methods for estimating SEMs with missing data. Section~\ref{NonGauSAR_sec:DIC} discusses Bayesian model comparison. Sections~\ref{sec:simulationstudy-1} and~\ref{sec:simulationstudy-2} present two simulation studies, and Section~\ref{sec:real} applies the SEMs to a real-world dataset. Section~\ref{sec:conclusion} summarises the main findings and outlines directions for future work. The paper also has an online supplement with additional technical details. The R code to reproduce all simulations and real-world analyses is available at~\href{https://github.com/AnjanaWijayawardhana/Non-Gaussian-SAR-with-Missing-Data-VB}{https://github.com/AnjanaWijayawardhana/Non-Gaussian-SAR-with-Missing-Data-VB}.

\section{Simultaneous Autoregressive
Models}
\label{sec:models}



In this section, we use our proposed framework to extend conventional Gaussian SAR models to non-Gaussian settings. We showcase the extension using spatial error models (SEMs), a widely adopted form of SAR-type models.

Let $\textbf{y}^\ast=(y^\ast_1,y^\ast_2,...,y^\ast_n)^\top$ be the $n \times 1$ vector of response variable observed at $n$ spatial locations $\textbf{s}_1, \hdots,\textbf{s}_n$, $\textbf{X}$ be the $n\times (r+1)$ design matrix containing the covariates, and $\textbf{W}$ be the $n\times n$ spatial weight matrix. The SEM is given by
\begin{equation}
\label{eq:SEM}
\textbf{y}^\ast=\textbf{X}\boldsymbol{\beta}+(\textbf{I}_n-\rho\textbf{W})^{-1}\textbf{e},
\end{equation}
\noindent where  $\textbf{e}$ is the $n \times 1$ vector of error terms, $\boldsymbol{\beta}=(\beta_0, \beta_1, \hdots, \beta_r)^\top$ is the $(r+1) \times 1$ vector of fixed effects parameters, $\rho$ is the spatial autocorrelation parameter, which measures the strength and direction of spatial dependence~\citep{lesage2009introduction}, and $\textbf{I}_n$ is the $n \times n$ identity matrix. Section~\ref{sec:models_withoutyj} discusses SEMs with Gaussian and Student's $t$ errors. 
Section \ref{sec:SEM_YJ} discusses two SEMs with the Yeo-Johnson transformation~\citep{yeo2000new}. 

\subsection{SEMs with Gaussian and Student's $t$ errors}
\label{sec:models_withoutyj}




The conventional SEM assumes Gaussian error terms $\textbf{e}\sim N(\textbf{0},\sigma_{\textbf{e}}^2\textbf{I}_n)$ in the model presented in Equation~\eqref{eq:SEM}, where $\sigma_{\textbf{e}}^2$ is the error variance parameter. This assumption implies that the distribution of $\textbf{y}^\ast$ is also Gaussian, with the mean vector $\textbf{X}\boldsymbol{\beta}$ and covariance matrix $\sigma^2_{\textbf{e}}(\textbf{A}^\top\textbf{A})^{-1}$, where $\textbf{A}=\textbf{I}_n-\rho\textbf{W}$. We call this model the SEM-Gau.

We now discuss SEM-t, which assumes that the error terms $\textbf{e}$ in Equation~\eqref{eq:SEM} follow a Student's $t$-distribution.
Student's $t$-distribution has been widely used to model heavy-tailed data in popular statistical models, such as regression \citep{Lange01121989, fernandez1999multivariate} and mixed-effects models \citep{Pinheiro01062001, wang2014multivariate}. \cite{huang2021robust} applied Student's $t$ errors to spatial autoregressive scalar-on-function regression models.

Let $e_i\sim t_\nu(0,\sigma^2_{e})$, where $e_i$ is the $i^{\textrm{th}}$ element of the error vector $\textbf{e}$ in Equation~\eqref{eq:SEM} and $t_\nu(0,\sigma^2_{e})$ is the Student's $t$ distribution with $\nu$ degrees of freedom ($\nu \geq 3$), mean zero, and scale parameter $\sigma^2_{e}$. These Student's $t$ errors, $e_i$ for $i = 1, \ldots, n$, can be expressed as a scale mixture of normals~\citep{Chan_Koop_Poirier_Tobias_2019}:
\begin{equation}
    \label{eq:terrors.scl}
    \begin{split}
        e_i & \mid \tau_i, \sigma^2_{e}~\sim N(0,\sigma^2_{e}\tau_i),\\
        \tau_i & \mid \nu  \sim IG\left(\frac{\nu}{2},\frac{\nu}{2}\right), ~~~~i=1, \hdots n
    \end{split}
\end{equation}
\noindent where $IG(a,b)$ denotes the inverse gamma distribution with shape parameter $a$ and rate parameter $b$.  The conditional distribution of errors presented in Equation~\eqref{eq:terrors.scl}, can be written in vector notation as  
\begin{equation}
\label{eq:terrors.vec}
        \textbf{e}\mid \boldsymbol{\tau}, \sigma^2_{e}~\sim N(0,\sigma^2_{e}\boldsymbol{\Sigma}_{\boldsymbol{\tau}}),
\end{equation}

\noindent where the matrix $\boldsymbol{\Sigma}_{\boldsymbol{\tau}}=\textrm{diag}(\tau_1, \hdots, \tau_n)$. We now incorporate the error structure from Equation~\eqref{eq:terrors.vec} into the SEM described in Equation~\eqref{eq:SEM}. As a result, the distribution of $\textbf{y}^\ast$ conditional on the latent vector $\boldsymbol{\tau}$ is also Gaussian, with the mean vector $\textbf{X}\boldsymbol{\beta}$ and covariance matrix $\sigma^2_{\textbf{e}}(\textbf{A}^\top\boldsymbol{\Sigma}^{-1}_{\boldsymbol{\tau}}\textbf{A})^{-1}$, where $\textbf{A}=\textbf{I}_n-\rho\textbf{W}$. This model is referred to as the SEM-t. The log-likelihood function of $\textbf{y}^\ast$ for both the SEM-Gau and SEM-t is given by:
\begin{equation}
\label{eq:log.like.SEM.Gau}
    \text{log}~p_{\textbf{y}^\ast}(\textbf{y}^\ast \mid \boldsymbol{\xi} )=-\frac{n}{2}\textrm{log}(2\pi)-\frac{n}{2}\textrm{log}(\sigma^2_{\boldsymbol{e}})+\frac{1}{2}\textrm{log}|\textbf{M}|-\frac{1}{2\sigma^2_{\boldsymbol{e}}}\textbf{r}^\top\textbf{M}\textbf{r},
\end{equation}

\noindent where for the SEM-Gau, $\boldsymbol{\xi} = \boldsymbol{\phi}$, while for the SEM-t, $\boldsymbol{\xi} = (\boldsymbol{\phi}^\top, \boldsymbol{\tau}^\top)^\top$. The definitions for $\textbf{r}$, $\textbf{M}$, and $\boldsymbol{\phi}$ are given in Table~\ref{tab:new.models}.

\begin{table}[h]
\centering
\caption{The definitions of $\textbf{r}$, $\textbf{M}$, and the parameters $\boldsymbol{\phi}$ for SEM-Gau, SEM-t, YJ-SEM-Gau, and YJ-SEM-t. For SEM-Gau and SEM-t, the density of the response vector conditional on $\boldsymbol{\xi}$ follows a multivariate Gaussian distribution with mean $\textbf{X}\boldsymbol{\beta}$ and covariance matrix $\sigma^2_{\textbf{e}}\textbf{M}^{-1}$, where $\textbf{A}=\textbf{I}_n-\rho\textbf{W}$.} 
\label{tab:new.models}
\begin{tabular}{ccccc}
\hline
{Model} &  $\boldsymbol{\phi}$ & \textbf{r} &  $\textbf{M}$\\ \hline
SEM-Gau &   $(\boldsymbol{\beta}^\top, \sigma^2_{\textbf{e}}, \rho)^\top$ & $\textbf{y}-\textbf{X}\boldsymbol{\beta}$& $\textbf{A}^\top\textbf{A}$\\
SEM-t   &  $ (\boldsymbol{\beta}^\top, \sigma^2_{\textbf{e}}, \rho, \nu)^\top$ & $\textbf{y}-\textbf{X}\boldsymbol{\beta}$   & $\textbf{A}^\top\boldsymbol{\Sigma}^{-1}_{\boldsymbol{\tau}}\textbf{A}$   \\ 
YJ-SEM-Gau     &  $ (\boldsymbol{\beta}^\top, \sigma^2_{\textbf{e},}, \rho, \gamma)^\top$ & $t_{\gamma}(\textbf{y})-\textbf{X}\boldsymbol{\beta}$ &  $\textbf{A}^\top\textbf{A}$ \\ 

YJ-SEM-t   & $ (\boldsymbol{\beta}^\top, \sigma^2_{\textbf{e}}, \rho, \nu,\gamma)^\top$    &  $t_{\gamma}(\textbf{y})-\textbf{X}\boldsymbol{\beta}$& $\textbf{A}^\top\boldsymbol{\Sigma}^{-1}_{\boldsymbol{\tau}}\textbf{A}$ \\
\hline
\end{tabular}
\end{table}

\subsection{SEMs with transformations}
\label{sec:SEM_YJ}


This section extends the SEM-Gau and SEM-t in Equation~\eqref{eq:SEM} by applying the YJ transformation element-wise to the response vector $\textbf{y}^\ast$.

Let $y_i$ be the $i^{\textrm{th}}$ element of the response vector in the SEM with YJ transformation. We assume that $y_i$ is modelled as
\begin{equation}
    {y}_i=t_{{\gamma}}^{-1}(y^\ast_i),\quad i=1,...,n,
\end{equation}
\noindent where $y^\ast_i$ is the $i^{\textrm{th}}$ element of the response vector in the SEM given in Equation~\eqref{eq:SEM}, with either Gaussian or Student’s $t$-distributed errors, and $t_{{\gamma}}^{-1}(\cdot)$ is the inverse function of the YJ transformation. For $0 < \gamma < 2$, it is defined as:
\begin{equation}
\label{eq:YJ_inv}
y_i=t_{{\gamma}}^{-1}(y^\ast_i) =
\begin{cases} 
(y^\ast_i \gamma + 1)^{1/\gamma} - 1 & \text{if } y^\ast_i \geq 0 \\
1 - (-(2 - \gamma)y^\ast_i + 1)^{1/(2 - \gamma)} & \text{if }  y^\ast_i < 0. \\
\end{cases}
\end{equation}

\noindent As the YJ transformation makes the data more symmetric and less skewed, applying its inverse transformation moves the data away from symmetry. This implies that $ \textbf{y} = t_{{\gamma}}^{-1}(\textbf{y}^\ast) = (t^{-1}_{{\gamma}}(y^\ast_1), t^{-1}_{{\gamma}}(y^\ast_2), \dots, t^{-1}_{{\gamma}}(y^\ast_n))^{\top} $ becomes an asymmetric, non-Gaussian response variable. In vector notation, the SEM with the YJ transformation (YJ-SEM) is defined as:
\begin{equation}
\label{eq:YJ-SEM}
\textbf{y}=t_{{\gamma}}^{-1}(\textbf{y}^\ast)=t_{{\gamma}}^{-1}(\textbf{X}\boldsymbol{\beta}+(\textbf{I}_n-\rho\textbf{W})^{-1}\textbf{e}).\\
\end{equation}
Using the standard technique for transforming random variables, it is straightforward to show that the density of $\textbf{y}$ for the YJ-SEM is:
\begin{equation}
\label{eq:density_y}
  p_\textbf{y}{(\textbf{y} \mid \boldsymbol{\xi} )}=p_{\textbf{y}^\ast}({t_{\boldsymbol{\gamma}}(\textbf{y})}) \prod_{i=1}^n \left( \frac{dt_{\gamma}(y_i)}{dy_i}  \right) ,~~~~i=1, \hdots n,
\end{equation}



\noindent where $p_{\textbf{y}^\ast}(\cdot)$ is the density of $\textbf{y}^\ast$, conditional on the parameter vector $\boldsymbol{\xi}$; see Section~\ref{sec:models_withoutyj}, and $\frac{dt_{\gamma}(y_i)}{dy_i}$ is the derivative of the YJ transformation with respect to $y_i$; see Equation~\eqref{eq:der_yj_by_yi} in \textcolor{black}{Section~\ref{sec:YJ_derivatives}} of the online supplement. The vector $\boldsymbol{\xi}$ denotes the parameters of the YJ-SEM defined in Equation~\eqref{eq:YJ-SEM}.


The YJ-SEM with Gaussian errors (YJ-SEM-Gau) assumes that the error term in Equation~\eqref{eq:YJ-SEM} follows a normal distribution: $\textbf{e} \sim N(0, \sigma_{\textbf{e}}^2 \textbf{I}_n)$, where $ \sigma_{\textbf{e}}^2$ is a variance parameter. The YJ-SEM with Student’s $t$-distributed errors (YJ-SEM-t) assumes that the error term follows the structure given in Equation~\eqref{eq:terrors.vec}. The log-likelihood function of $\textbf{y}$ for YJ-SEM-Gau and YJ-SEM-t is expressed as:
\begin{equation}
\label{eq:log.like.YJ.SEM.t}
    \begin{split}
           \text{log}~p_{\textbf{y}}(\textbf{y} \mid \boldsymbol{\xi} )&=-\frac{n}{2}\textrm{log}(2\pi)-\frac{n}{2}\textrm{log}(\sigma^2_{\boldsymbol{e}})+\frac{1}{2}\textrm{log}|\textbf{M}|-\frac{1}{2\sigma^2_{\boldsymbol{e}}}\textbf{r}^\top\textbf{M}\textbf{r} \\ &+
           \sum_{i=1}^n \text{log}\left(  \frac{dt_{\gamma}(y_i)}{dy_i} \right),
    \end{split}
\end{equation}


\noindent where for the YJ-SEM-Gau, $\boldsymbol{\xi} = \boldsymbol{\phi}$, while for the YJ-SEM-t, $\boldsymbol{\xi} = (\boldsymbol{\phi}^\top, \boldsymbol{\tau}^\top)^\top$. The definitions for $\textbf{r}$, $\textbf{M}$, and $\boldsymbol{\phi}$ are given in Table~\ref{tab:new.models}. Note that $y_i=y^{\ast}_{i}$ for $i=1,...,n$, corresponds to an identity transformation in both SEM-Gau and SEM-t.

The three SEMs, SEM-Gau, SEM-t, and YJ-SEM-Gau, are special cases of the YJ-SEM-t. The SEM-Gau is a YJ-SEM-t where the degrees of freedom parameter ($\nu$) equals
 $\infty$ and the YJ parameter ($\gamma$) equals 1, while the YJ-SEM-Gau is a YJ-SEM-t with $\nu =\infty$ and $\gamma \neq 1$. Finally, the SEM-t is also a YJ-SEM-t with $\gamma=1$.

\section{SEMs with missing responses}
\label{sec:SEMmissing}

This section develops the formulation of SEMs in settings where response values are missing and introduces the underlying missing data mechanisms, focusing in particular on the missing not at random (MNAR) mechanism.

Consider that the response vector $\textbf{y}$ of an SEM presented in Sections~\ref{sec:models_withoutyj} and~\ref{sec:SEM_YJ} contains missing values. Let $\textbf{y}_o$ denote the subset of $\textbf{y}$ with $n_o$ observed units, and $\textbf{y}_u$ denote the subset with $n_u$ unobserved (missing) units. The complete response vector can then be written as $\textbf{y} = (\textbf{y}_o^\top, \textbf{y}_u^\top)^\top$.

We define a binary missingness indicator vector $\textbf{m} = (m_1, \ldots, m_n)^\top$, where
\[
m_i = 
\begin{cases}
1, & \text{if } y_i \text{ is missing,}\\
0, & \text{if } y_i \text{ is observed.}
\end{cases}
\]

\noindent In the presence of missing responses, the vector $\textbf{r}$ and the matrices $\textbf{X}$ and $\textbf{M}$, as defined in Table~\ref{tab:new.models}, are partitioned into corresponding observed and unobserved components as follows:
\begin{equation}
\label{mat:portions_of_xwM}
\textbf{r}=
\begin{pmatrix}
    \textbf{r}_o\\
   {\textbf{r}_u}
\end{pmatrix},
~\textbf{X}=
\begin{pmatrix}
    \textbf{X}_o\\
   {\textbf{X}_u}
\end{pmatrix},
~\textbf{M}=
\begin{pmatrix}
    \textbf{M}_{oo}  &  \textbf{M}_{ou}\\
    \textbf{M}_{uo} & \textbf{M}_{uu}
\end{pmatrix},
\end{equation}


\noindent where $\textbf{r}_o$ and $\textbf{r}_u$ are subvectors of $\textbf{r}$ corresponding to the observed and unobserved responses, respectively. Similarly, $\textbf{X}_o$ and $\textbf{X}_u$ represent the design matrices associated with the observed and unobserved responses, while $\textbf{M}_{oo}$, $\textbf{M}_{ou}$, $\textbf{M}_{uo}$, and $\textbf{M}_{uu}$ denote the corresponding submatrices of $\textbf{M}$.

\subsection{Missing data mechanism and MNAR assumption}





The missing data mechanism is characterised by the joint probability mass function of the missingness indicator vector $\textbf{m} = (m_1, \ldots, m_n)^\top$, where each element $m_i$ denotes whether the corresponding response $y_i$ is missing ($m_i = 1$) or observed ($m_i = 0$). Let $p(m_i \mid y_i, \textbf{x}_i^*, \boldsymbol{\psi})$ denote the probability that $y_i$ is missing, possibly depending on itself and on a set of covariates $\textbf{x}_i^* = (1, x_{i1}, \ldots, x_{iq})^\top$. The parameter vector $\boldsymbol{\psi}=(\boldsymbol{\psi}_\textbf{x}^\top,\psi_{\textbf{y}})^\top$ consists of the fixed effect parameter vector associated with covariates $\textbf{x}^{*}_i$, denoted as $\boldsymbol{\psi}_\textbf{x} = (\psi_0, \psi_1, \psi_2, \hdots, \psi_{q})^\top$, and the fixed effect parameter corresponding to $\textbf{y}$, denoted as $\psi_{\textbf{y}}$. 


Three main types of missing-data mechanisms are commonly distinguished following \citet{rubin1976inference}: 
(i) \textit{missing completely at random} (MCAR), where the probability of missingness is independent of both the observed and unobserved data and on observed covariates, that is, $p(m_i\mid y_i, \mathbf{x}_i^{*}, \boldsymbol{\psi}) = p(m_i \mid \boldsymbol{\psi})$; 
(ii) \textit{missing at random} (MAR), where the probability of missingness may depend on observed covariates but not on the observed and unobserved responses, so that $p(m_i \mid y_i, \mathbf{x}_i^{*}, \boldsymbol{\psi}) = p(m_i \mid \mathbf{x}_i^{*}, \boldsymbol{\psi})$; and 
(iii) \textit{missing not at random} (MNAR), where the probability of missingness depends on the response variable itself and observed covariates. Under MNAR, the term $p(m_i \mid y_i, \mathbf{x}_i^{*}, \boldsymbol{\psi})$ cannot be simplified further. 

The conditional probability mass function of $m_i$ under MNAR, $p(m_i \mid y_i, \textbf{x}_i^*, \boldsymbol{\psi})$ can be modelled using any binary-response model, such as a logistic or probit regression. In this paper, we model the probability that the $i^{\text{th}}$ response is missing via a logistic regression:
\begin{equation}
\label{eq:joint.logistic.SEM_}
p(m_i = 1 \mid y_i, \textbf{x}_i^*, \boldsymbol{\psi})
= \frac{\exp(\textbf{x}_i^{* \top}\boldsymbol{\psi}_{\textbf{x}} + y_i \psi_y)}{1 + \exp(\textbf{x}_i^{* \top}\boldsymbol{\psi}_{\textbf{x}} + y_i \psi_y)},
\end{equation}
so that the joint probability mass function of vector $\textbf{m}$ is
\begin{equation}
\label{eq:joint.logistic.SEM}
p(\textbf{m} \mid \textbf{y}, \textbf{X}^*, \boldsymbol{\psi})
= \prod_{i=1}^{n} 
\frac{\exp\{(\textbf{x}_i^{* \top}\boldsymbol{\psi}_{\textbf{x}} + y_i \psi_y) m_i\}}
{1 + \exp(\textbf{x}_i^{* \top}\boldsymbol{\psi}_{\textbf{x}} + y_i \psi_y)},
\end{equation}

\noindent where $\textbf{X}^\ast=(\textbf{x}^{\ast\top}_1,\dots,\textbf{x}^{\ast\top}_n)^{\top}$ denotes the design matrix for the missingness model, with dimension $n \times (q+1)$.


\subsection{Joint modelling of $\textbf{y}$ and $\textbf{m}$}
\label{sec:joint_mod_y_m}

In our study, the primary objective is the estimation of the parameters of the proposed SEMs, denoted by $\boldsymbol{\xi}$, which govern the distribution of the response vector $\textbf{y}$. However, under the MNAR mechanism, the missingness process is not ignorable, and valid inference requires modelling the joint distribution of $(\textbf{y}, \textbf{m})$ or, equivalently, the joint estimation of $\boldsymbol{\xi}$ and $\boldsymbol{\psi}$~\citep{wijayawardhana2025variational}. The joint distribution of $\textbf{y}$ and $\textbf{m}$ is denoted by $p(\textbf{y}, \textbf{m} \mid \boldsymbol{\xi},\boldsymbol{\psi})$. Using the selection model factorisation~\citep{little2019statistical}, $p(\textbf{y}, \textbf{m} \mid \boldsymbol{\xi},\boldsymbol{\psi})$ can be factorised as
\begin{equation}
\label{eq:selectionmodels}
p(\textbf{y}, \textbf{m} \mid \boldsymbol{\xi},\boldsymbol{\psi}) =p(\textbf{m} \mid \textbf{y}, \boldsymbol{\psi}) p(\textbf{y} \mid \boldsymbol{\xi}),
\end{equation}
\noindent where $p(\textbf{y} \mid \boldsymbol{\xi})$ represents the likelihood of the SEM, and $p(\textbf{m} \mid \textbf{y}, \boldsymbol{\psi})$ denotes the joint probability mass function of missingness indicators defined in Equation~\eqref{eq:joint.logistic.SEM}. The log-likelihood of $\mathbf{y}$ for the different SEMs, along with their full model specifications, is provided in Section~\ref{sec:models}.

It is important to note that the factorisation in Equation~\eqref{eq:selectionmodels} is a standard and widely applicable decomposition for jointly modelling a missingness mechanism and a statistical model with missing data, and it applies to any SAR-type model; see Section~\ref{online_sec:ext_SAR} of the online supplement. This factorisation enables practical inference for SAR models with missing values under MNAR. 
~In Section~\ref{sec:methods}, we discuss the estimation of the joint posterior distribution of the model parameters ($\boldsymbol{\xi}, \boldsymbol{\psi}$) and $\textbf{y}_u$ using the proposed variational Bayes method. 


\section{Variational Bayes inference for SEMs with missing data}
\label{sec:methods}

Section~\ref{sec:variationalBayesInference} provides an overview of the variational Bayes (VB) inference framework for estimating SEMs with missing responses. Section~\ref{sec:HVB} then introduces an efficient hybrid variational Bayes (HVB) algorithm. For completeness, Section~\ref{sec:VB_full} of the online supplement presents the VB algorithm for SEMs with fully observed data.

\subsection{Variational Bayes inference under missing responses \label{sec:variationalBayesInference}}

Variational Bayes (VB) inference approximates a posterior distribution by formulating it as an optimisation problem, offering a more computationally efficient alternative to the computationally intensive Markov chain Monte Carlo (MCMC) methods for complex statistical models. The standard VB approach for SEMs with missing responses is described below.

We assume that some values in the response vector of an SEM are missing under the MNAR mechanism, as described in Section~\ref{sec:SEMmissing}. Under this assumption, Bayesian inference requires estimating the joint posterior distribution of the model parameters $\boldsymbol{\xi}$ and $\boldsymbol{\psi}$, based on the joint likelihood defined in Equation~\eqref{eq:selectionmodels}. Since the subvector $\textbf{y}_u$ of $\textbf{y} = (\textbf{y}_o^\top, \textbf{y}_u^\top)^\top$ is unobserved, it must also be inferred. Consequently, Bayesian inference involves estimating the joint posterior distribution of $\boldsymbol{\xi}$, $\boldsymbol{\psi}$, and $\textbf{y}_u$. For the SEM-Gau and YJ-SEM-Gau, $\boldsymbol{\xi} = \boldsymbol{\phi}$, while for the SEM-t and YJ-SEM-t, $\boldsymbol{\xi} = (\boldsymbol{\phi}^\top, \boldsymbol{\tau}^\top)^\top$; see Table~\ref{tab:new.models} for the definition of $\boldsymbol{\phi}$ for each SEM.


Let the joint posterior distribution of $\boldsymbol{\xi}$, $\boldsymbol{\psi}$, and $\textbf{y}_u$ be denoted as $p(\boldsymbol{\xi}, \boldsymbol{\psi}, \textbf{y}_u \mid \textbf{y}_o, \textbf{m})$. We denote the prior distributions of $\boldsymbol{\xi}$ and $\boldsymbol{\psi}$ by $p(\boldsymbol{\xi})$ and $p(\boldsymbol{\psi})$, respectively. Using the selection model factorisation in Equation~\eqref{eq:selectionmodels}, the joint posterior distribution ${p(\boldsymbol{\xi}, \boldsymbol{\psi}, \textbf{y}_u \mid \textbf{y}_o, \textbf{m})}$ is given by
\begin{equation} 
\label{eq:joint_post_miss}
p(\boldsymbol{\xi}, \boldsymbol{\psi}, \textbf{y}_u \mid \textbf{y}_o, \textbf{m}) \propto p(\textbf{m}\mid {\textbf{y}},\boldsymbol{\psi})p({\textbf{y}}\mid \boldsymbol{\xi})p(\boldsymbol{\xi})p(\boldsymbol{\psi}).
\end{equation}

\noindent We define $h(\boldsymbol{\xi}, \boldsymbol{\psi}, \textbf{y}_u)=p(\textbf{y} \mid \boldsymbol{\xi}) p(\textbf{m} \mid \textbf{y}, \boldsymbol{\psi}) p(\boldsymbol{\xi}) p(\boldsymbol{\psi})$. The term $h(\boldsymbol{\xi}, \boldsymbol{\psi}, \textbf{y}_u)$ for each SEM with missing responses, along with the total number of parameters to be estimated ($s$) for each model, are provided in Table~\ref{tab:post_miss}.

\begin{table}[h]
\centering
\caption{The term $h(\boldsymbol{\xi}, \boldsymbol{\psi}, \textbf{y}_u)=p(\textbf{y} \mid \boldsymbol{\xi}) p(\textbf{m} \mid \textbf{y}, \boldsymbol{\psi}) p(\boldsymbol{\xi})p(\boldsymbol{\psi})$, and the total number of model parameters $s$ (including the length of the latent vector $\boldsymbol{\tau}$ if exists) for the SEM-Gau, SEM-t, YJ-SEM-Gau, and YJ-SEM-t with missing responses. {The density ${p(\boldsymbol{\tau}\mid \boldsymbol{\phi})=\prod_{i=1}^np(\tau_i\mid \nu)}$ represents the $n$ independent inverse gamma latent variables used in SEMs with Student-$t$ errors (see Equation~\eqref{eq:terrors.scl}).}}

\label{tab:post_miss}
\begin{tabular}{ccc}
\hline
{Model} &  $h(\boldsymbol{\xi},\boldsymbol{\psi},\textbf{y}_u)$ & $s$\\ \hline
SEM-Gau & $p({\textbf{y}}\mid \boldsymbol{\phi})p(\textbf{m}\mid {\textbf{y}},\boldsymbol{\psi})p(\boldsymbol{\phi})p(\boldsymbol{\psi})$ & $r+q+5$\\
SEM-t   & $p(\textbf{y} \mid \boldsymbol{\tau}, \boldsymbol{\phi})p(\textbf{m}\mid {\textbf{y}},\boldsymbol{\psi})p(\boldsymbol{\tau}\mid \boldsymbol{\phi}) p(\boldsymbol{\phi}) p(\boldsymbol{\psi})$     & $r+q+6+n$\\ 
YJ-SEM-Gau     & $p(\textbf{y} \mid \boldsymbol{\phi})p(\textbf{m}\mid {\textbf{y}},\boldsymbol{\psi})p(\boldsymbol{\phi})p(\boldsymbol{\psi})$& $r+q+6$\\ 
YJ-SEM-t   & $p(\textbf{y} \mid \boldsymbol{\tau}, \boldsymbol{\phi})p(\textbf{m}\mid {\textbf{y}},\boldsymbol{\psi})p(\boldsymbol{\tau}\mid \boldsymbol{\phi}) p(\boldsymbol{\phi}) p(\boldsymbol{\psi})$ & $r+q+7+n$ \\
\hline
\end{tabular}
\end{table}

We consider the variational approximation $q_{\boldsymbol{\lambda}}(\boldsymbol{\xi}, \boldsymbol{\psi}, \textbf{y}_u)$, indexed by the variational parameter $\boldsymbol{\lambda}$ to approximate the joint posterior $p(\boldsymbol{\xi}, \boldsymbol{\psi}, \textbf{y}_u \mid \textbf{y}_o, \textbf{m})$ in Equation~\eqref{eq:joint_post_miss}. The VB approach approximates this posterior distribution by minimising the Kullback-Leibler (KL) divergence
between $q_{\boldsymbol{\lambda}}(\boldsymbol{\xi}, \boldsymbol{\psi}, \textbf{y}_u)$ and $p(\boldsymbol{\xi}, \boldsymbol{\psi}, \textbf{y}_u \mid \textbf{y}_o, \textbf{m})$, which is defined as
\begin{equation}
    \label{eq:ELBO}
    \begin{split}
     \text{KL}(\boldsymbol{\lambda})&=\text{KL}\left(q_{\boldsymbol{\lambda}}(\boldsymbol{\xi},\boldsymbol{\psi},\textbf{y}_u)\mid \mid p(\boldsymbol{\xi}, \boldsymbol{\psi},\textbf{y}_u \mid \textbf{y}_o, \textbf{m}) \right)\\
           & =\int   \text{log} \left(\frac{q_{\boldsymbol{\lambda}}(\boldsymbol{\xi}, \boldsymbol{\psi},\textbf{y}_u)}{p\left(\boldsymbol{\xi}, \boldsymbol{\psi},\textbf{y}_u|\textbf{y}_o, \textbf{m}\right)} \right) q_{\boldsymbol{\lambda}}(\boldsymbol{\xi}, \boldsymbol{\psi},\textbf{y}_u)d\boldsymbol{\xi}d\boldsymbol{\psi}d\textbf{y}_u. \\
    \end{split}
\end{equation}

Minimising the KL divergence in Equation~\eqref{eq:ELBO} is equivalent to maximising the evidence lower bound (ELBO) on the logarithm of the marginal likelihood, $\text{log}~p(\textbf{y}_o, \textbf{m})$, denoted by $\mathcal{L}(\boldsymbol{\lambda})$, where $p(\textbf{y}_o, \textbf{m})=\int ~p(\textbf{y}_o, \textbf{m}\mid \boldsymbol{\xi}, \boldsymbol{\psi},  \textbf{y}_u)p(\textbf{y}_u \mid \boldsymbol{\xi}, \boldsymbol{\psi})p\left(\boldsymbol{\xi}, \boldsymbol{\psi}\right)d\boldsymbol{\boldsymbol{\xi} }d\boldsymbol{\psi}d\textbf{y}_u$~\citep{blei2017variational}. The ELBO can be written as an expectation with respect to the variational distribution and is given by
\begin{equation}
    \label{eq:ELBO.expectation_wrt_q}
            \mathcal{L}(\boldsymbol{\lambda})=\mathbb{E}_{{q}_{\boldsymbol{\lambda}}}\left [\text{log}~h(\boldsymbol{\xi},\boldsymbol{\psi},\textbf{y}_u)-\text{log}~  q_{\boldsymbol{\lambda}}(\boldsymbol{\xi}, \boldsymbol{\psi}, \textbf{y}_u)\right],
\end{equation}

\noindent where $\mathbb{E}_{{q}_{\boldsymbol{\lambda}}}\left[\cdot\right]$ denotes the expectation with respect to $q_{\boldsymbol{\lambda}}$. Table~\ref{tab:post_miss} provides expressions for $h(\boldsymbol{\xi}, \boldsymbol{\psi}, \textbf{y}_u)$ for different SEMs.  


To maximise ELBO given in Equation~\eqref{eq:ELBO.expectation_wrt_q} with respect to variational parameters, $\boldsymbol{\lambda}$, the stochastic gradient ascent (SGA) method is used~(\citeauthor{nott2012regression},~\citeyear{nott2012regression};~\citeauthor{pmlr-v32-titsias14},~\citeyear{pmlr-v32-titsias14},~\citeyear{NIPS2015_1373b284}). 
The  SGA method updates the initial value for $\boldsymbol{\lambda}$ (say $\boldsymbol{\lambda}^{(0)}$) according to the iterative scheme, 
\begin{equation}
    \label{eq:SGA}    \boldsymbol{\lambda}^{(t+1)}=\boldsymbol{\lambda}^{(t)}+\mathbcal{a}^{(t)}\circ\widehat{\nabla_{\boldsymbol{\lambda}}\mathcal{L}(\boldsymbol{\lambda}^{(t)})},
\end{equation}

\noindent where 
$\widehat{\nabla_{\boldsymbol{\lambda}}\mathcal{L}(\boldsymbol{\lambda})}$ is an unbiased estimate of the true gradient 
$\nabla_{\boldsymbol{\lambda}}\mathcal{L}(\boldsymbol{\lambda})$ of the ELBO in Equation~\eqref{eq:ELBO.expectation_wrt_q}, $\mathbcal{a}^{(t)}$ ($t=0,1,\hdots$) is a vector of adaptive step sizes, obtained using the ADADELTA method \citep{Zeiler2012ADADELTAAA}, described in Section~\ref{sec:sup:ADADELTA} of the online supplement to facilitate rapid convergence of the SGA algorithm. The symbol $\circ$ represents the element-wise product of two vectors. The updating of Equation~\eqref{eq:SGA} is done until a stopping criterion is satisfied. Further details on the SGA method are provided in Section~\ref{sec:VB_full} of the online supplement.

In standard VB inference, the missing data vector, $\textbf{y}_u$, is treated as a set of parameters, and the joint posterior distribution of $\boldsymbol{\xi}$, $\boldsymbol{\psi}$, and $\textbf{y}_u$ is approximated using the joint variational distribution $q_{\boldsymbol{\lambda}}(\boldsymbol{\xi}, \boldsymbol{\psi},\textbf{y}_u)$. However, \citet{wijayawardhana2025variational} demonstrated that this joint approximation is often insufficiently flexible to capture the posterior distributions of model parameters and $\textbf{y}_u$ simultaneously, due to the complex dependencies among the spatially correlated elements of $\textbf{y}_u$. To address this limitation, Section~\ref{sec:HVB} introduces an alternative hybrid algorithm that effectively combines VB optimisation with MCMC sampling.

\subsection{Hybrid variational Bayes \label{sec:HVB}}

This section extends the hybrid VB (HVB) algorithm of~\citet{wijayawardhana2025variational} to efficiently handle missing values $\textbf{y}_u$ in non-Gaussian SEMs proposed in Section \ref{sec:models}. The approach employs a Gaussian variational approximation with a factor covariance structure~\citep{ong2018gaussian} to approximate the posterior distribution of the parameters, while employing an MCMC sampler to generate samples of $\textbf{y}_u$ from its conditional distribution, given $\textbf{y}_o$, $\textbf{m}$, $\boldsymbol{\xi}$ and $\boldsymbol{\psi}$. We first factorise the joint variational distribution that approximates $\boldsymbol{\xi}, \boldsymbol{\psi}$, and $\textbf{y}_u$, $q_{\boldsymbol{\lambda}}(\boldsymbol{\xi}, \boldsymbol{\psi}, \textbf{y}_u)$  as follows:
\begin{equation}
    \label{eq:q}
    q_{\boldsymbol{\lambda}}(\boldsymbol{\xi}, \boldsymbol{\psi}, \textbf{y}_u)=p({\textbf{y}}_u \mid {\textbf{y}}_o, \textbf{m},\boldsymbol{\xi,\boldsymbol{\psi}})q_{\boldsymbol{\lambda}}^{0}(\boldsymbol{\xi},\boldsymbol{\psi}),
\end{equation}
\noindent where $p(\textbf{y}_u \mid \textbf{y}_o, \textbf{m}, \boldsymbol{\xi}, \boldsymbol{\psi})$ represents the conditional distribution of $\textbf{y}_u$, given the observed data ($\textbf{y}_o$, $\textbf{m}$), and the model parameters ($\boldsymbol{\xi}$, $\boldsymbol{\psi}$). The term $q_{\boldsymbol{\lambda}}^{0}(\boldsymbol{\xi},\boldsymbol{\psi})$ is the variational approximation used to approximate the posterior distribution of $\boldsymbol{\xi}$ and $\boldsymbol{\psi}$.
The ELBO in Equation~\eqref{eq:ELBO.expectation_wrt_q} can be rewritten by substituting the factorised variational distribution $q_{\boldsymbol{\lambda}}(\boldsymbol{\xi}, \boldsymbol{\psi}, \textbf{y}_u)$ from Equation~\eqref{eq:q}. By applying Bayes’ rule and simplifying, we obtain 
\begin{equation} 
\mathcal{L}(\boldsymbol{\lambda}) = \mathbb{E}_{{q}_{\boldsymbol{\lambda}}}\left[\text{log}~p({\textbf{y}}_o,\textbf{m}\mid \boldsymbol{\xi},\boldsymbol{\psi})+\text{log}~p(\boldsymbol{\xi},\boldsymbol{\psi})-\text{log}~  q_{\boldsymbol{\lambda}}^0(\boldsymbol{\xi},\boldsymbol{\psi}) \right]=  \mathcal{L}^0(\boldsymbol{\lambda}),
\label{eq:ELBO.expectation_wrt_qaug}
\end{equation}

\noindent where $\mathcal{L}^0(\boldsymbol{\lambda})$ is the ELBO resulting from approximating only the posterior distribution of parameters ($\boldsymbol{\xi}$, $\boldsymbol{\psi}$), that is, $p(\boldsymbol{\xi}, \boldsymbol{\psi} \mid \textbf{y}_o, \textbf{m})$, directly using the variational distribution $q_{\boldsymbol{\lambda}}^0(\boldsymbol{\xi},\boldsymbol{\psi})$; see~\citet{wijayawardhana2025variational} for the complete proof.

We assume that the variational distribution, $q_{\boldsymbol{\lambda}}^0(\boldsymbol{\xi},\boldsymbol{\psi})$, follows a multivariate Gaussian distribution with a factor covariance structure~\citep{ong2018gaussian}, 
${q_{\boldsymbol{\lambda}}^{0}(\boldsymbol{\xi}, \boldsymbol{\psi}) \sim N\left( (\boldsymbol{\xi}^\top, \boldsymbol{\psi}^\top)^\top; \boldsymbol{\mu}, \textbf{B} \textbf{B}^\top + \textbf{D}^2 \right)}$, where $\boldsymbol{\mu}$ is an $s \times 1$ vector of variational means, $\textbf{B}$ is an $s \times p$ matrix with the upper triangular elements set to zero ($p << s$), and $\textbf{D}$ is an $s \times s$ diagonal matrix with positive diagonal elements $\textbf{d} = (d_1, \dots, d_{s})^{\top}$. For all SEMs with missing values, the value of $s$ is provided in Table~\ref{tab:post_miss}. The vector of variational parameters is given by $\boldsymbol{\lambda} = \left( \boldsymbol{\mu}^\top, \text{vech}(\textbf{B})^\top, \textbf{d}^\top \right)^\top$, where the 'vech' operator vectorises a matrix by stacking its columns from left to right while removing all the elements above the diagonal (the super-diagonal elements) of the matrix. 


We now describe how the SGA method is used in the proposed HVB algorithm. At each iteration, an unbiased estimate of the ELBO gradient, $\nabla_{\boldsymbol{\lambda}}\mathcal{L}(\boldsymbol{\lambda})$, denoted $\widehat{\nabla_{\boldsymbol{\lambda}}\mathcal{L}(\boldsymbol{\lambda})}$, needs to be computed for the SGA update given in Equation~\eqref{eq:SGA}. This is achieved using the reparameterisation trick~\citep{kingma2013auto}. First, samples are generated from the variational distribution $q_{\boldsymbol{\lambda}}^0(\boldsymbol{\xi}, \boldsymbol{\psi})$ by drawing ${\boldsymbol{\delta}^0} = (\boldsymbol{\eta}^0, \boldsymbol{\epsilon}^0)\sim N(\mathbf{0}, \mathbf{I}_{s + p})$, where $\boldsymbol{\eta}^0$ is a $p$-dimensional vector and $\boldsymbol{\epsilon}^0$ is an $s$-dimensional vector. Next, we compute $(\boldsymbol{\xi}^\top, \boldsymbol{\psi}^\top)^\top = t^0(\boldsymbol{\delta}^0, \boldsymbol{\lambda}) = \boldsymbol{\mu} + \textbf{B} \boldsymbol{\eta}^0 + \textbf{d} \circ \boldsymbol{\epsilon}^0$.
The density of $\boldsymbol{\delta}^0$ is denoted by $f_{\boldsymbol{\delta}^0}(\boldsymbol{\delta}^0)$. Let $\boldsymbol{\delta} = ({\boldsymbol{\delta}^0}^\top, \textbf{y}_u^\top)^\top$, where the product density is expressed as ${f_{\boldsymbol{\delta}}(\boldsymbol{\delta}) = f_{\boldsymbol{\delta}^0}(\boldsymbol{\delta}^0) p({\textbf{y}}_u \mid {\textbf{y}}_o, \textbf{m}, t^0(\boldsymbol{\delta}^0, \boldsymbol{\lambda}))}$. Finally, the transformation from $\boldsymbol{\delta}$ to the parameter and augmented missing value space is given by $((\boldsymbol{\xi}^\top,\boldsymbol{\psi}^\top)^\top,{\textbf{y}}_u^\top)^\top=t(\boldsymbol{\delta},\boldsymbol{\lambda})=(t^0(\boldsymbol{\delta}^0,\boldsymbol{\lambda})^\top,\textbf{y}_u^\top)^\top=((\boldsymbol{\mu}+\textbf{B}\boldsymbol{\eta}^0+\textbf{d}\circ \boldsymbol{\epsilon}^0)^\top,\textbf{y}_u^\top)^\top$.

The reparameterisation gradient of the ELBO in Equation~\eqref{eq:ELBO.expectation_wrt_qaug} is derived by differentiating under the integral sign as shown below:
\begin{equation}
    \label{eq:ELBO.grad.lamda.aug}
                    \nabla_{\boldsymbol{\lambda}}\mathcal{L}(\boldsymbol{\lambda})=\mathbb{E}_{f_{\boldsymbol{\delta}}}\left[\frac{dt^0(\boldsymbol{\delta}^0,\boldsymbol{\lambda})^\top}{d\boldsymbol{\lambda}}(\nabla_{(\boldsymbol{\xi}^\top,\boldsymbol{\psi}^\top)^\top}\text{log}~h(\boldsymbol{\xi},\boldsymbol{\psi},\textbf{y}_u)-\nabla_{(\boldsymbol{\xi}^\top,\boldsymbol{\psi}^\top)^\top}\text{log}~  q_{\boldsymbol{\lambda}}^0(\boldsymbol{\xi},\boldsymbol{\psi}) )\right],
\end{equation}

\noindent where $\frac{dt^0(\boldsymbol{\delta}^0,\boldsymbol{\lambda})}{d\boldsymbol{\lambda}}$ is the derivative of the transformation $t^0(\boldsymbol{\delta}^0,\boldsymbol{\lambda})=\boldsymbol{\mu}+\textbf{B}\boldsymbol{\eta}^0+\textbf{d}\circ \boldsymbol{\epsilon}^0$ with respect to the variational parameters $\boldsymbol{\lambda}=(\boldsymbol{\mu}^\top,\text{vech}(\textbf{B})^\top,\textbf{d}^\top)^\top$; see Section~\ref{sec:der.hvb} of the online supplement for the proof. The expressions for $\frac{dt^0(\boldsymbol{\delta}^0,\boldsymbol{\lambda})^\top}{d\boldsymbol{\lambda}}$ and $\nabla_{(\boldsymbol{\xi}^\top,\boldsymbol{\psi}^\top)^\top}\text{log}~  q_{\boldsymbol{\lambda}}^0(\boldsymbol{\xi},\boldsymbol{\psi})$ can be found in Section~\ref{nongousar_appendix:deri_t_log_q} of the online supplement. In addition, the expressions for $\nabla_{(\boldsymbol{\xi}^\top,\boldsymbol{\psi}^\top)^\top}\text{log}~h(\boldsymbol{\xi},\boldsymbol{\psi},\textbf{y}_u)$  are provided in Section~\ref{sec:grad_with_miss} of the online supplement for different SEMs. 


Algorithm \ref{alg:HVB} outlines the HVB algorithm. For all four SEMs with missing data, the conditional distribution of missing data, $p(\mathbf{y}_u \mid \mathbf{y}_o, \mathbf{m},\boldsymbol{\xi},\boldsymbol{\psi})$ is not available in closed form. To sample from $p(\mathbf{y}_u \mid \mathbf{y}_o, \mathbf{m}, \boldsymbol{\xi}, \boldsymbol{\psi})$ in step 5 of Algorithm~\ref{alg:HVB}, we use the MCMC steps outlined in Algorithm~\ref{alg:AugvbMCMCstep}.




\begin{algorithm}[h]
  \caption{Hybrid variational Bayes (HVB) algorithm.}
  \begin{algorithmic}[1]
  \label{alg:HVB}
   \STATE Initialise $\boldsymbol{\lambda}^{(0)}=(\boldsymbol{\mu}^{\top (0)},\textrm{vech}{(\textbf{B})}^{\top (0)},\textbf{d}^{\top (0)})^{\top}$ and set $t=0$ 
  \REPEAT 
      \STATE Generate $({\boldsymbol{\eta}^0}^{(t)},{\boldsymbol{\epsilon}^0}^{(t)})\sim N(\textbf{0},\textbf{I}_{s+p})$
      \STATE Generate $(\boldsymbol{\xi}^{(t)^\top},\boldsymbol{\psi}^{(t)^\top})^\top\sim q_{\boldsymbol{\lambda}^{(t)}}^0(\boldsymbol{\xi},\boldsymbol{\psi})$ using its reparameterised representation.
      \STATE Generate $\textbf{y}_u^{(t)}~\sim p(\textbf{y}_u\mid \textbf{y}_o, \textbf{m}, \boldsymbol{\xi}^{(t)}, \boldsymbol{\psi}^{(t)})$
      \STATE Construct unbiased estimates $\widehat{{\nabla_{\boldsymbol{\mu}}\mathcal{L}(\boldsymbol{\lambda})}},\widehat{{\nabla_{\textrm{vech}{(\textbf{B})}}\mathcal{L}(\boldsymbol{\lambda})}}, \text{and}\widehat{{\nabla_{\textbf{d}}\mathcal{L}(\boldsymbol{\lambda})}}$ using Equations ~\eqref{eq:grad_wrt_mu_vb2},~\eqref{eq:grad_wrt_B_vb2} and ~\eqref{eq:grad_wrt_d_vb2} in Section~\ref{nongousar_appendix:deri_t_log_q} of the online supplement.
      \STATE Set adaptive learning rates  $\mathbcal{a}^{(t)}_{\boldsymbol{\mu}}$ , $\mathbcal{a}^{(t)}_{\text{vech}(\textbf{B})}$ and $\mathbcal{a}_{\textbf{d}}^{(t)}$, using ADADELTA described in Section~\ref{sec:sup:ADADELTA} of the online supplement.
      \STATE Set $\boldsymbol{\mu}^{(t+1)} = \boldsymbol{\mu}^{(t)} + \mathbcal{a}^{(t)}_{\boldsymbol{\mu}} \circ \widehat{\nabla_{\boldsymbol{\mu}}  \mathcal{L}(\boldsymbol{\lambda}^{(t)})}$.
      \STATE  Set $\textrm{vech}(\textbf{B})^{(t+1)} = \textrm{vech}(\textbf{B})^{(t)} + \mathbcal{a}_{\text{vech}(\textbf{B})}^{(t)} \circ \widehat{\nabla_{\textrm{vech}(\textbf{B})} \mathcal{L}(\boldsymbol{\lambda}^{(t)})}$.
      \STATE Set $\textbf{d}^{(t+1)} = \textbf{d}^{(t)} + \mathbcal{a}_{\textbf{d}}^{(t)} \circ  \widehat{\nabla _{\textbf{d}}  \mathcal{L}(\boldsymbol{\lambda}^{(t)})}$.
      \STATE Set $\boldsymbol{\lambda}^{(t+1)} = (\boldsymbol{\mu}^{\top (t+1)}, \textrm{vech}(\textbf{B})^{\top (t+1)}, \textbf{d}^{\top (t+1)})^{\top}$, and $t = t + 1$
      \UNTIL {some stopping rule is satisfied}
  \end{algorithmic}
\end{algorithm}

\begin{algorithm}
  \caption{MCMC steps within the $t^{th}$ iteration of the HVB algorithm.}
  \begin{algorithmic}[1]
  \label{alg:AugvbMCMCstep}
  \STATE Initialise missing values $\textbf{y}_{u,0}~\sim p(\textbf{y}_u\mid \boldsymbol{\xi}^{(t)}, \textbf{y}_o)$
  \FOR{$i=1, \dots, N_1$} 
      \STATE Sample $\widetilde{\textbf{y}}_u$ from the proposal distribution $p(\widetilde{\textbf{y}}_u\mid \boldsymbol{\xi}^{(t)},\textbf{y}_o)$. 
      \STATE Sample $u$ from uniform distribution, $u~\sim \mathcal{U}(0,1)$
      \STATE Calculate $a=\text{min}\left (1,\frac{p(\textbf{m} \mid \widetilde{\textbf{y}},\boldsymbol{\psi}^{(t)})}{p(\textbf{m} \mid \textbf{y}_{{i-1}},\boldsymbol{\psi}^{(t)})}\right)$, where $\widetilde{\textbf{y}}=(\textbf{y}_o^{\top},{\widetilde{\textbf{y}}_u}^\top)^\top$ and $\textbf{y}_{i-1}=(\textbf{y}^\top_o,{\textbf{y}^\top_{u,{i-1}}})^\top$
      \IF{$a>u$}
        \STATE $\textbf{y}_{u,i}=\widetilde{\textbf{y}}_u$
      \ELSE
        \STATE $\textbf{y}_{u,i}=\textbf{y}_{u,i-1}$
      \ENDIF
  \ENDFOR
\STATE Output $\textbf{y}^{(t)}_{u}=\textbf{y}_{u,N_1}$
  \end{algorithmic}
\end{algorithm}

{ The MCMC steps in Algorithm~\ref{alg:AugvbMCMCstep} generate samples from the proposal distribution $p(\widetilde{\textbf{y}}_u\mid \boldsymbol{\xi}^{(t)},\textbf{y}_o)$, which follows a multivariate Gaussian with the mean vector given by $ \textbf{X}_u\boldsymbol{\beta}-\textbf{M}_{uu}^{-1}\textbf{M}_{uo}\textbf{r}_o$ and the covariance matrix given by $\sigma^2_{\textbf{e}}\textbf{M}_{uu}^{-1}$, for both SEM-Gau and SEM-t; see Table~\ref{tab:new.models} and Equation~\eqref{mat:portions_of_xwM} for details on the partitioning of $\textbf{r}$, $\textbf{X}$ and $\textbf{M}$ across different SEMs.}


{For YJ-SEM-Gau and YJ-SEM-t, the proposals are generated in two steps. First, we sample $\widetilde{\textbf{y}}_u^*$ from the conditional distribution $p({\widetilde{\textbf{y}}_u^*}\mid \boldsymbol{\xi}^{(t)},\textbf{y}_o^*)$, which follows a multivariate Gaussian with the mean vector given by $ \textbf{X}_u\boldsymbol{\beta}-\textbf{M}_{uu}^{-1}\textbf{M}_{uo}(\textbf{y}_o^*-\textbf{X}_o\boldsymbol{\beta})$ and the covariance matrix given by $\sigma^2_{\textbf{e}}\textbf{M}_{uu}^{-1}$, where $\textbf{y}_o^*=t_{\gamma}(\textbf{y}_o)$. Then, we apply the inverse Yeo-Johnson (YJ) transformation to obtain the final proposal: \( \widetilde{\textbf{y}}_u = t_{\gamma}^{-1}(\widetilde{\textbf{y}}_u^*) \).} 

As $n$ and $n_u$ increase, the HVB algorithm implemented using the MCMC scheme outlined in Algorithm~\ref{alg:AugvbMCMCstep} fails to estimate the parameters accurately due to a low acceptance rate. To address this, we partition $\mathbf{y}_u$ into $k$ blocks and update one block at a time. The MCMC steps for sampling the missing values one block at a time are detailed in Algorithm~\ref{alg:AugvbMCMCstepB} in Section \ref{sec:HVB_All} of the online supplement. We refer to the modified HVB algorithm with block-wise updates as HVB-AllB (which uses MCMC steps in Algorithm~\ref{alg:AugvbMCMCstepB}), and the original version without block-wise updates as HVB-NoB (which uses MCMC steps in Algorithm~\ref{alg:AugvbMCMCstep}) in the following sections.




\section{Bayesian model comparison\label{sec:DIC}}
\label{NonGauSAR_sec:DIC}

To assess the model fit of different SEMs in the subsequent sections of this study, we employ extended versions of the standard Deviance Information Criterion (DIC; \citet{spiegelhalter2002bayesian}) as proposed by \citet{10.1214/06-BA122}. This section first outlines DIC methods for models without missing data, followed by those for models with missing data.

\subsection{Model comparison with full data}

For full data SEMs, both $ \text{DIC}_1 $ and $ \text{DIC}_2$ are computed to evaluate model fit, with lower values indicating better fit. $ \text{DIC}_1$ is defined as:
\begin{equation}
    \label{eq:DIC1}
        \text{DIC}_1 =-4\mathbb{E}_{\boldsymbol{\phi}}[\text{log}~p(\textbf{y} \mid \boldsymbol{\phi})]+
        2~\text{log}~p(\textbf{y}\mid \overline{\boldsymbol{\phi}}),
\end{equation}

\noindent where $\overline{\boldsymbol{\phi}}$ is the posterior mean of ${\boldsymbol{\phi}}$, and $p(\textbf{y} \mid \boldsymbol{\phi})$ denotes the density of $\textbf{y}$ given the model parameters $\boldsymbol{\phi}$. Both the expectation and the posterior mean $\overline{\boldsymbol{\phi}}$ are computed using samples generated from the approximate posterior distribution estimated by the proposed VB method in Section~\ref{sec:VB_full} of the online supplement. For all SEMs, expressions for $\text{log}~p(\textbf{y} \mid \boldsymbol{\phi})$ are presented in \textcolor{black}{Section~\ref{sec:online_model_comp}} of the online supplement.

\noindent $\text{DIC}_2$ is defined as:
\begin{equation}
    \label{eq:DIC2}
        \text{DIC}_2 =-4\mathbb{E}_{\boldsymbol{\phi}}[\text{log}~p(\textbf{y} \mid \boldsymbol{\phi})]+
        2~\text{log}~p(\textbf{y}\mid \widehat{\boldsymbol{\phi}}),
\end{equation}

\noindent where $\widehat{\boldsymbol{\phi}}$ is the value of ${\boldsymbol{\phi}}$ that maximises the function $p(\textbf{y} \mid \boldsymbol{\phi})p(\boldsymbol{\phi})$, estimated using samples drawn from the approximate posterior obtained by the proposed VB method. The expectation is also computed using the samples from the approximate posterior.

\subsection{Model comparison with missing responses}

$\text{DIC}_5$ is used to assess model fit for SEMs with missing data and is defined as:
\begin{equation}
    \label{eq:DIC5}
        \text{DIC}_5 =-4\mathbb{E}_{\boldsymbol{\phi},\boldsymbol{\psi},\textbf{y}_u}[\text{log}~p(\textbf{y},\textbf{m} \mid \boldsymbol{\phi},\boldsymbol{\psi})]+
        2~\text{log}~p(\textbf{y}_o, \widehat{\textbf{y}}_u, \textbf{m} \mid \widehat{\boldsymbol{\phi}}, \widehat{\boldsymbol{\psi}}),
\end{equation}

\noindent where $p(\textbf{y}, \textbf{m} \mid \boldsymbol{\phi}, \boldsymbol{\psi})$ is the likelihood of $\textbf{y}$ and $\textbf{m}$, conditional on model parameters $\boldsymbol{\phi}$ and $\boldsymbol{\psi}$. The values $\widehat{\boldsymbol{\phi}}$, $\widehat{\boldsymbol{\psi}}$, and $\widehat{\textbf{y}}_u$ are the posterior samples of $\boldsymbol{\phi}$, $\boldsymbol{\psi}$ and $\textbf{y}_u$ that maximise the function $p(\textbf{y}, \textbf{m} \mid \boldsymbol{\phi}, \boldsymbol{\psi})p(\boldsymbol{\phi}, \boldsymbol{\psi})$, where the samples are drawn from the approximate posterior distribution estimated by the proposed HVB method in Section~\ref{sec:HVB}. The expectation is also computed using these variational posterior samples. Note that $p(\textbf{y}, \textbf{m} \mid \boldsymbol{\phi}, \boldsymbol{\psi}) = p(\textbf{y} \mid \boldsymbol{\phi}) p(\textbf{m} \mid \textbf{y}, \boldsymbol{\psi})$; see the selection model factorisation in Equation~\eqref{eq:selectionmodels} in Section~\ref{sec:joint_mod_y_m}.


It is worth noting that the density function $p(\textbf{y} \mid \boldsymbol{\phi})$, which appears in all DIC expressions in Equations~\eqref{eq:DIC1}, \eqref{eq:DIC2}, and \eqref{eq:DIC5}, represents the density of the response vector $\textbf{y}$ conditional on $\boldsymbol{\phi}$. For the SEM-Gau and YJ-SEM-Gau, this distribution is equivalent to $p(\textbf{y} \mid \boldsymbol{\xi})$, as for these two models $\boldsymbol{\xi}=\boldsymbol{\phi}$ (see Section~\ref{sec:models}). In contrast, for the SEM-t and YJ-SEM-t, $p(\textbf{y} \mid \boldsymbol{\phi})$ needs to be derived. This is because for the latter two models, $\boldsymbol{\xi}=(\boldsymbol{\phi}^\top,\boldsymbol{\tau}^\top)^\top$, so $p(\textbf{y} \mid \boldsymbol{\phi})$ represents the marginal density of $\textbf{y}$ given $\boldsymbol{\phi}$, with the latent vector $\boldsymbol{\tau}$ integrated out (see \textcolor{black}{Section~\ref{sec:online_model_comp}} of the online supplement). Further details, including the logarithm of the density $p(\textbf{y} \mid \boldsymbol{\phi})$ for all four models, are provided in \textcolor{black}{Section~\ref{sec:online_model_comp}} of the online supplement.

\section{Simulation study: Assessing the accuracy of VB methods}
\label{sec:simulationstudy-1}


This section compares the accuracy of the proposed VB methods (with full data and with missing data) against the Hamiltonian Monte Carlo (HMC)~\citep{neal2011HMCchapter} method. The HMC method is implemented using the \texttt{RStan} interface~\citep{stanpkg}, specifically employing the No-U-Turn Sampler (NUTS) algorithm~\citep{hoffman2014no}. The posterior densities estimated using HMC are considered as ground truth to assess the accuracy of the VB methods. A relatively small simulated dataset ($n = 625$) is used for this comparison, as HMC becomes computationally expensive for SAR models with larger datasets, particularly in the presence of missing data~\citep{wijayawardhana2025variational}. The simulation study presented in Section~\ref{sec:simulationstudy-2} and the real data application in Section~\ref{sec:real} use moderately large datasets ($n = 4{,}378$).

The prior distributions used for both the VB and HMC algorithms throughout this paper are described next. To map the parameters $\sigma^2_{\textbf{e}}$, $\rho$, $\nu$, and $\gamma$ onto the real line, we apply the following transformations: $\omega^\prime = \text{log}(\sigma^2_{\textbf{e}})$, $\rho^\prime = \text{log}(1+\rho) - \text{log}(1-\rho)$, $\nu^\prime=\text{log}(\nu-3)$, and $\gamma^\prime=\text{log}(\gamma)-\text{log}(2-\gamma)$. The prior distributions are then specified on the transformed parameter space. Table~\ref{NonGauSAR_tab:priors} summarises the priors and hyperparameters. We use diffuse (non-informative) priors for all model parameters so that posterior inference is driven primarily by the data. For example, the fixed effects $\boldsymbol{\beta}$ are assigned weakly informative normal priors, $\boldsymbol{\beta} \sim N(\textbf{0}, 100\textbf{I})$, where \textbf{I} is an identity matrix of appropriate dimension (see Table~\ref{NonGauSAR_tab:priors}). For the relatively large datasets analysed in this paper, the choice of these non-informative hyperparameters has little impact on the parameter estimates or on the convergence behaviour of the VB and HMC algorithms. 



\begin{table}[ht]
    \centering
    \setlength{\tabcolsep}{3pt} 
        \caption{Prior distributions of model parameters}
    \begin{tabular}{ccccccc}
   \bottomrule
    \\{Parameter} & $\boldsymbol{\beta}$ & $\omega^\prime$ & $\rho^\prime$ & $\nu^\prime$ & $\gamma^\prime$ & $\boldsymbol{\psi}$ \\
    \bottomrule
    \\{Prior distribution} & $N(\textbf{0}, \sigma^2_{\boldsymbol{\beta}} \textbf{I})$ & $N(0, \sigma^2_{\omega^\prime})$ & $N(0, \sigma^2_{\rho^\prime})$ & $N(0, \sigma^2_{\nu^\prime})$ & $N(0, \sigma^2_{\gamma^\prime})$ & $N(\textbf{0}, \sigma^2_{\boldsymbol{\psi}} \textbf{I})$ \\
    \\{Hyperparameters} & $\sigma^2_{\boldsymbol{\beta}} = 10^2$ & $\sigma^2_{\omega^\prime} = 10^2$ & $\sigma^2_{\rho^\prime} = 10^2$ & $\sigma^2_{\nu^\prime} = 10^2$& $\sigma^2_{\gamma^\prime} = 10^2$ & $\sigma^2_{\boldsymbol{\psi}} = 10^2$ \\
    \bottomrule
    \end{tabular}
    \label{NonGauSAR_tab:priors}
\end{table}

The comparison in this section is based on a simulated dataset generated from the YJ-SEM-Gau. The dataset contains 625 observations with five covariates, all generated from standard normal distributions. The spatial weight matrix, $\textbf{W}$, is based on a $25 \times 25$ regular lattice, where neighbouring units are defined using the rook neighbourhood~\citep{lloyd2010spatial}. The model parameters are specified as follows: the fixed effects ($\boldsymbol{\beta}$) are randomly drawn from a discrete uniform distribution over \(-3\) to \(3\) (excluding 0). The error variance is set to $\sigma^2_{\mathbf{e}} = 1$ and the spatial autocorrelation parameter is $\rho = 0.8$. The Yeo–Johnson parameter is set to $\gamma = 1.25$, introducing moderate left skewness into the simulated response variable. See Figure~\ref{fig:HMCvsVB_density} in Section~\ref{sec:sim_comp_online} of the online supplement for the kernel density plot of the response variable. Section~\ref{NonGauSAR_sec:sim_comparison_full} compares the proposed VB method with HMC using this dataset without missing values, while Section~\ref{NonGauSAR_sec:sim_results_miss_accu} compares the proposed HVB method with HMC in the presence of missing responses by introducing missing values into this dataset.


\subsection{SEMs with full data}
\label{NonGauSAR_sec:sim_comparison_full}

We now assess the accuracy of the VB method for SEMs with full data. The VB algorithm, outlined in Algorithm~\ref{alg:SAGfull}  in Section \ref{sec:VB_full} of the online supplement, and the HMC algorithm are used to fit the YJ-SEM-Gau to the simulated dataset. The initial values for the VB algorithm are set as follows: the variational mean vector ($\boldsymbol{\mu}$) is initialised using the estimates of $\boldsymbol{\beta}$, $\sigma^2_{\textbf{e}}$, and $\rho$ obtained from fitting the SEM-Gau via maximum likelihood (ML) estimation method. The variational mean corresponding to the parameter $\gamma$ is initialised to 1. For the variational covariance matrix parameters, the elements of $\textbf{B}$ and the diagonal elements of $\textbf{D}$ are all initialised to 0.01. A value of $p=4$ factors is used. The results do not improve when the number of factors is increased. The HMC algorithm is initialised in the same way, with its initial parameter values set to the initial variational means described above. Both the VB and HMC algorithms are run for 10,000 iterations, at which point convergence is achieved. For the HMC algorithm, the first 5,000 iterations are treated as burn-in and discarded. Convergence for the VB algorithm is assessed visually by inspecting the plots of variational means over iterations. For the HMC algorithms, convergence is assessed by inspecting trace plots of the model parameters (see Section~\ref{sec:sim_conv_analysis_6} of the online supplement).

The HMC algorithm directly generates posterior samples of the model parameters. In contrast, the VB algorithm first estimates the variational parameters, $\boldsymbol{\lambda}$, by running the variational optimisation. Then, 10,000 posterior draws of the model parameters, ${\boldsymbol{\xi}=\boldsymbol{\phi}=(\boldsymbol{\beta}^\top,{\sigma}^2_{e},\rho,\gamma)^\top}$, are generated from the variational distribution $q_{\boldsymbol{\lambda}}(\boldsymbol{\xi})$; see {Section}~\ref{sec:VB_full} of the online supplement.

\begin{figure}[htbp]
    \centering
    \includegraphics[width=0.8\textwidth]{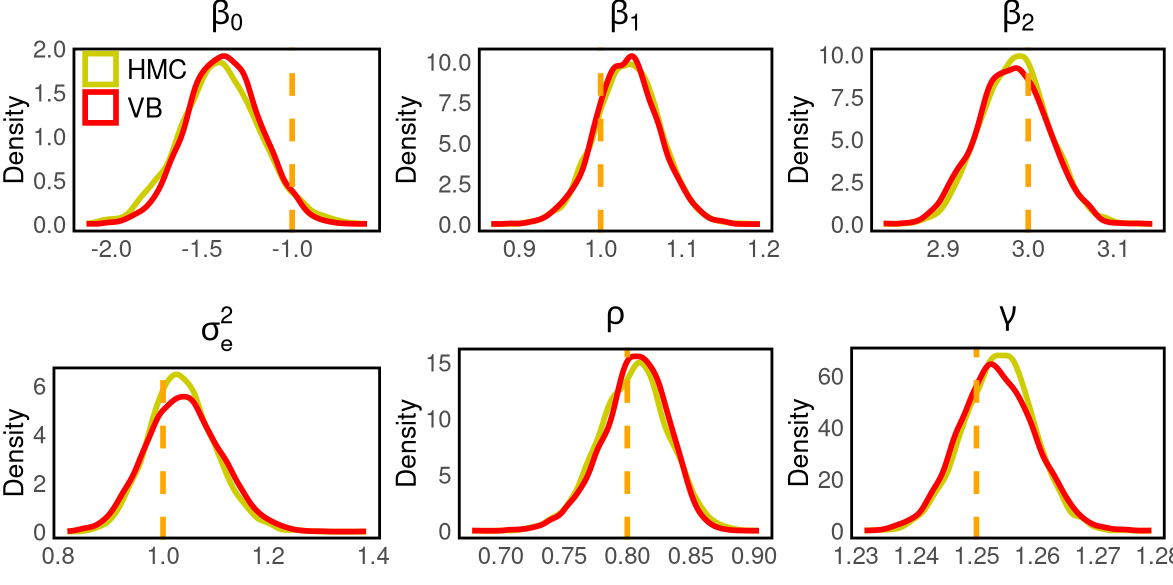} 
    \caption{Comparison of posterior densities of selected parameters in the YJ-SEM-Gau, obtained using the HMC and VB methods on the simulated dataset without missing values. The vertical line represents the true parameter value.}
    \label{fig:HMCvsVBfull}
\end{figure}

Figure~\ref{fig:HMCvsVBfull} compares the posterior densities of selected model parameters estimated using the VB and HMC algorithms. For all model parameters, the posterior densities obtained from both methods are almost indistinguishable and successfully recover the true values.


\subsection{SEMs with missing responses}
\label{NonGauSAR_sec:sim_results_miss_accu}


Now, missing values are introduced into the response variable $\textbf{y}$ under the MNAR mechanism using the logistic model described in Equation~\eqref{eq:joint.logistic.SEM_}, in order to compare the performance of the HVB algorithm presented in Section~\ref{sec:HVB} with HMC. To serve as a covariate in the logistic model, an additional variable $\textbf{x}^\ast$ is generated from a log-normal distribution with log-scale mean 0 and log-scale standard deviation 1. This simulated covariate, and the response variable $\textbf{y}$ from the YJ-SEM-Gau, are used as the predictors in the missing value model defined in Equation~\eqref{eq:joint.logistic.SEM_}. The coefficients are set as follows: $\psi_0 = -1.0$ (intercept), $\psi_{\textbf{x}^\ast} = 0.5$ (coefficient of $\textbf{x}^\ast$), and $\psi_{\textbf{y}} = -0.1$ (coefficient of $\textbf{y}$). This configuration yields approximately $50\%$ missing values in $\textbf{y}$, corresponding to $ n_u= 320$.

The HVB-NoB algorithm (without block-wise update) described in Algorithm~\ref{alg:HVB} and the HMC algorithm are used to estimate the YJ-SEM-Gau with missing data. The tuning parameter, $N_1$ of the HVB-NoB algorithm, is set to 10. The initial values for the algorithm are set as follows: the variational mean vector ($\boldsymbol{\mu}$) is initialised using the estimates of $\boldsymbol{\beta}$, $\sigma^2_{\mathbf{e}}$, and $\rho$ obtained by fitting the SEM-Gau via ML to the complete cases only, i.e., ignoring locations with missing responses. 
The initial value of the YJ transformation parameter $\gamma$ is set to 1. All fixed-effect parameters of the missing value model ($\boldsymbol{\psi}$) are initialised to 0.1. For the variational covariance matrix, all elements of $\textbf{B}$ and the diagonal elements of $\textbf{D}$ are initialised to 0.01. The missing responses ($\textbf{y}_u$) are initialised by drawing samples from the conditional distribution $p(\textbf{y}_u \mid \boldsymbol{\xi}^{(0)}, \textbf{y}_o)$, where $\boldsymbol{\xi}^{(0)}$ denotes the vector of initial parameter values, set to the corresponding initial variational means described above. We use $p = 4$ factors in the HVB-NoB algorithm, as increasing the number of factors did not lead to improved performance. The HMC algorithm is initialised similarly: initial parameter values are set in the same way as the variational means, and missing values are initialised using draws from $p(\textbf{y}_u \mid \boldsymbol{\xi}^{(0)}, \textbf{y}_o)$. Both the HVB-NoB and HMC algorithms are run for 10,000 iterations. Convergence diagnostics plots are provided in \textcolor{black}{Section~\ref{sec:sim_conv_analysis_6}} of the online supplement.

Now, we briefly explain how posterior samples of the model parameters, 
$\boldsymbol{\xi}$, $\boldsymbol{\psi}$, and the missing values 
$\boldsymbol{y}_u$, are generated in the HVB-NoB method. Once the algorithm 
converges, the set of variational parameters 
$\boldsymbol{\lambda} = \left( \boldsymbol{\mu}^\top, \mathrm{vech}(\mathbf{B})^\top, 
\mathbf{d}^\top \right)^\top$ is obtained. Given these variational parameters, 
10{,}000 draws from the variational distribution
\[
q_{\boldsymbol{\lambda}}^{0}(\boldsymbol{\xi}, \boldsymbol{\psi}) 
= N\!\left( (\boldsymbol{\xi}^\top, \boldsymbol{\psi}^\top)^\top;\, 
\boldsymbol{\mu},\, \mathbf{B}\mathbf{B}^\top + \mathbf{D}^2 \right)
\]
are generated, forming the posterior samples of $\boldsymbol{\xi}$ and 
$\boldsymbol{\psi}$. Conditional on these parameter samples, posterior samples 
of $\mathbf{y}_{u}^{(i)}$ are drawn from
\[
p(\mathbf{y}_u \mid \mathbf{y}_o, \mathbf{m}, 
\boldsymbol{\xi}^{(i)}, \boldsymbol{\psi}^{(i)}),
\]
using the MCMC steps described in Algorithm~\ref{alg:AugvbMCMCstep}, where 
$\boldsymbol{\xi}^{(i)}$ and $\boldsymbol{\psi}^{(i)}$ denote the 
$i^{\text{th}}$ posterior draws of the model parameters, for 
$i = 1, \dots, 10{,}000$.

\begin{figure}[htbp]
    \centering
    \includegraphics[width=0.8\textwidth]{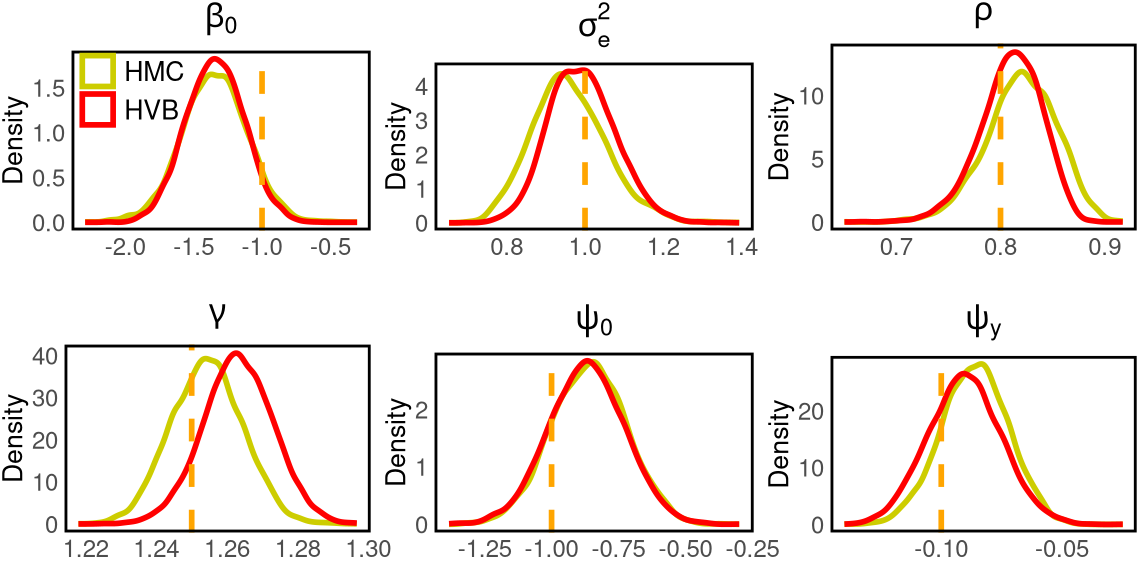} 
    \caption{Comparison of posterior densities for selected parameters of the YJ-SEM-Gau and missing data models, estimated using the HMC and HVB-NoB methods on the simulated dataset with missing values. The vertical line represents the true parameter value.}
    \label{fig:HMCvsVBmiss}
\end{figure}

Figure~\ref{fig:HMCvsVBmiss} compares the posterior densities of selected model 
parameters from the YJ-SEM-Gau and the missing data model, estimated using HVB-NoB 
and HMC on the simulated dataset with missing values. For most parameters, the 
posterior densities produced by the two methods are nearly identical. The 
posterior density of $\gamma$ estimated using HVB-NoB is slightly different 
compared to HMC, which is expected since HVB is an approximate inference method, 
whereas HMC yields asymptotically exact posterior samples. Importantly, this 
minor loss in accuracy is offset by the substantial computational efficiency 
gained with HVB-NoB, as discussed in Section~\ref{sec:sim1_summary}.

\subsection{Summary and computation time comparison}
\label{sec:sim1_summary}

Overall, the simulation results demonstrate that the proposed VB methods, both with full data and with missing data, provide accurate approximations to the posterior distributions obtained via HMC. For all key model parameters, the VB-based posterior densities closely match those from HMC, with only a slight overestimation observed for the transformation parameter $\gamma$ in the missing data setting. 

Notably, the computation times per iteration of the VB and HVB-NoB algorithms are significantly lower than those of the HMC algorithm. The computation time per iteration for the VB algorithm is approximately 0.0361 seconds, while for the HMC algorithm, it is 0.4680 seconds when estimating the YJ-SEM-Gau without missing data. The computation time per iteration for the HVB-NoB algorithm is approximately 0.0542 seconds, while for the HMC algorithm, it is 41.1658 seconds for estimating the YJ-SEM-Gau with missing data. These results demonstrate that the proposed VB and HVB-NoB methods provide accurate parameter estimates while offering substantial computational savings relative to HMC.


\section{Simulation study: Assessing the accuracy and robustness of proposed SEMs}  
\label{sec:simulationstudy-2}

This section evaluates the robustness and accuracy of the proposed SEMs and the VB methods using moderately large simulated datasets. The objective of this simulation study is to assess how well the proposed SEMs capture key characteristics of non-Gaussian spatial data, such as skewness and heavy-tailed distributions, both with full data and in the presence of missing values.

We generate datasets using a spatial weight matrix from the \texttt{spData} \texttt{R} package~\citep{spData}, based on $25,357$ houses sold between 1993 and 1998 in Lucas County, Ohio, USA (see Section~\ref{sec:real} for further details). We simulate two datasets from the YJ-SEM-t using the spatial weight matrix $\textbf{W}_{1998}$, corresponding to 4,378 houses sold in 1998:

\begin{enumerate}
    \item Dataset 1: $\gamma = 0.5$ (right skew) and $\nu = 4$ (heavy tails).
    \item Dataset 2: $\gamma = 1$ (no skew) and $\nu = 30$ (light tails).
\end{enumerate}

\begin{figure}[H]
    \centering
    \begin{minipage}{0.5\textwidth}
        \centering
        \includegraphics[width=\linewidth, height=0.2\textheight]{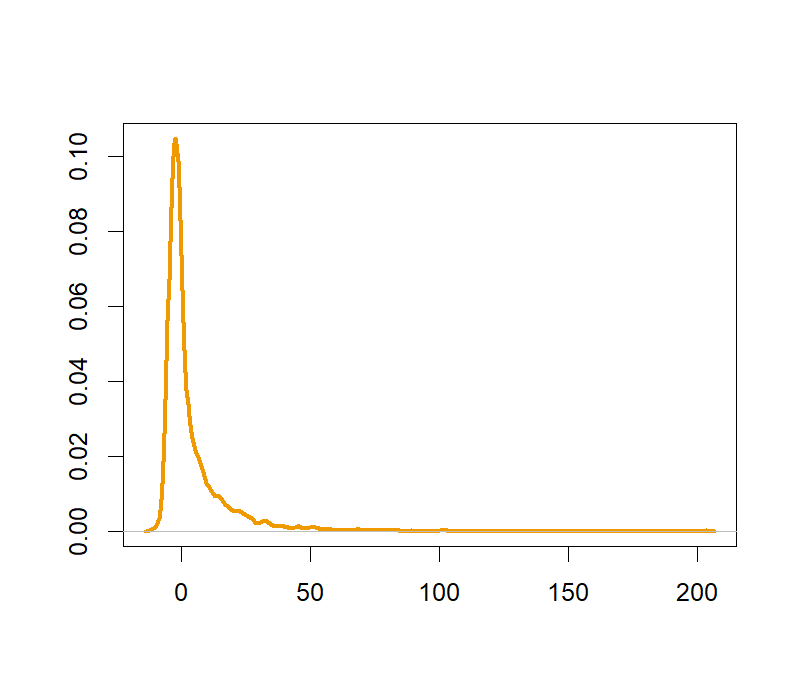} 
    \end{minipage}%
    \begin{minipage}{0.5\textwidth}
        \centering
        \includegraphics[width=\linewidth, height=0.2\textheight]{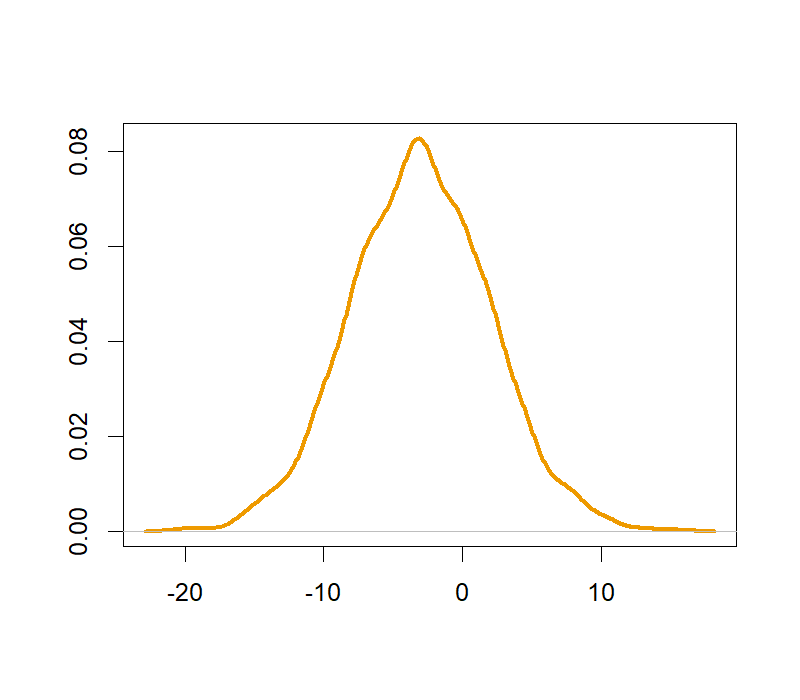} 
    \end{minipage}%
    \caption{Kernel density plots of the simulated data: The left panel displays the kernel density of the response variable from the  YJ-SEM-t with $\nu=4$ and $\gamma=0.5$, displaying right skewness and heavy tails (dataset 1). The right panel displays the kernel density of the response variable from the YJ-SEM-t with $\nu=30$ and $\gamma=1$, which appears nearly symmetric with light tails (dataset 2).}
    \label{fig:densities.sim.data}
\end{figure}

The remaining parameters for both datasets are set as follows: six fixed effect parameters ($\boldsymbol{\beta}$) are randomly drawn from a discrete uniform distribution between -3 and 3 (excluding 0), $\sigma^2_{\textbf{e}} = 0.5$, $\rho = 0.8$, and all covariates are generated from a standard normal distribution, $N(0,1)$. Figure~\ref{fig:densities.sim.data} displays kernel density plots of the response variables for both datasets.

\label{sec:sim_miss}

Section~\ref{sec:sim_full} of the online supplement fits the SEM-Gau, SEM-t, YJ-SEM-Gau, and YJ-SEM-t to simulated datasets 1 and 2 with complete data and evaluates how effectively these models, along with their VB estimation methods, capture the underlying characteristics of the datasets. In Section~\ref{sec:sim_miss_results}, missing values are introduced, and the performance of the SEMs along with the HVB estimation method (described in Section~\ref{sec:HVB}) is assessed under missing data.

\subsection{SEMs with missing responses}
\label{sec:sim_miss_results}

This section evaluates the robustness and accuracy of the proposed SEMs and the HVB method in the presence of missing responses, using simulated datasets 1 and 2. First, missing values are introduced in the response variable $\textbf{y}$ of each simulated dataset under the MNAR mechanism using the logistic regression model in Equation~\eqref{eq:joint.logistic.SEM_}. To construct this model, an additional covariate $\mathbf{x}^*$ is generated from a log-normal distribution with a log-scale mean of 0 and a log-scale standard deviation of 1, and the model is regressed on $\mathbf{x}^*$ and the response variable $\textbf{y}$. For the dataset 1, the logistic regression parameters are set to  $\psi_0 = 1$ (intercept), $\psi_{\textbf{x}^{*}} = -1$ (the coefficient for $\textbf{x}^\ast$), and $\psi_{\textbf{y}} = -0.1$ (the coefficient for $\textbf{y}$). This configuration results in approximately $40\%$ missing responses, leading to $n_u=1,729$. For the dataset 2, the intercept is set to $\psi_0 = 0.5$ (with the other coefficients unchanged), leading to approximately 40\% missing responses ($n_u = 1,774$).

We fit SEM-Gau, SEM-t, YJ-SEM-Gau, and YJ-SEM-t to these simulated datasets with missing values. As $n$ and $n_u$ are large, the HVB-AllB algorithm, which employs the MCMC steps detailed in Algorithm~\ref{alg:AugvbMCMCstepB} in Section~\ref{sec:HVB_All} of the online supplement, is utilised. The tuning parameters for the HVB-AllB algorithm are set as follows: the number of MCMC iterations ($N_1$) is set to 10, and the block size ($k^*$) is defined as $10\%$ of $n_u$, resulting in 11 blocks. 
The initial values for the variational mean vector $\boldsymbol{\mu}$ corresponding to $\boldsymbol{\beta}$, $\sigma^2_{\textbf{e}}$, $\rho$, $\gamma$, as well as the initial values for the missing responses $\mathbf{y}_u$, are specified using the procedure described in Section~\ref{NonGauSAR_sec:sim_results_miss_accu}.
For models with Student's $t$ errors, the degrees of freedom parameter $\nu$ and the set of latent variables $\boldsymbol{\tau} = (\tau_1, \dots, \tau_n)^\top$ must also be initialised. The variational mean corresponding to $\nu$ is initialised at 4. The variational means corresponding to the latent variables are initialised by sampling from the prior distribution $p(\tau_i \mid \nu^{(0)}) \sim \text{IG}(\frac{\nu^{(0)}}{2}, \frac{\nu^{(0)}}{2})$ for $i = 1, \dots, n$, where $\nu^{(0)} = 4$ denotes the initial value of $\nu$. For the variational covariance matrix, all elements of $\textbf{B}$ and the diagonal elements of $\textbf{D}$ are initialised to 0.01. 


We run the HVB-AllB algorithms for 10,000 iterations for the SEM-Gau and YJ-SEM-Gau across both simulated datasets. For the SEM-t and YJ-SEM-t, convergence generally requires more iterations due to the additional complexity introduced by the latent variables $\boldsymbol{\tau}$ and the parameter $\nu$. However, the number of required iterations varies across datasets and models. For simulated dataset 1, SEM-t is run for 20,000 iterations, whereas YJ-SEM-t achieves convergence within 10,000 iterations. For simulated dataset 2, SEM-t requires 30,000 iterations, while YJ-SEM-t is run for 20,000 iterations. 
Convergence diagnostic plots are provided in Section~\ref{sec:sim_conv_analysis_7} of the online supplement.

\subsubsection{Results for the simulated dataset 1}

This section presents the results for dataset 1 with missing values. Table~\ref{tab:sim_miss_dset1} presents the posterior means and 95\% credible intervals for some of the model parameters of the SEM-Gau, YJ-SEM-Gau, SEM-t, and YJ-SEM-t obtained using the HVB-AllB method applied to the simulated dataset 1 with missing values, along with the computation cost for one iteration. The table also includes the $\text{DIC}_5$ values for each model, calculated using the formula in Equation~\eqref{eq:DIC5}. Corresponding posterior density plots of the parameters are given in Figure~\ref{fig:densities_sk_miss} in \textcolor{black}{Section~\ref{sec:comp_rob_para_den}} of the online supplement.

We start by comparing the estimated values of the fixed effects ($\beta_0$ and ${\beta}_1$ for the SEMs and $\psi_0$ and ${\psi}_1$ for the missing data model), the variance parameter $ \sigma_{\textbf{e}}^2 $, and the spatial autocorrelation parameter $ \rho $, as these parameters are common to all four SEMs. The posterior means for $ \beta_0 $, $ \beta_1 $, $\psi_0$, $\psi_{\textbf{y}}$, and $ \rho $ are nearly identical for both YJ-SEM-Gau and YJ-SEM-t, closely matching the true values.
~In contrast, the estimates from SEM-Gau and SEM-t show significant deviations from the true values. The estimated posterior means of $ \sigma_{\textbf{e}}^2$ from SEM-Gau, SEM-t, and YJ-SEM-Gau differ considerably from the true value, while the posterior mean from YJ-SEM-t is much closer to the true value. The models incorporating the YJ transformation (YJ-SEM-Gau and YJ-SEM-t) successfully recover the true value of the parameter $\gamma$. The estimates of $\nu$ obtained from the SEM-t and YJ-SEM-t are slightly inaccurate. Additionally, the posterior means of the estimated missing values from YJ-SEM-t and YJ-SEM-Gau align more closely with the true missing values compared to those from SEM-Gau and SEM-t; see Figure~\ref{fig:sem_hsk_mean_sd_yu} in \textcolor{black}{Section~\ref{sec:comp_rob_missing_den}} of the online supplement.

\begin{table}[ht]
\centering
\small
\caption{Posterior means and 95\% credible intervals for selected parameters of different SEMs based on the simulated dataset 1 with missing data, along with the computation time (CT) in seconds for one VB iteration. The table also presents the $\text{DIC}_5$ values for each model. Parameters labelled as 'NA' indicate that they are not applicable to the corresponding model.}
\begin{tabular}{ccccc}
\hline
 & SEM-Gau & SEM-t & YJ-SEM-Gau   & YJ-SEM-t \\
\hline
${\beta}_0=-1$ & \makecell{1.9268\\ (1.6342, 2.2270)} & \makecell{1.2929\\ (1.1456, 1.4382)} & \makecell{-0.9862 \\ (-1.0256, -0.9462)}   &  \makecell{-0.9680\\ (-1.0238, -0.9123)}\\

${\beta}_1=1$ & \makecell{1.8848\\ (1.6344, 2.1381)} & \makecell{1.3230\\ (1.1675, 1.4797)} & \makecell{0.9938 \\ (0.9554, 1.0332)}  &  \makecell{0.9835 \\ (0.9338, 1.0324)}\\

$\sigma^2_{\textbf{e}}=0.5$ & \makecell{77.4912\\ (74.2316, 80.7856)} & \makecell{24.4549\\ (23.2015, 25.7503)} & \makecell{0.9994 \\ (0.9331, 1.0701)}  &  \makecell{0.7028\\ (0.6615, 0.7451)}  \\

$\rho=0.8$ & \makecell{0.2839\\ (0.2527, 0.3144)} & \makecell{0.1948\\ (0.1722, 0.2177)} & \makecell{0.8007 \\ (0.7858, 0.8149)}  &  \makecell{0.8044 \\ (0.7922, 0.8159)} \\

$\nu=4$ & NA & \makecell{3.3400\\ (3.2319, 3.4772)}  & NA & \makecell{9.5082 \\ (9.0327, 10.0109)} \\

$\gamma=0.5$ & NA & NA  & \makecell{0.4965\\ (0.4822, 0.5113)}  & \makecell{0.4990 \\ (0.4882, 0.5099)} \\

$\psi_0=1$ & \makecell{0.8251\\ (0.7048, 0.9441)} & \makecell{0.9171\\ (0.8126, 1.0203)} & \makecell{0.9337 \\ (0.8193, 1.0442)}   &  \makecell{0.9515 \\ (0.8525, 1.0522)}\\
${\psi}_{\textbf{y}}= -0.1$ & \makecell{-0.0793\\ (-0.0907, -0.0679)} & \makecell{-0.0997 \\ (-0.1178, -0.0812)} & \makecell{ -0.0794\\ (-0.0924, -0.0663)}  &  \makecell{ -0.0790\\ (-0.0891, -0.0686)}\\
\hline
$\text{DIC}_5$ & 36009.5 & 36302.7 &  18278.55 & {17855.59} \\
CT &0.5894 & 1.2250 & 0.6000 & 1.339 \\
\hline
\end{tabular}
\label{tab:sim_miss_dset1}
\end{table}

For the simulated dataset 1 with missing values in the response variable, the YJ-SEM-t provides the best fit, as indicated by the lowest $\text{DIC}_5$ value, among all four models. This result is expected, as the dataset is generated from the YJ-SEM-t, which incorporates both heavy tails and skewness. The second-best performer is YJ-SEM-Gau, which has the next lowest $\text{DIC}_5$ value. Furthermore, similar to the case without missing values discussed in \textcolor{black}{Section~\ref{sec:sim_full}} of the online supplement, the relatively small difference in the $\text{DIC}_5$ values between YJ-SEM-t and YJ-SEM-Gau suggests that YJ-SEM-Gau provides a comparably good fit for dataset 1.

\begin{figure}[H]
    \centering

    \begin{minipage}{0.49\textwidth}
        \centering
        \begin{minipage}{0.49\textwidth}
            \centering
            \includegraphics[width=\linewidth]{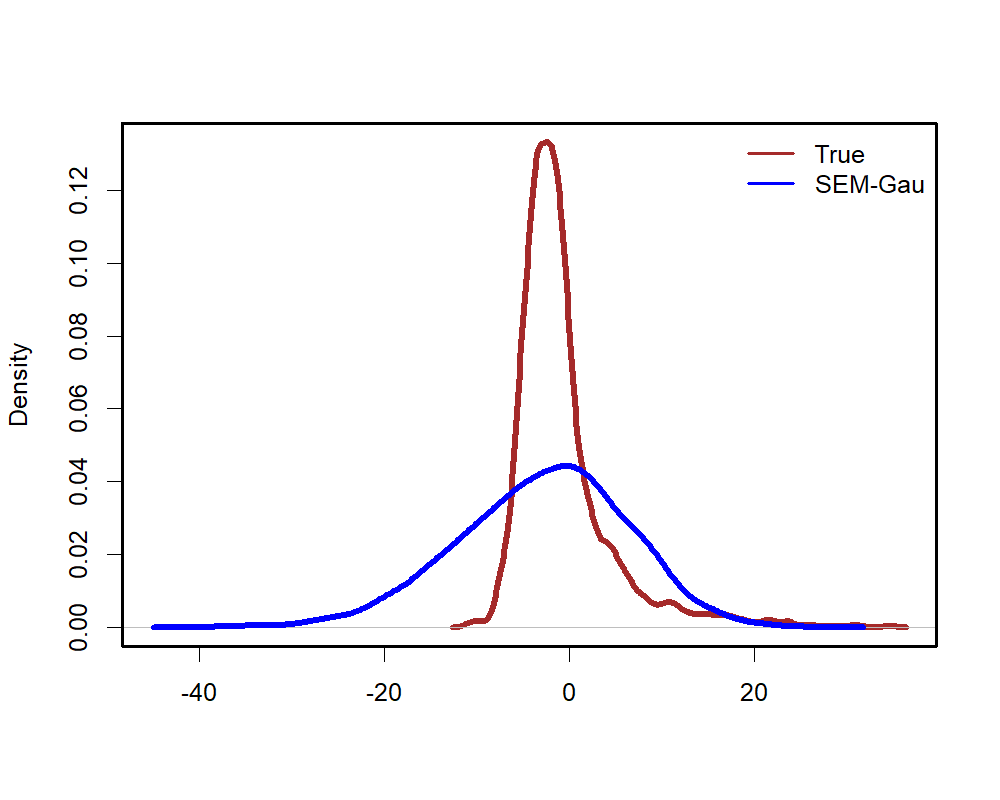}
        \end{minipage}%
        \begin{minipage}{0.49\textwidth}
            \centering
            \includegraphics[width=\linewidth]{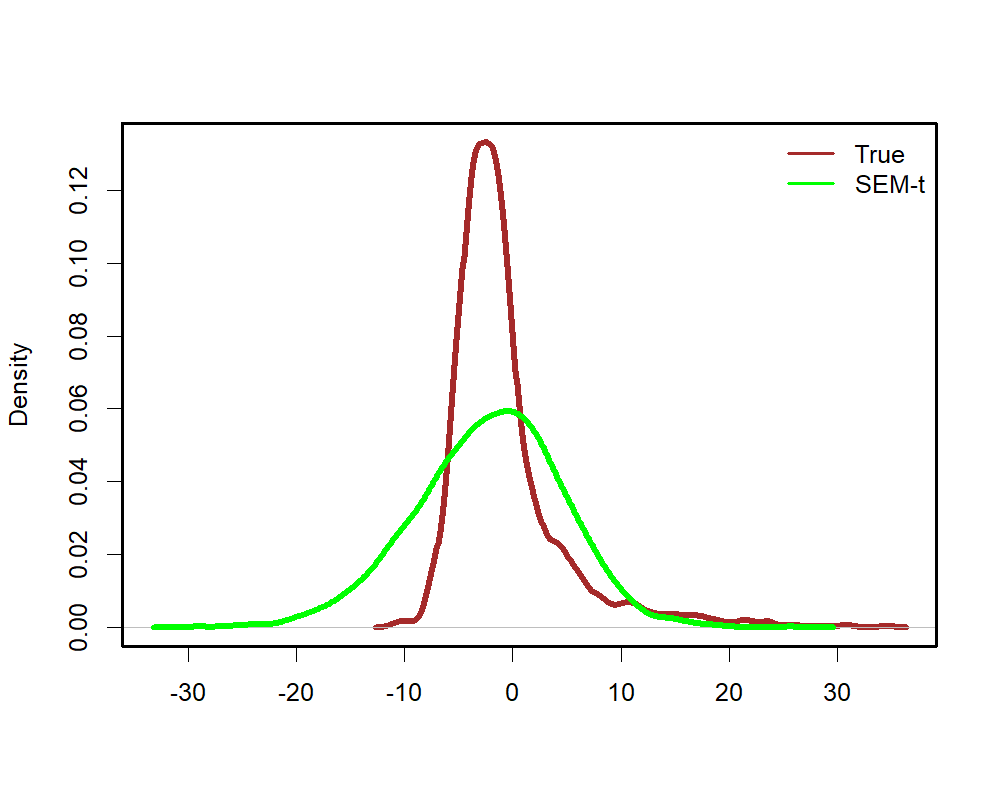}
        \end{minipage}

        \vspace{0.25cm}

        \begin{minipage}{0.49\textwidth}
            \centering
            \includegraphics[width=\linewidth]{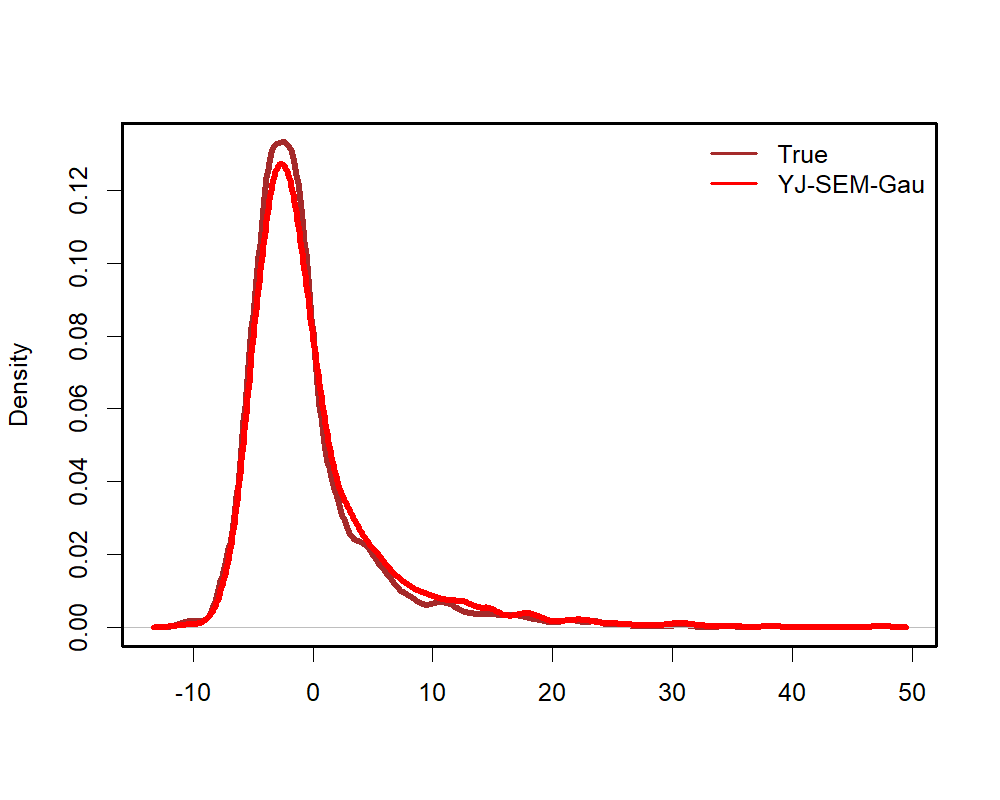}
        \end{minipage}%
        \begin{minipage}{0.49\textwidth}
            \centering
            \includegraphics[width=\linewidth]{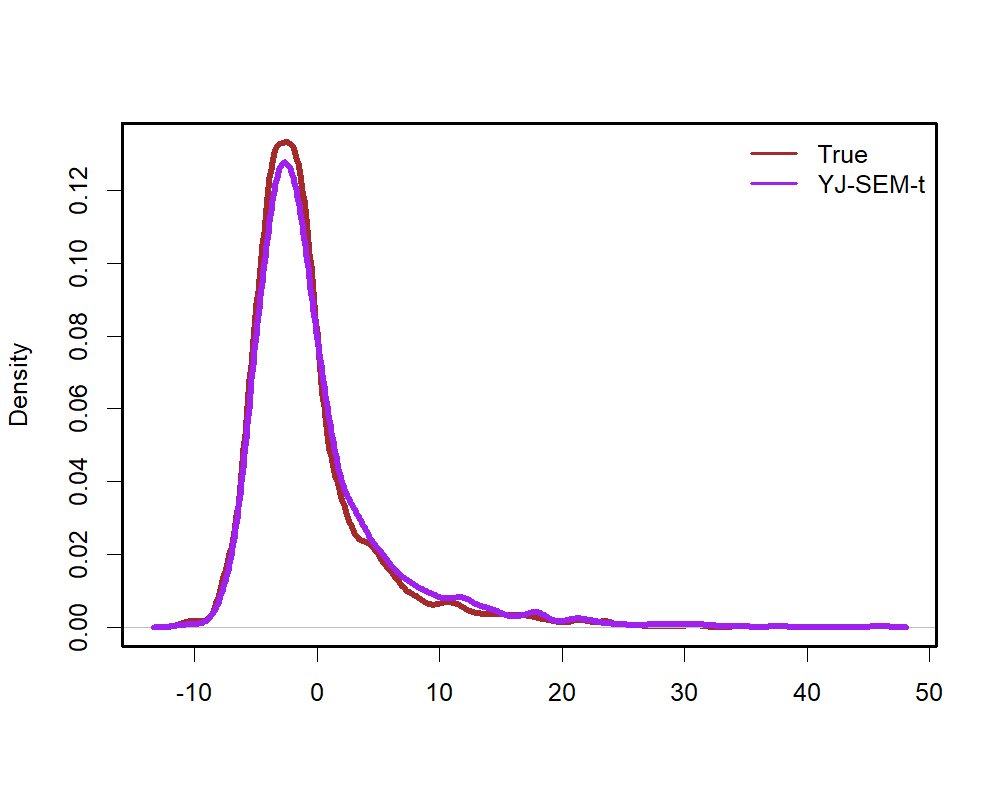}
        \end{minipage}
    \end{minipage}%
    \hfill
    \begin{minipage}{0.49\textwidth}
        \centering
        \includegraphics[width=\linewidth]{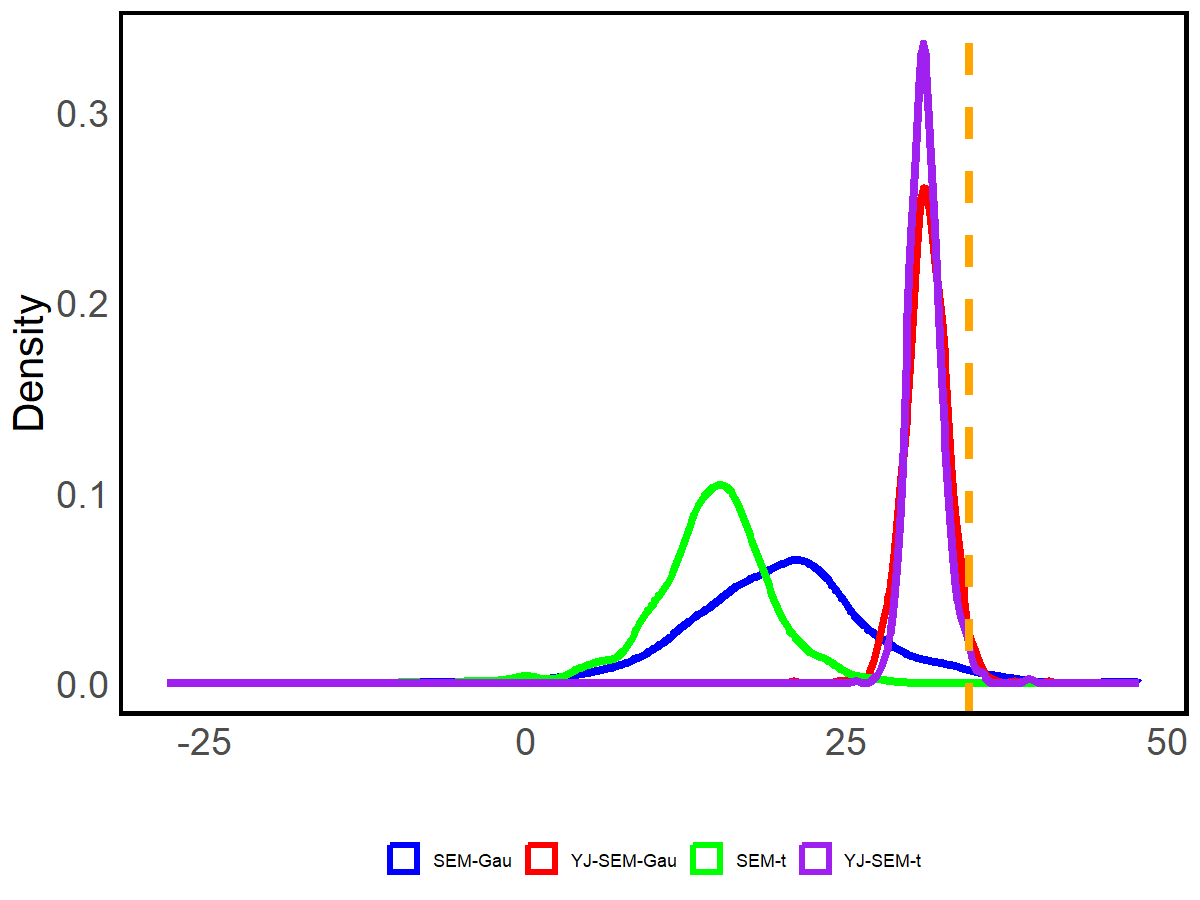}
    \end{minipage}

    \caption{
        Left panel: The kernel density of the true missing values ($\textbf{y}_u$) and the kernel densities of posterior means of the missing values obtained using the HVB-AllB method for different SEMs for the simulated dataset 1 with missing values. Right panel: The posterior density of the maximum missing value, $\textrm{max}(\textbf{y}_u)$, obtained by different SEMs for the simulated dataset 1 with missing values. The true value of the maximum missing value is indicated by the vertical line. 
    }
    \label{fig:sim_sk_miss_densities}
\end{figure}

The left four panels of Figure~\ref{fig:sim_sk_miss_densities} compare the kernel density of the true missing values ($\textbf{y}_u$) with the kernel densities of the posterior means of the missing values estimated under different SEMs for the simulated dataset 1. The densities from both YJ-SEM-Gau and YJ-SEM-t are nearly identical and closely align with the density of the true missing values. The right panel of Figure~\ref{fig:sim_sk_miss_densities} presents the posterior density of the maximum missing value, $\textrm{max}(\textbf{y}_u)$, estimated under each SEM. The true value of the maximum missing value lies closer to the posterior densities produced by YJ-SEM-Gau and YJ-SEM-t than to those produced by SEM-Gau and SEM-t. This behaviour is expected, as YJ-SEM-Gau and YJ-SEM-t provide parameter estimates that are closer to the true values and yield more comparable $\text{DIC}_5$ values (see Table~\ref{tab:sim_miss_dset1}).


\subsubsection{Results for the simulated dataset 2}

This section presents the results for dataset 2 with missing values. Table~\ref{tab:sim_miss_dset2} reports the posterior means and $95\%$ credible intervals for some of the model parameters of the SEM-Gau, SEM-t, YJ-SEM-Gau, and YJ-SEM-t, obtained using the HVB-AllB method on the simulated dataset 2 with missing values, along with the computation cost for one VB iteration. The table also presents the corresponding $\text{DIC}_5$ values for each model. Posterior density plots for these parameters are provided in Figure~\ref{fig:densities_norm_miss} in \textcolor{black}{Section~\ref{sec:comp_rob_para_den}} of the online supplement. The estimated posterior means for $\beta_0$, $\beta_1$,  $\sigma^2_{\textbf{e}}$, $\rho$, $\psi_0$, and $\psi_{\textbf{y}}$ across all four models closely match their true values. Similarly, the estimated posterior mean of $\gamma$ for the YJ-SEM-Gau and YJ-SEM-t is also close to the true value. The parameter $\nu$ is accurately estimated for both the SEM-t and YJ-SEM-t. In addition, all four models produce posterior means for the estimated missing values that closely match the true values; see Figure~\ref{fig:sem_norm_mean_sd_yu}  in \textcolor{black}{Section~\ref{sec:comp_rob_missing_den}} of the online supplement.

\begin{table}[ht] 
\centering
\small
\caption{Posterior means and 95\% credible intervals for selected parameters of different SEMs based on the simulated dataset 2 with missing data, along with the computation time (CT) in seconds for one VB iteration. The table also presents the $\text{DIC}_5$ values for each model. Parameters labelled as 'NA' indicate that they are not applicable to the corresponding model.}
\begin{tabular}{ccccc}
\hline
 & SEM-Gau & SEM-t & YJ-SEM-Gau   & YJ-SEM-t \\
\hline
${\beta}_0= -3$ & \makecell{-3.0131 \\ (-3.0495, -2.9764)} & \makecell{-3.0150 \\ (-3.0497, -2.9792)} & \makecell{ -2.9511\\ (-2.9820, -2.9199)}  &  \makecell{-2.9467\\ (-2.9767, -2.9161)} \\

${\beta}_1=2$ &\makecell{ 1.9942\\ (1.9582, 2.0309)} & \makecell{1.9884\\ (1.9605, 2.0165)} & \makecell{ 1.9803\\ (1.9621, 1.9979)}  &  \makecell{1.9834\\ (1.9568, 2.0096)}\\

$\sigma^2_{\textbf{e}}=0.5$ & \makecell{0.5528 \\ (0.5222, 0.5849)} & \makecell{0.5143\\ (0.4892, 0.5406)} & \makecell{ 0.5450\\ (0.5155, 0.5759)}  &  \makecell{0.5096\\ (0.4826, 0.5377 )} \\

$\rho=0.8$ & \makecell{0.8002 \\ (0.7857, 0.8141)} & \makecell{0.7990\\ (0.7874, 0.8101)} & \makecell{ 0.8003\\ (0.7892, 0.8113)}  &  \makecell{0.7971\\ (0.7841, 0.8095)}  \\

$\nu=30$ & NA & \makecell{30.2880\\ (29.1096, 31.4718)}   & NA & \makecell{29.7195\\ (27.9820, 31.6032)}   \\

$\gamma=1$ & NA & NA  & \makecell{1.0032\\ (0.9933, 1.0129)}   & \makecell{1.0017\\ (0.9961, 1.0074)}  \\

$\psi_0=0.5$ & \makecell{ 0.4949\\ (0.3847, 0.6044)} & \makecell{ 0.5036\\ (0.4219,  0.5841)} & \makecell{ 0.5477\\ (0.4319, 0.6619)}  &  \makecell{0.5467\\ (0.4742, 0.6204)}  \\
${\psi}_{\textbf{y}}=-0.1$ &\makecell{-0.0956\\ (-0.1210, -0.0706)} & \makecell{-0.1090\\ (-0.1320, -0.0864)} & \makecell{-0.0965 \\ (-0.1106, -0.0827)}  &  \makecell{-0.0972\\ (-0.1137, -0.0809)}\\
\hline
$\text{DIC}_5$ & 14713.89  &  14221.78 &  14695.32   & {14164.86} \\
CT & 0.5755 & 1.2166& 0.6048 & 1.3486 \\
\hline
\end{tabular}
\label{tab:sim_miss_dset2}
\end{table}

Based on the $\text{DIC}_5$ values, the YJ-SEM-t provides the best fit for simulated dataset 2 with missing values, while the SEM-t has the second-lowest $\text{DIC}_5$, indicating it is the second-best performing model.

The kernel densities of the posterior means of the missing values ($\textbf{y}_u$) estimated under all four SEMs closely match the kernel density of the true missing values for simulated dataset 2; see left panel of Figure~\ref{fig:sim_norm_miss_densities} in \textcolor{black}{Section~\ref{sec:comp_rob_missing_den}} of the online supplement. Similarly, for each SEM, the posterior density of the maximum missing value, 
$\max(\mathbf{y}_u)$, contains the true value, 
and the densities across the different SEMs are nearly identical; see the right 
panel of Figure~\ref{fig:sim_norm_miss_densities} in 
Section~\ref{sec:comp_rob_missing_den} of the online supplement. These results are consistent with expectations, as all four models produce similar parameter estimates and $\text{DIC}_5$ values, with estimates generally close to the true parameter values (see Table~\ref{tab:sim_miss_dset2}).

\subsection{Summary and computation time comparison}

This simulation study evaluates how well the proposed SEMs capture key characteristics of non-Gaussian spatial data, both with fully observed data and in the presence of missing values, using two simulated datasets: dataset 1 exhibits skewness and heavy tails, while dataset 2 is closer to Gaussian. This section summarises the findings of the simulation study with missing data, while Section~\ref{online_sec:summary_sim} of the online supplement summarises the corresponding results for the full data.


As shown in Tables~\ref{tab:sim_miss_dset1} and~\ref{tab:sim_miss_dset2}, the HVB-AllB algorithm applied to the SEM-t and YJ-SEM-t, both of which assume Student's $t$-distributed errors, requires more computation time per iteration compared to SEM-Gau and YJ-SEM-Gau. This is due to the added complexity of VB optimisation: these models introduce an additional latent variable vector of length $n$ and include the degrees of freedom parameter $\nu$, increasing the dimensionality of the optimisation. In addition, SEM-t and YJ-SEM-t typically require more VB iterations to converge. Convergence diagnostics plots in Section~\ref{sec:sim_conv_analysis_7} of the online supplement illustrate this behaviour. Consequently, HVB-AllB for SEM-t and YJ-SEM-t is generally more time-consuming than for SEM-Gau and YJ-SEM-Gau, both because each iteration is slower and more iterations are needed to achieve convergence.


In terms of performance, YJ-SEM-Gau and YJ-SEM-t are particularly effective 
for modelling skewed and heavy-tailed data. For simulated dataset~1, 
YJ-SEM-t attains the lowest $\textrm{DIC}_5$ value, indicating the best fit, 
while YJ-SEM-Gau achieves the second-lowest value, which is close to that of 
YJ-SEM-t. For simulated dataset~2, which is closer to Gaussian, all four 
models recover the true parameter values accurately and yield nearly identical 
$\textrm{DIC}_5$ values, with YJ-SEM-t again achieving the lowest. Although 
estimating YJ-SEM-t incurs substantially higher computational cost compared to 
SEM-Gau (the conventional SEM with Gaussian errors), YJ-SEM-Gau maintains a 
computation time comparable to SEM-Gau while providing considerable improvements 
when modelling non-Gaussian data. Similar patterns are observed in the analysis 
of the fully observed datasets, as presented in Section~\ref{sec:sim_full} of 
the online supplement.

\section{Real data application}
\label{sec:real}

We apply the proposed SEMs and VB methods to a dataset of single-family homes sold in Lucas County, Ohio, USA. This dataset is available in the \texttt{R} package \texttt{spData}~\citep{spData}. The data set includes various house characteristics: house age, the lot size in square feet, the number of rooms, the total living area in square feet, the number of bedrooms, and binary indicators for each year from 1993 to 1998 to represent the year of house sale. For our analysis, we focus on the house price data from 1998, which contains $4,378$ observations. We call this dataset Lucas-1998-HP.
House prices are typically positive and often exhibit skewness. To transform house prices into the entire real line and reduce skewness, researchers commonly apply logarithmic transformations before modelling~\citep{lesage2004models,math12233870}. Accordingly, we apply the natural logarithmic transformation to the house prices in the Lucas-1998-HP dataset. However, even after the transformation, the resulting log-prices remain left-skewed; see Figure~\ref{fig:densities_realdata} in Section~\ref{sec:online_real} of the online supplement.


First, the proposed VB algorithm (Algorithm~\ref{alg:SAGfull} in Section~\ref{sec:VB_full} of the online supplement) is used to estimate all four SEMs using the full Lucas-1998-HP dataset. Next, missing values are introduced into the response variable, and the SEMs are re-estimated using the proposed HVB-AllB algorithm (Algorithm~\ref{alg:HVB}, together with the MCMC steps in Algorithm~\ref{alg:AugvbMCMCstepB} in Section~\ref{sec:HVB_All} of the online supplement). This design enables a direct comparison between estimates obtained from the full dataset and those obtained in the presence of missing values, with the full-data estimates treated as the ground truth. The set of predictor variables includes various powers of house age ($\text{age}$, $\text{age}^2$, and $\text{age}^3$), the natural logarithm of the lot size in square feet ($\ln(\text{lotsize})$), the number of rooms (\text{rooms}), the natural logarithm of the total living area in square feet ($\ln(\text{LTA})$), and the number of bedrooms (beds). The natural logarithm of house price in hundreds of thousands of US dollars is selected as the response variable. Results for the complete dataset (without missing values) are presented in Section~\ref{sec:real_full}, while Section~\ref{sec:real_miss} provides the findings with missing data.

\subsection{Full data SEMs}
\label{sec:real_full}

This section presents the results of fitting all four SEMs to the full Lucas-1998-HP dataset. The initial values for the VB algorithms are set following the same approach used in the simulation studies in Sections~\ref{sec:simulationstudy-1} and~\ref{sec:simulationstudy-2}, and we use $p=4$ factors. Consistent with the simulation results with full data (see Section~\ref{sec:sim_full} of the online supplement), the models with Student’s $t$ errors require more iterations to achieve convergence. Accordingly, the VB algorithms are run for 20,000 iterations for SEM-Gau and YJ-SEM-Gau, and 75,000 iterations for SEM-t and YJ-SEM-t. Convergence is assessed visually using the trajectories of the variational means, with the diagnostic plots provided in Figure~\ref{fig:con_real_full} in Section~\ref{sec:real_conv_analysis} of the online supplement.


The estimated posterior means and $95\%$ credible intervals for selected model parameters across different SEMs, along with their corresponding DIC values, are summarised in Table~\ref{tab:real_full}. The posterior density plots for these parameters are provided in Figure~\ref{fig:para_densities_real_full} in Section~\ref{sec:online_real} of the online supplement. According to the $\text{DIC}_1$ and $\text{DIC}_2$ values, the YJ-SEM-t is the most appropriate SEM for the Lucas-1998-HP dataset, followed by the YJ-SEM-Gau as the second most suitable.

\begin{table}[ht]
\centering
\small
\caption{Posterior means and 95\% credible intervals for selected model parameters of different SEMs for the Lucas-1998-HP dataset (without missing values). The table also includes DIC values for each model. Parameters labelled as 'NA' indicate that they are not applicable to the corresponding model.  
}
\begin{tabular}{ccccc}
\hline
 & SEM-Gau & SEM-t & YJ-SEM-Gau   & YJ-SEM-t \\
\hline
$intercept$ & \makecell{ -0.4336\\ (-0.4469, -0.4207)} & \makecell{ -0.3865\\ (-0.3979, -0.3752)}  & \makecell{-0.3020\\ (-0.3278, -0.2765)}   &  \makecell{-0.2849 \\ (-0.3087, -0.2612)}\\
$\beta_{\text{age}}$ & \makecell{ 0.1650\\ (0.0557, 0.2742)} & \makecell{-0.4197 \\ (-0.5236, -0.3162)}  & \makecell{-0.1643 \\ (-0.2926, -0.0321)} & \makecell{-0.3201 \\ (-0.3984, -0.2405)}  \\

$\beta_{\text{rooms}}$ & \makecell{0.0068\\(-0.0176, 0.0313)} & \makecell{0.0188\\(-0.0018, 0.0395)}   & \makecell{ 0.0311\\(-0.0281, 0.0893)}&\makecell{0.0238\\(-0.0056, 0.0524)} \\
$\beta_{\text{log(lotsize)}}$ & \makecell{0.1614 \\(0.1480, 0.1752)} & \makecell{0.1193\\(0.1033, 0.1350)}  & \makecell{0.1441\\(0.1068, 0.1815)} &\makecell{0.1306\\(0.1192, 0.1420)} \\
$\sigma^2_{\textbf{e}}$ & \makecell{0.1577\\ (0.1516, 0.1641)} & \makecell{  0.0670\\ (0.0633, 0.0708)}  & \makecell{0.1031\\ (0.0967, 0.1099)} & \makecell{  0.0695\\ (0.0647, 0.0744)}  \\

$\rho$ & \makecell{0.6291 \\ (0.5605, 0.6902)}  & \makecell{ 0.5475\\ (0.4900, 0.6015)} & \makecell{ 0.6019\\ (0.5236, 0.6712)}  & \makecell{ 0.5990\\ (0.5261, 0.6651)}  \\

$\nu$ & NA & \makecell{ 3.0249\\ (3.0000, 3.1324)} & NA & \makecell{ 7.9482\\ (7.5327, 8.3940)}  \\

$\gamma$ & NA & NA  & \makecell{1.5264 \\ (1.4783, 1.5728)}  & \makecell{ 1.5529 \\ (1.5145, 1.5904)} \\
\hline

$\text{DIC}_1$ & 4470.316 &4878.664 &  4329.431 & {3720.687}\\
$\text{DIC}_2$ & 4469.655  & 5031.204  & 4332.07     & {3766.714} \\

\hline
\end{tabular}
\label{tab:real_full}
\end{table}


{We now compare estimates of key parameters across different SEMs. A crucial aspect of SEM estimation is identifying the underlying spatial correlation, represented by the parameter $\rho$. For the SEM-Gau, YJ-SEM-Gau, and YJ-SEM-t, the estimated values of $\rho$ are relatively close, with posterior mean values of 0.6291, 0.6019, and 0.5990, respectively. This suggests a moderate spatial correlation in the Lucas-1998-HP dataset. However, for the SEM-t, the estimated $\rho$ is noticeably lower, with a posterior mean value of 0.5475, compared to the other three models.
}

Similar to linear regression models, the fixed effects $\boldsymbol{\beta}$ in SEMs represent the influence of covariates on the response variable, which in this case is the logarithm of the house price. {An interesting disparity arises in the estimated posterior mean of the fixed effect for the covariate 'age' ($\beta_{\text{age}}$). For the SEM-Gau, $\beta_{\text{age}}$ is positive, while for the other three models, $\beta_{\text{age}}$ is negative. This discrepancy reflects a key limitation of SEM-Gau: it is misspecified for this dataset because it cannot adequately capture important features of the response variable, such as skewness and heavy tails. As a result, SEM-Gau produces misleading estimates for the effect of house age, suggesting an increase in price rather than the decrease indicated by the more flexible models.}
Similarly, the estimated posterior means of the fixed effect for the number of rooms ($\beta_{\text{rooms}}$) are 0.0068, 0.0188, 0.0311, and 0.0238 for SEM-Gau, SEM-t, YJ-SEM-Gau, and YJ-SEM-t, respectively, showing that the effect of the number of rooms on house prices varies across models. Notably, YJ-SEM-Gau shows the strongest positive influence, while SEM-Gau exhibits the weakest, further reflecting the consequences of misspecification.

Additionally, the SEM-t, YJ-SEM-Gau, and YJ-SEM-t provide further insights into the characteristics of the dataset. The posterior means of $\nu=3.0249$ for SEM-t and $7.9482$ for YJ-SEM-t suggest that the log-transformed house price distribution exhibits heavy tails. Furthermore, the posterior means of $\gamma$, $1.5264$ for YJ-SEM-Gau and $1.5529$ for YJ-SEM-t, indicate left skewness in the response variable, effectively capturing its asymmetry. These distributional features are visually evident in the kernel density plot of the response variable (see Figure~\ref{fig:densities_realdata} in Section~\ref{sec:online_real} of the online supplement).

\subsection{SEMs with missing responses}
\label{sec:real_miss}

This section analyses the Lucas-1998-HP dataset with missing values in the response variable (logarithm of the house price). Starting from the complete dataset, missing responses are generated using the logistic regression model in Equation~\eqref{eq:joint.logistic.SEM_}, with the age of the house and the response variable $\textbf{y}$ (from the SEMs) serving as covariates. The missing data model parameters are set to $\psi_0=0.1$, $\psi_{\textbf{x}^{*}}=0.1$, and $\psi_{\textbf{y}}=-0.1$. This configuration results in approximately 50\% of the responses being missing, yielding $n_u = 2{,}348$. 

All four SEMs are fitted to the Lucas-1998-HP dataset with missing values using the HVB method. Since both $n$ and $n_u$ are moderately large, the HVB-AllB algorithm is employed. The tuning parameters for the algorithm are set as follows: the number of MCMC iterations ($N_1$) is set to 10, and the block size ($k^\ast$) is chosen as 10\% of $n_u$, resulting in 11 blocks. The initial values for the HVB-AllB algorithm are set in the same manner as in the simulation study presented in Section~\ref{sec:sim_miss_results}. The number of factors is set to 4 (i.e., $p=4$). Similar to the simulation study with missing data described in Section~\ref{sec:sim_miss_results}, models with Student’s $t$ errors require a relatively high number of iterations to converge. Accordingly, the HVB-AllB algorithm is run for 10,000 iterations for both the SEM-Gau and YJ-SEM-Gau, 40,000 iterations for the SEM-t, and 20,000 iterations for the YJ-SEM-t. Convergence plots can be found in Figure~\ref{fig:con_real_miss} in~\textcolor{black} {Section~\ref{sec:real_conv_analysis}} of the online supplement.

The estimated posterior means and 95\% credible intervals for selected model parameters across different models, along with their corresponding $\text{DIC}_5$ values, are summarised in Table~\ref{tab:real_miss}. The posterior density plots for these parameters are provided in Figure~\ref{fig:para_densities_real_miss} in \textcolor{black}{Section~\ref{sec:online_real}} of the online supplement. According to the $\text{DIC}_5$ values, the YJ-SEM-t is the most appropriate SEM for the Lucas-1998-HP dataset with missing data, followed by the YJ-SEM-Gau as the second-best model.

\begin{table}[ht]
\centering
\small
\caption{Posterior means and 95\% credible intervals for selected parameters of different SEMs for the Lucas-1998-HP dataset with missing data. The table also includes $\text{DIC}_5$ values for each model. Parameters labelled as 'NA' indicate that they are not applicable to the corresponding model.   
}
\begin{tabular}{ccccc}
\hline
 & SEM-Gau & SEM-t & YJ-SEM-Gau   & YJ-SEM-t \\
\hline
$intercept$ & \makecell{-0.4189\\ (-0.4413, -0.3960)} &\makecell{-0.3721\\ (-0.3886, -0.3562)}   & \makecell{-0.2856\\ (-0.3008, -0.2703)}    &  \makecell{-0.2752\\ (-0.2956, -0.2544)} \\
$\beta_{\text{age}}$ & \makecell{0.1935\\ (0.0335, 0.3510)} &\makecell{-0.4640\\ (-0.5990, -0.3324)}   & \makecell{-0.1473\\ (-0.2598, -0.0345)}    &  \makecell{-0.3768\\ (-0.4870, -0.2687)} \\

$\beta_{\text{rooms}}$ & \makecell{0.0010\\ (-0.0409, 0.0432)} &\makecell{0.0179\\ (-0.0132, 0.0491)}   & \makecell{0.0341 \\ (0.0021, 0.0655)}    &  \makecell{0.0438\\ (0.0053, 0.0816)} \\

$\beta_{\text{log(lotsize)}}$ & \makecell{0.1360\\ (0.1097, 0.1620)} &\makecell{0.1124 \\ (0.0952, 0.1293)}   & \makecell{ 0.1173\\ (0.1020, 0.1323)}    &  \makecell{0.1219 \\ (0.0786, 0.1651)}  \\

$\sigma^2_{\textbf{e}}$ & \makecell{0.1463\\ (0.1378, 0.1553)} &\makecell{0.0696\\ (0.0649, 0.0746)}   & \makecell{0.0894\\ (0.0847, 0.0943)}    &  \makecell{0.0657 \\ (0.0617, 0.0699)} \\

$\rho$ & \makecell{ 0.5872\\ (0.4810, 0.6818)} &\makecell{0.5686\\ (0.4553, 0.6695)}   & \makecell{0.5494\\ (0.4337, 0.6502)}    &  \makecell{0.5377\\ (0.4325  0.6322)} \\

$\nu$ & NA & \makecell{4.1165\\ (3.9476, 4.3046)} & NA & \makecell{ 10.6394 \\ (10.2060, 11.1003)}   \\

$\gamma$ & NA & NA  & \makecell{ 1.5814\\ (1.5297, 1.6299)}   & \makecell{1.5998\\ (1.5552, 1.6415)} \\

$\psi_0=0.1$ & \makecell{0.1128 \\ (0.0193, 0.2064)} &\makecell{0.2075\\ (0.1361, 0.2806)}   & \makecell{0.1110\\ (0.0349, 0.1865)}    &  \makecell{0.1317\\ (0.0595, 0.2033)} \\
\hline
$\text{DIC}_5$ & 8074.382 &  7734.356  & 7309.394  & {6657.172}\\

\hline
\end{tabular}
\label{tab:real_miss}
\end{table}

The estimated parameters from different SEMs are now briefly discussed. For the spatial autocorrelation parameter, $\rho$, the posterior mean estimates across all SEMs are slightly lower than those obtained using the full dataset without missing values, except under the SEM-t (see Table~\ref{tab:real_full}). However, these estimates still indicate a moderate spatial correlation in the {Lucas-1998-HP} dataset, consistent with the conclusions drawn from SEMs fitted without missing values in Section~\ref{sec:real_full}. The posterior mean estimates for the fixed effects ($\boldsymbol{\beta}$) in the SEMs lead to the same conclusions as those drawn from the Lucas-1998-HP dataset without missing values. For example, the posterior mean estimate of $\beta_{\text{age}}$ in the SEM-Gau suggests a positive effect of age on house prices, whereas the estimates from the other three models indicate a negative effect. Moreover, the estimated posterior means of intercept in the missing data model, ${\psi}_0$, are close to the true value of $0.1$ for all models except SEM-t. The posterior mean estimates of $\nu$ for the SEM-t and YJ-SEM-t indicate moderate to strong heavy-tailed behaviour. Similarly, the estimated posterior mean values of $\gamma$ for the YJ-SEM-Gau and YJ-SEM-t suggest left skewness in the response. These conclusions are consistent with the findings from the analysis without missing data discussed in Section~\ref{sec:real_full}.





\section{Conclusion\label{sec:conclusion}}

Our article introduces three novel simultaneous autoregressive (SAR) models for non-Gaussian spatial data. Specifically, we extend the conventional spatial error model (SEM-Gau), a widely used SAR model that assumes Gaussian errors, to accommodate skewed and heavy-tailed features of real datasets. The novel SEMs are SEM-t, YJ-SEM-Gau, and YJ-SEM-t. The SEM-t assumes that the error terms follow a Student’s $t$ distribution, allowing for heavier tails than the Gaussian case. The YJ-SEM-Gau retains Gaussian errors but applies the Yeo–Johnson (YJ) transformation to handle asymmetric response variables. The YJ-SEM-t combines these approaches by assuming Student’s $t$ errors while incorporating the YJ transformation. These extensions improve the robustness of SAR models for analysing non-Gaussian spatial datasets. Additionally, we develop efficient variational Bayes (VB) methods to estimate these models with and without missing data. In cases with missing data, we assume that the responses are missing under the missing not at random (MNAR) mechanism. Standard VB methods are inadequate for accurately estimating these SEMs in the presence of missing data; therefore, we employ hybrid VB (HVB) algorithms to address this challenge. The posterior distributions obtained from the proposed VB and HVB methods are compared with those of the exact Hamiltonian Monte Carlo (HMC) method, demonstrating both accuracy and substantial computational gains. Although the proposed model extensions are illustrated using the SEM, both the modelling framework and the associated VB/HVB inference methods are general and can be readily adapted to other SAR-type models.

The empirical results show that: (1) The VB and HVB methods yield posterior density estimates for most model parameters that are similar to those obtained using the HMC method when estimating SEMs with and without missing values; (2) For models with Student’s $t$ errors (SEM-t and YJ-SEM-t), the posterior approximation of the degrees of freedom parameter ($\nu$) from the VB and HVB methods exhibits slight deviations compared to the true value; (3) For both VB and HVB algorithms, models with Student’s $t$ errors require more computation time per iteration than models with Gaussian errors (SEM-Gau and YJ-SEM-Gau); (4) In general, models with Gaussian errors tend to converge in fewer iterations compared to those with Student’s $t$ errors; (5) YJ-SEM-t is able to model complex response variables with skewness and heavy tails, and it provides the best fit to the Lucas-1998-HP dataset, both with and without missing data.

Future work could explore more flexible transformation families, such as Tukey’s g-and-h transformation~\citep{tukey1977modern,martinez1984gandh}, which jointly accommodate skewness and heavy tails. In addition, while the proposed non-Gaussian SEMs are developed for univariate spatial responses, extending the framework to multivariate and spatio-temporal settings would enable joint modelling of multiple non-Gaussian variables and dynamic processes. These extensions would substantially broaden the applicability of the models to a wider range of real-world spatial and spatio-temporal datasets.

\begin{singlespace}
\bibliographystyle{apalike}
\bibliography{ref.bib}
\end{singlespace}

\pagebreak

\renewcommand{\thefigure}{S\arabic{figure}} 

\renewcommand{\theequation}{S\arabic{equation}} 

\renewcommand{\thetable}{S\arabic{table}}

\renewcommand{\thealgorithm}{S\arabic{algorithm}}

\renewcommand{\thesection} 
{S\arabic{section}}

\doublespacing

\section*{Online Supplement for Bayesian Inference for Non-Gaussian Simultaneous Autoregressive Models with Missing Data}

\setcounter{page}{1} 
\setcounter{section}{0} 

\setcounter{equation}{0} 
\setcounter{table}{0} 
\setcounter{figure}{0} 
\setcounter{algorithm}{0} 

We use the following notation in the online supplement. Equation~(1), Table~1,
Figure~1, and Algorithm~1, etc, refer to the main paper, while Equation~(S1),
Table~S1, Figure~S1,  and Algorithm~S1, etc, refer to the online supplement.


\section{Yeo and Johnson transformation and its derivatives}
\label{sec:YJ_derivatives}
This section provides the Yeo and Johnson (YJ) transformation~\citep{yeo2000new} and its derivatives, which are used to construct the YJ-SEM-Gau and YJ-SEM-t discussed in Section~\ref{sec:SEM_YJ} of the main paper.

Let the random variable $y_i$ be asymmetric. The YJ transformation, denoted $t_{\gamma}(y_i)$, is often applied to reduce asymmetry and make $y_i$ more symmetric. It is defined as:

\begin{equation}
y_i^*=t_{{\gamma}}(y_i) =
\begin{cases} 
\frac{(y_i + 1)^\gamma - 1}{\gamma} & \text{if } y_i \geq 0, \\
-\frac{(-y_i + 1)^{2 - \gamma} - 1}{2 - \gamma} & \text{if } y_i < 0,
\end{cases}
\end{equation}
\noindent where $0 < \gamma < 2$.

The density function of SEMs with YJ transformations includes the derivative of the YJ transformation with respect to $y_i$, as shown in Equation~\eqref{eq:density_y} of the main paper. This derivative is given by: 
\begin{equation}
\label{eq:der_yj_by_yi}
\frac{d t_{{\gamma}}(y_i)}{d y_i} =
\begin{cases} 
{(y_i + 1)^{\gamma - 1}}& \text{if } y_i \geq 0, \\
(-y_i + 1)^{1 - \gamma}  & \text{if } y_i < 0.
\end{cases}
\end{equation}


\section{Variational Bayes approximation for SEMs with full data}
\label{sec:VB_full}

In this section, we present the variational Bayes (VB) algorithm for estimating the SEMs without missing values introduced in Section \ref{sec:models} of the main paper.

\begin{table}[h]
\centering
\caption{The definitions of $\textbf{r}$, $\textbf{M}$, and the parameters $\boldsymbol{\phi}$ for SEM-Gau, SEM-t, YJ-SEM-Gau, and YJ-SEM-t. For SEM-Gau and SEM-t, the density of the response vector conditional on $\boldsymbol{\xi}$ follows a multivariate Gaussian distribution with mean $\textbf{X}\boldsymbol{\beta}$ and covariance matrix $\sigma^2_{\textbf{e}}\textbf{M}^{-1}$, where $\textbf{A}=\textbf{I}_n-\rho\textbf{W}$.}
\label{tab:propertis_full_sem}
\begin{tabular}{ccccc}
\hline
{Model} &  $\boldsymbol{\phi}$ & \textbf{r} &  $\textbf{M}$\\ \hline
SEM-Gau &   $(\boldsymbol{\beta}^\top, \sigma^2_{\textbf{e}}, \rho)^\top$ & $\textbf{y}-\textbf{X}\boldsymbol{\beta}$& $\textbf{A}^\top\textbf{A}$\\
SEM-t   &  $ (\boldsymbol{\beta}^\top, \sigma^2_{\textbf{e}}, \rho, \nu)^\top$ & $\textbf{y}-\textbf{X}\boldsymbol{\beta}$   & $\textbf{A}^\top\boldsymbol{\Sigma}^{-1}_{\boldsymbol{\tau}}\textbf{A}$   \\ 
YJ-SEM-Gau     &  $ (\boldsymbol{\beta}^\top, \sigma^2_{\textbf{e},}, \rho, \gamma)^\top$ & $t_{\gamma}(\textbf{y})-\textbf{X}\boldsymbol{\beta}$ &  $\textbf{A}^\top\textbf{A}$ \\ 

YJ-SEM-t   & $ (\boldsymbol{\beta}^\top, \sigma^2_{\textbf{e}}, \rho, \nu,\gamma)^\top$    &  $t_{\gamma}(\textbf{y})-\textbf{X}\boldsymbol{\beta}$& $\textbf{A}^\top\boldsymbol{\Sigma}^{-1}_{\boldsymbol{\tau}}\textbf{A}$ \\
\hline
\end{tabular}
\end{table}

Consider Bayesian inference for the parameter vector of SEMs, $\boldsymbol{\xi}$. For SEM-Gau and YJ-SEM-Gau, we have $\boldsymbol{\xi} = \boldsymbol{\phi}$, while for SEM-t and YJ-SEM-t, $\boldsymbol{\xi} = (\boldsymbol{\phi}^\top, \boldsymbol{\tau}^\top)^\top$. See Table~\ref{tab:propertis_full_sem} for the definition of $\boldsymbol{\phi}$ for each SEM. Let the prior distribution of $\boldsymbol{\xi}$ be denoted by $p(\boldsymbol{\xi})$. Then, the posterior distribution of $\boldsymbol{\xi}$ given the data $\textbf{y}$ is denoted by $p(\boldsymbol{\xi} \mid \textbf{y})$ and is given by:
\begin{equation}
    \label{eq:post.full}
    p(\boldsymbol{\xi} \mid \textbf{y})\propto p(\textbf{y} \mid \boldsymbol{\xi}) p(\boldsymbol{\xi}),
\end{equation}
\noindent where the term $p(\textbf{y} \mid \boldsymbol{\xi})$ represents the likelihood, and the log-likelihood for different SEMs are given in Section~\ref{sec:models} of the main paper. We define $h(\boldsymbol{\xi})=p(\textbf{y} \mid \boldsymbol{\xi}) p(\boldsymbol{\xi})$. The terms $h(\boldsymbol{\xi})$ for each SEM, along with the total length of the parameter vector to be estimated in each model (i.e., the length of the vector $\boldsymbol{\xi}$), are provided in Table~\ref{tab:post_full}.



\begin{table}[h]
\centering
\caption{The term $h(\boldsymbol{\xi})=p(\textbf{y} \mid \boldsymbol{\xi})p(\boldsymbol{\xi})$, and the total number of model parameters $s$ (including the length of the latent vector $\boldsymbol{\tau}$ if exists) for the SEM-Gau, SEM-t, YJ-SEM-Gau, and YJ-SEM-t. The density $p(\boldsymbol{\tau}\mid \boldsymbol{\phi})=\prod_{i=1}^np(\tau_i\mid \nu)$ represents the $n$ independent inverse gamma latent variables used in SEMs with Student-$t$ errors (see Equation~\eqref{eq:terrors.scl} of the main paper).}

\label{tab:post_full}
\begin{tabular}{cccc}
\hline
{Model}  & $h(\boldsymbol{\xi})=p(\textbf{y} \mid \boldsymbol{\xi})p(\boldsymbol{\xi})$ &$s$ \\ \hline
SEM-Gau & $p(\textbf{y}\mid \boldsymbol{\phi})p(\boldsymbol{\phi})$ & $r+3$ \\
SEM-t     & $p(\textbf{y} \mid \boldsymbol{\tau}, \boldsymbol{\phi})p(\boldsymbol{\tau}\mid \boldsymbol{\phi}) p(\boldsymbol{\phi})$ & $r+4+n$    \\ 
YJ-SEM-Gau     &  $p(\textbf{y} \mid \boldsymbol{\phi})p(\boldsymbol{\phi})$& $r+4$\\ 
YJ-SEM-t  & $p(\textbf{y} \mid \boldsymbol{\tau}, \boldsymbol{\phi})p(\boldsymbol{\tau}\mid \boldsymbol{\phi}) p(\boldsymbol{\phi})$ & $r+5+n$\\
\hline
\end{tabular}
\end{table}

Now we explain how VB inference is used to approximate the posterior distribution in Equation~\eqref{eq:post.full}. We use the variational distribution $q_{\boldsymbol{\lambda}}(\boldsymbol{\xi})$, indexed by the variational parameter $\boldsymbol{\lambda}$ to approximate the posterior $p(\boldsymbol{\xi} \mid \textbf{y})$. The VB approach approximates this posterior distribution by minimising the Kullback-Leibler (KL) divergence between $q_{\boldsymbol{\lambda}}(\boldsymbol{\xi})$ and $p(\boldsymbol{\xi} \mid \textbf{y})$. The KL divergence between these two distributions is
\begin{equation}
    \label{eq:ELBO} 
    \begin{split}
     \text{KL}(\boldsymbol{\lambda})&=\text{KL}\left(q_{\boldsymbol{\lambda}}(\boldsymbol{\xi})\mid \mid p(\boldsymbol{\xi} \mid \textbf{y}) \right)\\
           & =\int   \text{log} \left(\frac{q_{\boldsymbol{\lambda}}(\boldsymbol{\xi})}{p(\boldsymbol{\xi} \mid \textbf{y})} \right) q_{\boldsymbol{\lambda}}(\boldsymbol{\xi})d\boldsymbol{\xi}.\\
    \end{split}
\end{equation}

Minimising KL divergence between $q_{\boldsymbol{\lambda}}(\boldsymbol{\xi})$ and $p(\boldsymbol{\xi} \mid \textbf{y})$ is equivalent to maximising evidence lower bound (ELBO) on the marginal log-likelihood, $\text{log}~p(\textbf{y})$, denoted by $\mathcal{L}(\boldsymbol{\lambda})$, with $p(\textbf{y})=\int ~p(\textbf{y}\mid \boldsymbol{\xi})p\left(\boldsymbol{\xi}\right)d\boldsymbol{\xi}$~\citep{blei2017variational}. The ELBO is
\begin{equation}
    \label{eq:ELBO}
    \begin{split}
            \mathcal{L}(\boldsymbol{\lambda})& =\int   \text{log} \left(\frac{h(\boldsymbol{\xi})}{q_{\boldsymbol{\lambda}}(\boldsymbol{\xi})} \right) q_{\boldsymbol{\lambda}}(\boldsymbol{\xi}) d\boldsymbol{\xi}, \\
    \end{split}
\end{equation}
\noindent where $h(\boldsymbol{\xi}) = p(\textbf{y} \mid \boldsymbol{\xi}) p(\boldsymbol{\xi})$. Table~\ref{tab:post_full} presents the expressions for $h(\boldsymbol{\xi})$ for each of the SEMs. The ELBO in Equation~\eqref{eq:ELBO} can be written as an expectation with respect to $q_{\boldsymbol{\lambda}}$,
\begin{equation}
    \label{eq:ELBO.expectation_wrt_q}
            \mathcal{L}(\boldsymbol{\lambda})=\mathbb{E}_{q}\left [\text{log}~h(\boldsymbol{\xi})-\text{log}~  q_{\boldsymbol{\lambda}}(\boldsymbol{\xi}) \right],
\end{equation}

\noindent where $\mathbb{E}_{q}\left[\cdot\right]$ denotes the expectation with respect to $q_{\boldsymbol{\lambda}}$.

We employ the Gaussian variational approximation, where we select $q_{\boldsymbol{\lambda}}(\boldsymbol{\xi})$ to be a multivariate Gaussian distribution. As a result, the variational parameters $\boldsymbol{\lambda}$ consist of both the mean vector and the distinct elements of the covariance matrix. We impose a factor covariance structure~\citep{ong2018gaussian} on the covariance matrix of $q_{\boldsymbol{\lambda}}(\boldsymbol{\xi})$, which reduces the number of distinct elements in the covariance matrix. Under a factor covariance structure for the covariance matrix, the variational distribution is parameterised as $q_{\boldsymbol{\lambda}}(\boldsymbol{\xi})\sim N(\boldsymbol{\xi}; \boldsymbol{\mu},\textbf{B}\textbf{B}^\top+\textbf{D}^2)$, where $\boldsymbol{\mu}$ is the $s \times 1$ mean vector, $\textbf{B}$ is an $s \times p$ full rank matrix with $p << s$, and $\textbf{D}$ is an $s \times s$ diagonal matrix with positive diagonal elements $\textbf{d}=(d_1, \hdots, d_{s})$. For all SEMs, the total number of parameters, $s$ is provided in Table~\ref{tab:post_full}. We further impose the restriction that the upper triangular elements of $\textbf{B}$ are all zero.

Since the ELBO in Equation~\eqref{eq:ELBO.expectation_wrt_q} does not have a closed-form solution for all four models, we use stochastic gradient ascent (SGA) methods~(\citeauthor{nott2012regression}, \citeyear{nott2012regression}; \citeauthor{pmlr-v32-titsias14}, \citeyear{pmlr-v32-titsias14}, \citeyear{NIPS2015_1373b284}) to maximise the ELBO with respect to the variational parameters, $\boldsymbol{\lambda}$. Let $\nabla_{\boldsymbol{\lambda}}\mathcal{L}(\boldsymbol{\lambda})$ be the gradient of the objective function to be optimised, $\mathcal{L}(\boldsymbol{\lambda})$, and $\widehat{\nabla_{\boldsymbol{\lambda}}\mathcal{L}(\boldsymbol{\lambda})}$ an unbiased estimate of $\nabla_{\boldsymbol{\lambda}}\mathcal{L}(\boldsymbol{\lambda})$.
The SGA method updates $\boldsymbol{\lambda}$ iteratively, starting from an initial value $\boldsymbol{\lambda}^{(0)}$, according to the following scheme:
\begin{equation}
    \label{eq:grad.ascent}    \boldsymbol{\lambda}^{(t+1)}=\boldsymbol{\lambda}^{(t)}+\mathbcal{a}^{(t)}\circ\widehat{\nabla_{\boldsymbol{\lambda}}\mathcal{L}(\boldsymbol{\lambda}^{(t)})},
\end{equation}

\noindent and continues until a stopping criterion is satisfied, where $\mathbcal{a}^{(t)}$ is a vector of element-wise learning rates at iteration $t$. The individual learning rates, denoted by $\mathcal{a}^{(t)}_i$, are typically chosen to satisfy the Robbins--Monro conditions: $\sum_t \mathcal{a}^{(t)}_i = \infty$ and $\sum_t (\mathcal{a}^{(t)}_i)^2 < \infty$, which ensure convergence of the sequence $\boldsymbol{\lambda}^{(t)}$ to a local optimum as $t \rightarrow \infty$, under regularity conditions~\citep{robbins1951stochastic, bottou2010large}. The symbol $\circ$ denotes the element-wise product of two vectors. To speed up convergence, we use the ADADELTA algorithm for adaptive learning rates~\citep{Zeiler2012ADADELTAAA}, detailed in Section~\ref{sec:sup:ADADELTA}. Further, minimising the variance of the gradient estimator, $\widehat{{\nabla_{\boldsymbol{\lambda}}\mathcal{L}(\boldsymbol{\lambda})}}$, is crucial for both stability and fast convergence of the SGA algorithm, which we address using the reparameterisation trick~\citep{kingma2013auto,pmlr-v32-rezende14}.

We now briefly explain how the reparameterisation trick, combined with a factor covariance structure for the Gaussian variational distribution, is used to obtain efficient gradient estimates in this paper. More detailed information can be found in~\citet{ong2018gaussian}.

To apply the reparameterisation trick, we begin by sampling from $q_{\boldsymbol{\lambda}}(\boldsymbol{\xi})$. This involves first drawing $\boldsymbol{\zeta} = (\boldsymbol{\eta}^\top, \boldsymbol{\epsilon}^\top)^\top$, where $\boldsymbol{\eta}$ is a $p$-dimensional vector and $\boldsymbol{\epsilon}$ is an $s$-dimensional vector, from a fixed distribution $f_{\boldsymbol{\zeta}}(\boldsymbol{\zeta})$ that is independent of the variational parameters. Next, we compute $\boldsymbol{\xi} = u(\boldsymbol{\zeta}, \boldsymbol{\lambda}) = \boldsymbol{\mu} + \textbf{B}\boldsymbol{\eta} + \textbf{d} \circ \boldsymbol{\epsilon}$. We let, $f_{\boldsymbol{\zeta}}(\boldsymbol{\zeta})$ to follow a standard normal distribution. i.e. $\boldsymbol{\zeta} = (\boldsymbol{\eta}^\top, \boldsymbol{\epsilon}^\top)^\top~\sim N(\textbf{0}, \textbf{I}_{s+p})$, where $\textbf{0}$ is the zero mean vector of size $s+p$ and $\textbf{I}_{s+p}$ is the identity matrix of the same size. Then, the expectation in Equation~\eqref{eq:ELBO.expectation_wrt_q} is expressed with respect to the distribution $f_{\boldsymbol{\zeta}}$ as
\begin{equation}
    \label{eq:ELBO.expectation_wrt_f}
    \begin{split}
                    \mathcal{L}(\boldsymbol{\lambda})&=\mathbb{E}_{q}\left[\text{log}~h(\boldsymbol{\xi})-\text{log}~  q_{\boldsymbol{\lambda}}(\boldsymbol{\xi}) \right]\\
                    &=\mathbb{E}_{f_{\boldsymbol{\zeta}}}\left[\text{log}~h(u(\boldsymbol{\zeta,\boldsymbol{\lambda}}))-\text{log}~  q_{\boldsymbol{\lambda}}(u(\boldsymbol{\zeta,\boldsymbol{\lambda}})) \right],
    \end{split}
\end{equation}

\noindent and differentiating $\mathcal{L}(\boldsymbol{\lambda})$  under the integral sign, we obtain
\begin{equation}
    \label{eq:ELBO.grad.lamda}
    \begin{split}
                    \nabla_{\boldsymbol{\lambda}}\mathcal{L}(\boldsymbol{\lambda})&=\mathbb{E}_{f_{\boldsymbol{\zeta}}}\left[\nabla_{\boldsymbol{\lambda}}~\text{log}~h(u(\boldsymbol{\zeta,\boldsymbol{\lambda}}))-\nabla_{\boldsymbol{\lambda}}~\text{log}~  q_{\boldsymbol{\lambda}}(u(\boldsymbol{\zeta,\boldsymbol{\lambda}})) \right],\\
                    &=\mathbb{E}_{f_{\boldsymbol{\zeta}}}\left[\frac{du(\boldsymbol{\zeta,\boldsymbol{\lambda}})^\top}{d\boldsymbol{\lambda}}\{\nabla_{\boldsymbol{\xi}}\text{log}~h(\boldsymbol{\xi})-\nabla_{\boldsymbol{\xi}}\text{log}~  q_{\boldsymbol{\lambda}}(\boldsymbol{\xi}) \}\right],\\
    \end{split}
\end{equation}

\noindent where $\frac{du(\boldsymbol{\zeta,\boldsymbol{\lambda}})}{d\boldsymbol{\lambda}}$ is the derivative of the transformation $u(\boldsymbol{\zeta,\boldsymbol{\lambda}})=\boldsymbol{\mu}+\textbf{B}\boldsymbol{\eta}+\textbf{d}\circ \boldsymbol{\epsilon}$ with respect to the variational parameters $\boldsymbol{\lambda}=(\boldsymbol{\mu}^\top
, \text{vech}(\textbf{B})^\top, \textbf{d}^\top)^\top$, where the ”vech” operator vectorises a matrix by stacking its columns
from left to right while removing all the elements above the diagonal (the superdiagonal
elements) of the matrix. The gradients $\frac{du(\boldsymbol{\zeta}, \boldsymbol{\lambda})}{d\boldsymbol{\lambda}}$ and $\nabla_{\boldsymbol{\xi}} \log q_{\boldsymbol{\lambda}}(\boldsymbol{\xi})$ are provided in Section~\ref{sec:grad_full_q_u}. The expressions for $\nabla_{\boldsymbol{\xi}} \log h(\boldsymbol{\xi})$ differ for each SEM and can be found in Section~\ref{sec:grad_with_full}.


Algorithm~\ref{alg:SAGfull} outlines the VB algorithm for estimating SEMs with full data. The unbiased estimate of the gradient, $\widehat{\nabla_{\boldsymbol{\lambda}} \mathcal{L}(\boldsymbol{\lambda})}$, in step 4 of Algorithm~\ref{alg:SAGfull} can be constructed using the reparameterisation trick described earlier, with a single sample drawn from $f_{\boldsymbol{\zeta}}(\boldsymbol{\zeta})$.

\begin{algorithm}
  \caption{Variational Bayes algorithm}
  \begin{algorithmic}[1]
  \label{alg:SAGfull}
   \STATE Initialise $\boldsymbol{\lambda}^{(0)}=(\boldsymbol{\mu}^{\top (0)},\textrm{vech}{(\textbf{B})}^{\top (0)},\textbf{d}^{\top (0)})^{\top}$ and set $t=0$ 
  \REPEAT 
      \STATE Generate $(\boldsymbol{\eta}^{(t)},\boldsymbol{\epsilon}^{(t)})\sim N(\textbf{0},\textbf{I}_{s+p})$
      \STATE Construct unbiased estimates $\widehat{{\nabla_{\boldsymbol{\mu}}\mathcal{L}(\boldsymbol{\lambda})}},\widehat{{\nabla_{\textrm{vech}(\textbf{B})}\mathcal{L}(\boldsymbol{\lambda})}}, \text{and}~ \widehat{{\nabla_{\textbf{d}}\mathcal{L}(\boldsymbol{\lambda})}}$ using Equations ~\eqref{eq:grad_wrt_mu},~\eqref{eq:grad_wrt_B} and ~\eqref{eq:grad_wrt_d} in Section~\ref{sec:grad_full_q_u} at $\boldsymbol{\lambda}^{(t)}$.
      \STATE Set adaptive learning rates for the variational means, $\mathbcal{a}^{(t)}_{\boldsymbol{\mu}}$ and the variational parameters $\textrm{vech}(\textbf{B})$ and $\textbf{d}$, $\mathbcal{a}_{\textrm{vech}{(\textbf{B})}}^{(t)}$ and $\mathbcal{a}_{\textbf{d}}^{(t)}$, respectively, using ADADELTA described in Section~\ref{sec:sup:ADADELTA} of the online supplement.
      \STATE Set $\boldsymbol{\mu}^{(t+1)} = \boldsymbol{\mu}^{(t)} + \mathbcal{a}^{(t)}_{\boldsymbol{\mu}} \circ \widehat{\nabla_{\boldsymbol{\mu}}  \mathcal{L}(\boldsymbol{\lambda}^{(t)})}$.
      \STATE  Set $\textrm{vech}(\textbf{B})^{(t+1)} = \textrm{vech}(\textbf{B})^{(t)} + \mathbcal{a}_{\text{vech}(\textbf{B})}^{(t)} \circ \widehat{\nabla_{\textrm{vech}(\textbf{B}) } \mathcal{L}(\boldsymbol{\lambda}^{(t)})}$.
      \STATE Set $\textbf{d}^{(t+1)} = \textbf{d}^{(t)} + \mathbcal{a}_{\textbf{d}}^{(t)} \circ  \widehat{\nabla _{\textbf{d}}  \mathcal{L}(\boldsymbol{\lambda}^{(t)})}$.
      \STATE Set $\boldsymbol{\lambda}^{(t+1)} = (\boldsymbol{\mu}^{\top (t+1)}, \textrm{vech}(\textbf{B})^{\top (t+1)}, \textbf{d}^{\top (t+1)})$, and $t = t + 1$
      \UNTIL {some stopping rule is satisfied}
  \end{algorithmic}
\end{algorithm}

\subsection{Derivation of $\frac{du(\boldsymbol{\zeta}, \boldsymbol{\lambda})}{d\boldsymbol{\lambda}}$, $\nabla_{\boldsymbol{\xi}} \log q_{\boldsymbol{\lambda}}(\boldsymbol{\xi})$ and $  \nabla_{\boldsymbol{\lambda}}\mathcal{L}(\boldsymbol{\lambda})$}

\label{sec:grad_full_q_u}

In the reparameterisation gradient of the ELBO given in Equation~\eqref{eq:ELBO.grad.lamda}, $\frac{du(\boldsymbol{\zeta,\boldsymbol{\lambda}})}{d\boldsymbol{\lambda}}$ is the derivative of the transformation $u(\boldsymbol{\zeta,\boldsymbol{\lambda}})=\boldsymbol{\mu}+\textbf{B}\boldsymbol{\eta}+\textbf{d}\circ \boldsymbol{\epsilon}$ with respect to the variational parameters $\boldsymbol{\lambda}=(\boldsymbol{\mu}^\top
, \text{vech}(\textbf{B})^\top, \textbf{d}^\top)^\top$. We write that ${u(\boldsymbol{\zeta,\boldsymbol{\lambda}})=\boldsymbol{\mu}+(\boldsymbol{\eta}^\top \otimes \textbf{I}_{s})\text{vech}(\textbf{B})+\textbf{d}\circ \boldsymbol{\epsilon}}$, where $\otimes$ represents the  Kronecker product, and $\textbf{I}_{s}$ is the identity matrix of size $s$. It can be shown that  ${\nabla_{\boldsymbol{\xi}}\text{log}~  q_{\boldsymbol{\lambda}}(\boldsymbol{\xi})=-(\textbf{B}\textbf{B}^\top+\textbf{D}^2)^{-1}(\boldsymbol{\xi}-\boldsymbol{\mu})}$, 
\begin{equation}
    \label{eq:du_by_dmu}
    \frac{\partial u(\boldsymbol{\zeta},\boldsymbol{\lambda})}{\partial \boldsymbol{\mu}}=\textbf{I}_{s}\\,~~~~~~~~~~
    \frac{\partial u(\boldsymbol{\zeta},\boldsymbol{\lambda})}{\partial\text{vech}(\textbf{B})}=\boldsymbol{\eta}^\top\otimes\textbf{I}_{s}\\,~~~\textrm{and}~~   \frac{\partial u(\boldsymbol{\zeta},\boldsymbol{\lambda})}{\partial \textbf{d}}=\boldsymbol{\epsilon}\\.
\end{equation}

The derivatives of the ELBO with respect to variational parameters are: 
\begin{equation}
    \label{eq:grad_wrt_mu}
    \begin{split}
            \nabla_{\boldsymbol{\mu}}\mathcal{L}(\boldsymbol{\lambda})&=\mathbb{E}_{f_{\boldsymbol{\zeta}}}[\nabla_{\boldsymbol{\xi}}~\text{log}~h(\boldsymbol{\mu}+\textbf{B}\boldsymbol{\eta}+\textbf{d}\circ\boldsymbol{\epsilon})\\
            &+(\textbf{B}\textbf{B}^\top+\textbf{D}^2)^{-1}(\textbf{B}\boldsymbol{\eta}+\textbf{d}\circ\boldsymbol{\epsilon})],
    \end{split}
\end{equation}
\begin{equation}
    \label{eq:grad_wrt_B}
    \begin{split}
        \nabla_{\text{vech}(\textbf{B})}\mathcal{L}(\boldsymbol{\lambda})&=\mathbb{E}_{f_{\boldsymbol{\zeta}}}[\nabla_{\boldsymbol{\xi}}\text{log}~h(\boldsymbol{\mu}+\textbf{B}\boldsymbol{\eta}+\textbf{d}\circ\boldsymbol{\epsilon})\boldsymbol{\eta}^\top\\
            &+(\textbf{B}\textbf{B}^\top+\textbf{D}^2)^{-1}(\textbf{B}\boldsymbol{\eta}+\textbf{d}\circ\boldsymbol{\epsilon})\boldsymbol{\eta}^\top],
    \end{split}
\end{equation}

and
\begin{equation}
    \label{eq:grad_wrt_d}
    \begin{split}
    \nabla_{\textbf{d}}\mathcal{L}(\boldsymbol{\lambda})&=\mathbb{E}_{f_{\boldsymbol{\zeta}}}[\text{diag}(\nabla_{\boldsymbol{\xi}}\text{log}~h(\boldsymbol{\mu}+\textbf{B}\boldsymbol{\eta}+\textbf{d}\circ\boldsymbol{\epsilon})\boldsymbol{\epsilon}^\top\\
            &+(\textbf{B}\textbf{B}^\top+\textbf{D}^2)^{-1}(\textbf{B}\boldsymbol{\eta}+\textbf{d}\circ\boldsymbol{\epsilon})\boldsymbol{\epsilon}^\top)],
    \end{split}
\end{equation}

\noindent where $\text{diag}(\cdot)$ is the vector of diagonal elements extracted from a square matrix. The analytical expressions for $\nabla_{\boldsymbol{\xi}}\text{log}~h(\boldsymbol{\mu}+\textbf{B}\boldsymbol{\eta}+\textbf{d}\circ\boldsymbol{\epsilon})=\nabla_{\boldsymbol{\xi}}\text{log}~h(\boldsymbol{\xi})$ in Equations~\eqref{eq:grad_wrt_mu}-\eqref{eq:grad_wrt_d} for different SEMs are provided in Section~\ref{sec:grad_with_full}.


\subsection{Derivation of $\nabla_{\boldsymbol{\xi}}\text{log}~h(\boldsymbol{\mu}+\textbf{B}\boldsymbol{\eta}+\textbf{d}\circ\boldsymbol{\epsilon})=\nabla_{\boldsymbol{\xi}}\text{log}~h(\boldsymbol{\xi})$}

\label{sec:grad_with_full}


The term $h(\boldsymbol{\xi})$ represents the likelihood multiplied by the prior distribution (see Section~\ref{sec:VB_full}). For different SEMs, the expressions for $h(\boldsymbol{\xi})$ are given in Table~\ref{tab:post_full}. In this section, we present the derivatives (gradients) of $\log h(\boldsymbol{\xi})$ with respect to the model parameter vector $\boldsymbol{\xi}$, denoted as $\nabla_{\boldsymbol{\xi}} \log h(\boldsymbol{\xi})$. These gradients are required to construct unbiased estimates of the reparameterisation gradient. 

To map the parameters $\sigma^2_{\textbf{e}}$, $\rho$, $\nu$, and $\gamma$ onto the real line, we apply the following transformations: $\omega^\prime = \text{log}(\sigma^2_{\textbf{e}})$, $\rho^\prime = \text{log}(1+\rho) - \text{log}(1-\rho)$, $\nu^\prime=\text{log}(\nu-3)$, and $\gamma^\prime=\text{log}(\gamma)-\text{log}(2-\gamma)$. Prior distributions are then specified on the transformed parameter space. Table~\ref{tab:priors} shows the prior distributions used for the model parameters.

\begin{table}[ht]
    \centering
    \setlength{\tabcolsep}{3pt} 
        \caption{Prior distributions of model parameters}
    \begin{tabular}{ccccccc}
   \bottomrule
    \\{Parameter} & $\boldsymbol{\beta}$ & $\omega^\prime$ & $\rho^\prime$ & $\nu^\prime$ & $\gamma^\prime$ & $\boldsymbol{\psi}$ \\
    \bottomrule
    \\{Prior distribution} & $N(\textbf{0}, \sigma^2_{\boldsymbol{\beta}} \textbf{I})$ & $N(0, \sigma^2_{\omega^\prime})$ & $N(0, \sigma^2_{\rho^\prime})$ & $N(0, \sigma^2_{\nu^\prime})$ & $N(0, \sigma^2_{\gamma^\prime})$ & $N(\textbf{0}, \sigma^2_{\boldsymbol{\psi}} \textbf{I})$ \\
    \\{Hyperparameters} & $\sigma^2_{\boldsymbol{\beta}} = 10^2$ & $\sigma^2_{\omega^\prime} = 10^2$ & $\sigma^2_{\rho^\prime} = 10^2$ & $\sigma^2_{\nu^\prime} = 10^2$& $\sigma^2_{\gamma^\prime} = 10^2$ & $\sigma^2_{\boldsymbol{\psi}} = 10^2$ \\
    \bottomrule
    \end{tabular}
    \label{tab:priors}
\end{table}

\subsubsection{Gradients for SEM-Gau}
\label{sec:grad_SEM_Gau_full}

For SEM-Gau, we set $\boldsymbol{\xi} = \boldsymbol{\phi}$, 
where $\boldsymbol{\phi} = (\boldsymbol{\beta}^\top, \sigma^2_{\mathbf{e}}, \rho)^\top$, 
so that 
$h(\boldsymbol{\xi}) = p(\mathbf{y} \mid \boldsymbol{\phi})\, p(\boldsymbol{\phi})$. The logarithm of $h(\boldsymbol{\xi})$ is $\text{log}~p(\textbf{y}\mid \boldsymbol{\phi})+ \text{log}~p(\boldsymbol{\phi})$. Note that, for $\sigma^2_{\textbf{e}}$, and $\rho$, we utilise the transformations described in Section~\ref{sec:grad_with_full}, and  place the priors on the transformed parameters as in Table~\ref{tab:priors}. This leads to 
\begin{equation} 
\label{eq:h_SEM_Gau_full}
\text{log}~h(\boldsymbol{\xi}) \propto -\frac{n}{2}\omega^\prime+\frac{1}{2}\textrm{log}|\textbf{M}|-\frac{e^{-\omega{^\prime}}}{2}\textbf{r}^\top\textbf{M}\textbf{r}-\frac{\boldsymbol{\beta}^\top\boldsymbol{\beta}}{2 \sigma^2_{\boldsymbol{\beta}}}-\frac{{\omega^\prime}^2}{2\sigma^2_{\omega^\prime}}-\frac{{\rho^\prime}^2}{2\sigma^2_{\rho^\prime}},
\end{equation} 

\noindent where the values of $\sigma^2_{\boldsymbol{\beta}}$, $\sigma^2_{\omega^\prime}$, and $\sigma^2_{\rho^\prime}$ are each set to 100, as detailed in Table~\ref{tab:priors}, while $\textbf{r}$ and $\textbf{M}$ are provided in Table~\ref{tab:propertis_full_sem}.

The derivative of $\text{log}~h(\boldsymbol{\xi})$ in Equation~\eqref{eq:h_SEM_Gau_full} with respect to $\boldsymbol{\beta}$  is

\begin{equation}
    \label{eq:SEM_Gau_grad_beta}
     \frac{\partial \text{log} h(\boldsymbol{\xi})}{\partial \boldsymbol{\beta}}=e^{-\omega^\prime}\textbf{r}^\top\textbf{M}\textbf{X}-\frac{\boldsymbol{\beta}^\top}{\sigma^2_{\boldsymbol{\beta}}},
\end{equation}

\noindent the derivative of $\text{log}~h(\boldsymbol{\xi})$ with respect to $\omega^\prime$ is
\begin{equation}
    \label{eq:SEM_Gau_grad_omegadash}
     \frac{\partial \text{log} h(\boldsymbol{\xi})}{\partial \omega^\prime}=-\frac{n}{2}+\frac{e^{-\omega^\prime}}{2}\textbf{r}^\top\textbf{M}\textbf{r}-\frac{\omega^\prime}{\sigma^2_{\omega^\prime}},
\end{equation}

\noindent and the derivative of $\text{log}~h(\boldsymbol{\xi})$ with respect to $\rho^\prime$ is
\begin{equation}
    \label{eq:SEM_Gau_grad_rhodash}
     \frac{\partial \text{log} h(\boldsymbol{\xi})}{\partial \rho^\prime}=\left(\frac{1}{2}\text{tr}\left\{(\textbf{A}^\top\textbf{A})^{-1}\left(\frac{\partial \textbf{A}^\top\textbf{A}}{\partial \rho}\right)\right\}-\frac{e^{-\omega^\prime}}{2}\textbf{r}^\top\left(\frac{\partial \textbf{A}^\top\textbf{A}}{\partial \rho}\right)\textbf{r}\right)\left(\frac{\partial \rho}{\partial \rho^\prime}\right)-\frac{\rho^\prime}{\sigma^2_{\rho^\prime}},
\end{equation}
\noindent where $\text{tr}\{\cdot\}$ denotes the trace operator, which computes the sum of the diagonal elements of a matrix,
\begin{equation*}
    \begin{split}
                \left(\frac{\partial \textbf{A}^\top\textbf{A}}{\partial \rho}\right) &=-(\textbf{W}^\top+\textbf{W})+2\rho\textbf{W}^\top\textbf{W},\\
            \left(\frac{\partial \rho}{\partial \rho^\prime}\right) &=\frac{2 e^{\rho^\prime}}{(1+e^{\rho^\prime})^2}.\\
    \end{split}
\end{equation*}

\subsubsection{Gradients for SEM-t}
\label{sec:grad_SEM_t_full}

For the SEM-t, we set $\boldsymbol{\xi} = (\boldsymbol{\phi}^{\top},\boldsymbol{\tau}^{\top})^{\top}$, 
where $\boldsymbol{\phi}=(\boldsymbol{\beta}^\top, \sigma^2_{\textbf{e}}, \rho, \nu)^\top$, 
so that 
${h(\boldsymbol{\xi})=p(\textbf{y} \mid \boldsymbol{\tau}, \boldsymbol{\phi})p(\boldsymbol{\tau}\mid \boldsymbol{\phi}) p(\boldsymbol{\phi})}$. The logarithm of $h(\boldsymbol{\xi})$ is $\text{log}~p(\textbf{y} \mid \boldsymbol{\tau}, \boldsymbol{\phi})+  \text{log}~p(\boldsymbol{\tau}\mid \boldsymbol{\phi}) +\text{log}~p(\boldsymbol{\phi})$. Note that, for $\sigma^2_{\textbf{e}}$, $\rho$ and $\nu$, we utilise the transformations described in Section~\ref{sec:grad_with_full}, and  place the priors on the transformed parameters as in Table~\ref{tab:priors}. This leads to 
\begin{equation} 
\label{eq:h_SEM_t_full}
\begin{split}
    \text{log}~h(\boldsymbol{\xi}) &\propto -\frac{n}{2}\omega^\prime+\frac{1}{2}\textrm{log}|\textbf{M}|-\frac{e^{-\omega{^\prime}}}{2}\textbf{r}^\top\textbf{M}\textbf{r}\\
    &+\sum_{i=1}^n\textrm{log}~p(\tau_i \mid \nu)+\textrm{log}\left|\frac{\partial\tau_i}{\partial \tau_i^\prime}\right|-\frac{\boldsymbol{\beta}^\top\boldsymbol{\beta}}{2 \sigma^2_{\boldsymbol{\beta}}}-\frac{{\omega^\prime}^2}{2\sigma^2_{\omega^\prime}}-\frac{{\rho^\prime}^2}{2\sigma^2_{\rho^\prime}}-\frac{{\nu^\prime}^2}{2\sigma^2_{\nu^\prime}},
\end{split}
\end{equation} 
\noindent where the values of $\sigma^2_{\boldsymbol{\beta}}$, $\sigma^2_{\omega^\prime}$, $\sigma^2_{\rho^\prime}$, and $\sigma^2_{\nu^\prime}$ are each set to 100, as detailed in Table~\ref{tab:priors}, while $\textbf{r}$ and $\textbf{M}$ are provided in Table~\ref{tab:propertis_full_sem}. The term $ \left| \frac{\partial \tau_i}{\partial \tau_i^\prime} \right|$ represents the derivative of the inverse transformation $\tau_i = e^{\tau_i^\prime}$, which maps the latent variable $\tau_i$, for $i=1,\hdots, n$, onto the real line using the transformation $\tau_i^\prime = \log(\tau_i)$.


The derivative of $\text{log}~h(\boldsymbol{\xi})$ in Equation~\eqref{eq:h_SEM_t_full} with respect to $\boldsymbol{\beta}$  is
\begin{equation}
    \label{eq:SEM_t_grad_beta}
     \frac{\partial \text{log} h(\boldsymbol{\xi})}{\partial \boldsymbol{\beta}}=e^{-\omega^\prime}\textbf{r}^\top\textbf{M}\textbf{X}-\frac{\boldsymbol{\beta}^\top}{\sigma^2_{\boldsymbol{\beta}}},
\end{equation}

\noindent the derivative of $\text{log}~h(\boldsymbol{\xi})$ with respect to $\omega^\prime$ is
\begin{equation}
    \label{eq:SEM_t_grad_omegadash}
     \frac{\partial \text{log} h(\boldsymbol{\xi})}{\partial \omega^\prime}=-\frac{n}{2}+\frac{e^{-\omega^\prime}}{2}\textbf{r}^\top\textbf{M}\textbf{r}-\frac{\omega^\prime}{\sigma^2_{\omega^\prime}},
\end{equation}

\noindent the derivative of $\text{log}~h(\boldsymbol{\xi})$ with respect to $\rho^\prime$ is
\begin{equation}
    \label{eq:SEM_t_grad_rhodash}
     \frac{\partial \text{log} h(\boldsymbol{\xi})}{\partial \rho^\prime}=\left(\frac{1}{2}\text{tr}\left\{(\textbf{A}^\top\textbf{A})^{-1}\left(\frac{\partial \textbf{A}^\top\textbf{A}}{\partial \rho}\right)\right\}+e^{-\omega^\prime}(\textbf{A}\textbf{r})^\top\boldsymbol{\Sigma}^{-1}_{\boldsymbol{\tau}}\textbf{W}\textbf{r}\right)\left(\frac{\partial \rho}{\partial \rho^\prime}\right)-\frac{\rho^\prime}{\sigma^2_{\rho^\prime}},
\end{equation}

\noindent and the derivative of $\text{log}~h(\boldsymbol{\xi})$ with respect to $\nu^\prime$ is
\begin{equation}
    \label{eq:SEM_t_grad_nudash}
\begin{split}
         \frac{\partial \text{log} h(\boldsymbol{\xi})}{\partial \nu^\prime}&=n\left(\frac{\partial~\textrm{log}~\Gamma\left(\frac{e^{\nu^\prime+3}}{2}\right)}{\partial\nu^\prime}\right)-\frac{e^{\nu^\prime}}{2}\sum_{i=1}^n\left(\tau_i^\prime+e^{-\tau_i^\prime}\right)-\frac{ne^{\nu^\prime}}{2}\textrm{log}\left(\frac{2}{e^{\nu^\prime}+3}\right)+\frac{ne^{\nu^\prime}}{2}\\
         &-\frac{\nu^\prime}{\sigma^2_{\nu^\prime}},\\
\end{split}
\end{equation}

\noindent where $\frac{\partial~\textrm{log}~\Gamma\left(\frac{e^{\nu^\prime+3}}{2}\right)}{\partial\nu^\prime}$ is the derivative of the log-gamma function, which is computed numerically using \texttt{R}.

The derivative of $\text{log}~h(\boldsymbol{\xi})$ with respect to $i^{\textrm{th}}$ element of the vector $\boldsymbol{\tau}^{\prime}$, $\tau_i^{\prime}$ is
\begin{equation}
    \label{eq:SEM_t_grad_taudash}
     \frac{\partial \text{log} h(\boldsymbol{\xi})}{\partial \tau_i^\prime}=-\frac{1}{2}+\frac{1}{2}e^{-\omega^\prime}({\textbf{r}_{\textbf{A}}}_i)^2e^{\tau_i^\prime}+\left(\frac{e^{\nu^\prime+3}}{2}\right)\left(-1+e^{-\tau_i^\prime}\right),
\end{equation}

\noindent where ${\textbf{r}_{\textbf{A}}}_i$ is the $i^{\textrm{th}}$ element of the vector $\textbf{A}\textbf{r}$.
 
\subsubsection{Gradients for YJ-SEM-Gau}
\label{sec:grad_YJ_SEM_Gau_full}

For the YJ-SEM-Gau, we set $\boldsymbol{\xi} = \boldsymbol{\phi}$, 
where $\boldsymbol{\phi}=(\boldsymbol{\beta}^\top, \sigma^2_{\textbf{e}}, \rho, \gamma)^\top$, 
so that 
${h(\boldsymbol{\xi}) = p(\mathbf{y} \mid \boldsymbol{\phi})\, p(\boldsymbol{\phi})}$. The logarithm of $h(\boldsymbol{\xi})$ is $\text{log}~p(\textbf{y}\mid \boldsymbol{\phi})+ \text{log}~p(\boldsymbol{\phi})$. Note that, for $\sigma^2_{\textbf{e}}$, $\rho$ and $\gamma$, we utilise the transformations described in Section~\ref{sec:grad_with_full}, and  place the priors on the transformed parameters as in Table~\ref{tab:priors}. This leads to 
\begin{equation} 
\label{eq:h_YJ_SEM_Gau_full}
\begin{split}
    \text{log}~h(\boldsymbol{\xi})& \propto -\frac{n}{2}\omega^\prime+\frac{1}{2}\textrm{log}|\textbf{M}|-\frac{e^{-\omega{^\prime}}}{2}\textbf{r}^\top\textbf{M}\textbf{r}+
           \sum_{i=1}^n \text{log}\left( \frac{dt_{\gamma}(y_i)}{dy_i} \right)\\
           &-\frac{\boldsymbol{\beta}^\top\boldsymbol{\beta}}{2 \sigma^2_{\boldsymbol{\beta}}}-\frac{{\omega^\prime}^2}{2\sigma^2_{\omega^\prime}}-\frac{{\rho^\prime}^2}{2\sigma^2_{\rho^\prime}}-\frac{{\gamma^\prime}^2}{2\sigma^2_{\gamma^\prime}},
\end{split}
\end{equation} 

\noindent where the values of $\sigma^2_{\boldsymbol{\beta}}$, $\sigma^2_{\omega^\prime}$, $\sigma^2_{\rho^\prime}$ , and $\sigma^2_{\gamma^\prime}$ are each set to 100, as detailed in Table~\ref{tab:priors}, while $\textbf{r}$ and $\textbf{M}$ are provided in Table~\ref{tab:propertis_full_sem}.

The derivative of $\text{log}~h(\boldsymbol{\xi})$ in Equation~\eqref{eq:h_YJ_SEM_Gau_full} with respect to $\boldsymbol{\beta}$  is
\begin{equation}
    \label{eq:YJSEMGau_grad_beta}
     \frac{\partial \text{log} h(\boldsymbol{\xi})}{\partial \boldsymbol{\beta}}=e^{-\omega^\prime}\textbf{r}^\top\textbf{M}\textbf{X}-\frac{\boldsymbol{\beta}^\top}{\sigma^2_{\boldsymbol{\beta}}},
\end{equation}

\noindent the derivative of $\text{log}~h(\boldsymbol{\xi})$ with respect to $\omega^\prime$ is
\begin{equation}
    \label{eq:YJSEMGau_grad_omegadash}
     \frac{\partial \text{log} h(\boldsymbol{\xi})}{\partial \omega^\prime}=-\frac{n}{2}+\frac{e^{-\omega^\prime}}{2}\textbf{r}^\top\textbf{M}\textbf{r}-\frac{\omega^\prime}{\sigma^2_{\omega^\prime}},
\end{equation}

\noindent the derivative of $\text{log}~h(\boldsymbol{\xi})$ with respect to $\rho^\prime$ is
\begin{equation}
    \label{eq:YJSEMGau_grad_rhodash}
     \frac{\partial \text{log} h(\boldsymbol{\xi})}{\partial \rho^\prime}=\left(\frac{1}{2}\text{tr}\left\{(\textbf{A}^\top\textbf{A})^{-1}\left(\frac{\partial \textbf{A}^\top\textbf{A}}{\partial \rho}\right)\right\}-\frac{e^{-\omega^\prime}}{2}\textbf{r}^\top\left(\frac{\partial \textbf{A}^\top\textbf{A}}{\partial \rho}\right)\textbf{r}\right)\left(\frac{\partial \rho}{\partial \rho^\prime}\right)-\frac{\rho^\prime}{\sigma^2_{\rho^\prime}},
\end{equation}

\noindent and the derivative of $\text{log}~h(\boldsymbol{\xi})$ with respect to $\gamma^\prime$ is
\begin{equation}
    \label{eq:YJSEMGau_grad_gammadash}
     \frac{\partial \text{log} h(\boldsymbol{\xi})}{\partial \gamma^\prime}=\left(-e^{-\omega^\prime}\textbf{r}^\top\textbf{M}\left(\frac{\partial t_\gamma(\textbf{y})}{\partial\gamma}\right)+\sum_{i=0}^n  \frac{\partial \textrm{log}\left(\frac{d t_\gamma(y_i)}{d y_i}\right)}{\partial \gamma}\right)\left(\frac{\partial \gamma}{\partial \gamma^\prime}\right)
     -\frac{\gamma^\prime}{\sigma^2_{\gamma^\prime}},
\end{equation}

\noindent where $ \frac{\partial t_\gamma(\textbf{y})}{\partial\gamma}= \left(\frac{\partial t_\gamma(y_{1})}{\partial\gamma}, \frac{\partial t_\gamma({y}_2)}{\partial\gamma}, \hdots, \frac{\partial t_\gamma({y}_n)}{\partial\gamma}\right)^{\top}$, with the $i^{\textrm{th}}$ elements of this vector is equal to
\begin{equation*}
\frac{\partial t_\gamma({y}_i)}{\partial\gamma} =
\begin{cases} 
\frac{(y_i + 1)^{\gamma}(\gamma \textrm{log}(y_i+1)-1)+1}{\gamma^2} & \text{if } y_i \geq 0, \\
\frac{(2-\gamma)(-y_i+1)^{(2-\gamma)}\textrm{log}(-y_i+1)-(-y_i + 1)^{2 - \gamma}+1}{(2 - \gamma)^2} & \text{if } y_i < 0,
\end{cases}
\end{equation*}

\begin{equation*}
\frac{\partial \textrm{log}\left(\frac{d t_\gamma(y_i)}{d y_i}\right)}{\partial \gamma} =
\begin{cases} 
\textrm{log}~(y_i+1) & \text{if } y_i \geq 0, \\
-\textrm{log}~(y_i+1) & \text{if } y_i < 0,
\end{cases}
\end{equation*}

\noindent and
\begin{equation*}
    \frac{\partial \gamma}{\partial \gamma^\prime}=\frac{2e^{\gamma^\prime}}{(1+e^{\gamma^\prime})^{2}}.
\end{equation*}


\subsubsection{Gradients for YJ-SEM-t}
\label{sec:grad_YJ_SEM_t_full}

For the YJ-SEM-t, we set $\boldsymbol{\xi} = (\boldsymbol{\phi}^{\top},\boldsymbol{\tau}^{\top})^{\top}$, 
where $\boldsymbol{\phi}=(\boldsymbol{\beta}^\top, \sigma^2_{\textbf{e}}, \rho, \nu, \gamma)^\top$, 
so that 
$h(\boldsymbol{\xi})=p(\textbf{y} \mid \boldsymbol{\tau}, \boldsymbol{\phi})p(\boldsymbol{\tau}\mid \boldsymbol{\phi}) p(\boldsymbol{\phi})$. The logarithm of $h(\boldsymbol{\xi})$ is $\text{log}~p(\textbf{y} \mid \boldsymbol{\tau}, \boldsymbol{\phi})+  \text{log}~p(\boldsymbol{\tau}\mid \boldsymbol{\phi}) +\text{log}~p(\boldsymbol{\phi})$. Note that, for $\sigma^2_{\textbf{e}}$, $\rho$, $\nu$, and $\gamma$, we utilise the transformations described in Section~\ref{sec:grad_with_full}, and  place the priors on the transformed parameters as in Table~\ref{tab:priors}. This leads to 
\begin{equation} 
\label{eq:h_YJ_SEM_t_full}
\begin{split}
\text{log}~h(\boldsymbol{\xi}) &\propto -\frac{n}{2}\omega^\prime+\frac{1}{2}\textrm{log}|\textbf{M}|-\frac{e^{-\omega{^\prime}}}{2}\textbf{r}^\top\textbf{M}\textbf{r}+ \sum_{i=1}^n \text{log}\left(  \frac{dt_{\gamma}(y_i)}{dy_i} \right)\\
    &+\sum_{i=1}^n\textrm{log}~p(\tau_i \mid \nu)+\textrm{log}\left|\frac{\partial\tau_i}{\partial \tau_i^\prime}\right|-\frac{\boldsymbol{\beta}^\top\boldsymbol{\beta}}{2 \sigma^2_{\boldsymbol{\beta}}}-\frac{{\omega^\prime}^2}{2\sigma^2_{\omega^\prime}}-\frac{{\rho^\prime}^2}{2\sigma^2_{\rho^\prime}}-\frac{{\nu^\prime}^2}{2\sigma^2_{\nu^\prime}},
\end{split}
\end{equation}

\noindent where the values of $\sigma^2_{\boldsymbol{\beta}}$, $\sigma^2_{\omega^\prime}$, $\sigma^2_{\rho^\prime}$, $\sigma^2_{\nu^\prime}$, and $\sigma^2_{\gamma^\prime}$ are each set to 100, as detailed in Table~\ref{tab:priors}, while $\textbf{r}$ and $\textbf{M}$ are provided in Table~\ref{tab:propertis_full_sem}.

The derivative of $\text{log}~h(\boldsymbol{\xi})$ in Equation~\eqref{eq:h_YJ_SEM_t_full} with respect to $\boldsymbol{\beta}$  is
\begin{equation}
    \label{eq:YJSEMt_grad_beta}
     \frac{\partial \text{log} h(\boldsymbol{\xi})}{\partial \boldsymbol{\beta}}=e^{-\omega^\prime}\textbf{r}^\top\textbf{M}\textbf{X}-\frac{\boldsymbol{\beta}^\top}{\sigma^2_{\boldsymbol{\beta}}},
\end{equation}

\noindent the derivative of $\text{log}~h(\boldsymbol{\xi})$ with respect to $\omega^\prime$ is
\begin{equation}
    \label{eq:YJSEMt_grad_omegadash}
     \frac{\partial \text{log} h(\boldsymbol{\xi})}{\partial \omega^\prime}=-\frac{n}{2}+\frac{e^{-\omega^\prime}}{2}\textbf{r}^\top\textbf{M}\textbf{r}-\frac{\omega^\prime}{\sigma^2_{\omega^\prime}},
\end{equation}

\noindent the derivative of $\text{log}~h(\boldsymbol{\xi})$ with respect to $\rho^\prime$ is
\begin{equation}
    \label{eq:YJSEMt_grad_rhodash}
     \frac{\partial \text{log} h(\boldsymbol{\xi})}{\partial \rho^\prime}=\left(\frac{1}{2}\text{tr}\left\{(\textbf{A}^\top\textbf{A})^{-1}\left(\frac{\partial \textbf{A}^\top\textbf{A}}{\partial \rho}\right)\right\}+e^{-\omega^\prime}(\textbf{A}\textbf{r})^\top\boldsymbol{\Sigma}^{-1}_{\boldsymbol{\tau}}\textbf{W}\textbf{r}\right)\left(\frac{\partial \rho}{\partial \rho^\prime}\right)-\frac{\rho^\prime}{\sigma^2_{\rho^\prime}},
\end{equation}

\noindent the derivative of $\text{log}~h(\boldsymbol{\xi})$ with respect to $\nu^\prime$ is
\begin{equation}
    \label{eq:YJSEMt_grad_nudash}
\begin{split}
         \frac{\partial \text{log} h(\boldsymbol{\xi})}{\partial \nu^\prime}&=n\left(\frac{\partial~\textrm{log}~\Gamma\left(\frac{e^{\nu^\prime+3}}{2}\right)}{\partial\nu^\prime}\right)-\frac{e^{\nu^\prime}}{2}\sum_{i=1}^n\left(\tau_i^\prime+e^{-\tau_i^\prime}\right)-\frac{ne^{\nu^\prime}}{2}\textrm{log}\left(\frac{2}{e^{\nu^\prime}+3}\right)+\frac{ne^{\nu^\prime}}{2}\\
         &-\frac{\nu^\prime}{\sigma^2_{\nu^\prime}},\\
\end{split}
\end{equation}

\noindent and the derivative of $\text{log}~h(\boldsymbol{\xi})$ with respect to $\gamma^\prime$ is
\begin{equation}
    \label{eq:YJSEMt_grad_gammadash}
     \frac{\partial \text{log} h(\boldsymbol{\xi})}{\partial \gamma^\prime}=\left(-e^{-\omega^\prime}\textbf{r}^\top\textbf{M}\left(\frac{\partial t_\gamma(\textbf{y})}{\partial\gamma}\right)+\sum_{i=0}^n  \frac{\partial \textrm{log}\left(\frac{d t_\gamma(y_i)}{d y_i}\right)}{\partial \gamma}\right)\left(\frac{\partial \gamma}{\partial \gamma^\prime}\right)
     -\frac{\gamma^\prime}{\sigma^2_{\gamma^\prime}}.
\end{equation}

The derivative of $\text{log}~h(\boldsymbol{\xi})$ with respect to $i^{\textrm{th}}$ element of the vector $\boldsymbol{\tau}^{\prime}$, $\tau_i^{\prime}$ is
\begin{equation}
    \label{eq:YJSEM_t_grad_taudash}
     \frac{\partial \text{log} h(\boldsymbol{\xi})}{\partial \tau_i^\prime}=-\frac{1}{2}+\frac{1}{2}e^{-\omega^\prime}({\textbf{r}_{\textbf{A}}}_i)^2e^{\tau_i^\prime}+\left(\frac{e^{\nu^\prime+3}}{2}\right)\left(-1+e^{-\tau_i^\prime}\right),
\end{equation}

\noindent where ${\textbf{r}_{\textbf{A}}}_i$ is the $i^{\textrm{th}}$ element of the vector $\textbf{A}\textbf{r}$.

\section{Calculate adaptive learning rates using ADADELTA}
\label{sec:sup:ADADELTA}

To ensure stable and efficient convergence of all VB algorithms presented in this paper, we use the ADADELTA algorithm~\citep{Zeiler2012ADADELTAAA} to compute adaptive learning rates $\mathbcal{a}^{(t)}$. ADADELTA assigns different step sizes to each element of the variational parameter vector $\boldsymbol{\lambda}$. 
A brief description of the ADADELTA algorithm is provided below.


The update for the $i^{\textrm{th}}$ element of $\boldsymbol{\lambda}$, $\lambda_i$ is
\begin{equation}
    \label{eq:ADADELTA.updating}
{\lambda}^{(t+1)}_i = {\lambda}^{(t)}_i + \Delta {\lambda}^{(t)}_i,
\end{equation}

\noindent where, the step size $\Delta {\lambda}^{(t)}_i$ is $a_i^{(t)} g_{{\lambda}_i}^{(t)}$. The term $g_{{\lambda}_i}^{(t)}$ denotes the $i^{th}$ component of $\widehat{\nabla_{\boldsymbol{\lambda}}\mathcal{L}(\boldsymbol{\lambda}^{(t)})}$ and $a_i^{(t)}$ is defined as:
\begin{equation}
    \label{eq:a_i_t}
    \mathcal{a}_i^{(t)}=\sqrt{\frac{\mathbb{E}\left({\Delta}_{{\lambda}_i}^2\right)^{(t-1)}+\alpha}{\mathbb{E}\left(g^2_{{\lambda}_i}\right)^{(t)}+\alpha}},
\end{equation}


\noindent where $\alpha$ is a small positive constant, $\mathbb{E}\left({\Delta}_{{\lambda_ i}}^2\right)^{(t)}$ and $\mathbb{E}\left(g^2_{{\lambda}_i}\right)^{(t)}$  are decayed moving average estimates of  ${{\Delta}{{\lambda_i}}^{(t)}}^{2}$ and ${g_{{\lambda}_i}^{(t)}}^{2}$, defined by
\begin{equation}
    \label{eq:E_Delta2}
    \mathbb{E}\left({\Delta}_{{\lambda }_i}^2\right)^{(t)}=\upsilon \mathbb{E}\left({\Delta}_{{\lambda }_i}^2\right)^{(t-1)} +(1-\upsilon) {{\Delta}{{\lambda_i}}^{(t)}}^{2},
\end{equation}
\noindent and
\begin{equation}
    \label{eq:E_g2}
    \mathbb{E}\left(g^2_{{\lambda}_i}\right)^{(t)}=\upsilon \mathbb{E}\left(g^2_{{\lambda}_i}\right)^{(t-1)} +(1-\upsilon) {g_{{\lambda}_i}^{(t)}}^{2},
\end{equation}

\noindent where the variable $\upsilon$ is a decay constant. We use the default tuning parameter choices $\alpha = 10^{-6}$ and $\upsilon = 0.95$, and initialise $ \mathbb{E}\left({\Delta}_{{\lambda }_i}^2\right)^{(0)}= \mathbb{E}\left(g^2_{{\lambda}_i}\right)^{(0)} = 0$.

\section{Selection model factorisation for SAR models}
\label{online_sec:ext_SAR}

This section outlines how the selection model factorisation for SEMs, discussed in Section~\ref{sec:joint_mod_y_m} of the main paper, can be straightforwardly extended to other simultaneous autoregressive (SAR)–type models.

A wide range of SAR-type models exists, including the spatial error model (SEM), the spatial autoregressive model (SAM), and the spatial Durbin model (SDM), the latter of which incorporates spatially lagged covariates~\citep{anselin1988spatial, cressie1993statistics}. The selection model factorisation is a highly general decomposition and can be applied to any of these SAR-type models with only minor modifications. For illustration, consider the SDM:
\begin{equation}
\label{lit_eq:SDM}
\textbf{y} = \rho \textbf{W} \textbf{y} + \textbf{X} \boldsymbol{\beta}_1 +   \textbf{X}_s\boldsymbol{\beta}_2 + \textbf{e},
\end{equation}

\noindent where $\boldsymbol{\beta}_1 = (\beta_{10}, \dots, \beta_{1r})^\top$ is the $(r+1)\times 1$ vector of fixed effects associated with the covariates in the design matrix $\textbf{X}$. The $n \times r$ matrix $\textbf{X}_s = \textbf{W} \textbf{X}^o$ represents the spatially lagged covariates, where $\textbf{X}^o$ contains the covariates excluding the intercept column.
The vector $\boldsymbol{\beta}_2 = (\beta_{21}, \dots, \beta_{2r})^\top$ is the $r \times 1$ vector of fixed effects corresponding to the spatially lagged covariates, and $\textbf{e} \sim N(0, \sigma^2_\textbf{e}\textbf{I}_n)$.

The parameter vector for the SDM with Gaussian errors is therefore ${\boldsymbol{\phi} = (\boldsymbol{\beta}_1^\top, \boldsymbol{\beta}_2^\top, \rho, \sigma^2_{\textbf{e}})^\top}$. In the Gaussian case, $\boldsymbol{\xi} = \boldsymbol{\phi}$. If we instead assume Student's-$t$ errors, $e_i \sim t_\nu(0, \sigma^2_{\textbf{e}})$, then the parameter vector becomes ${\boldsymbol{\phi} = (\boldsymbol{\beta}_1^\top, \boldsymbol{\beta}_2^\top, \rho, \sigma^2_{\textbf{e}}, \nu)^\top}$, and $\boldsymbol{\xi} = (\boldsymbol{\phi}^\top, \boldsymbol{\tau}^\top)^\top$, which is consistent with the extension of the SEMs to Student's-$t$ errors (SEM-t) described in Section~\ref{sec:models_withoutyj} of the main paper.

Importantly, the same selection model factorisation used in Equation~\eqref{eq:selectionmodels} of the main paper applies directly to both the Gaussian SDM and its Student’s-$t$ extension, with the only difference being the form of the parameter vector $\boldsymbol{\xi}$. Specifically, the joint density of the missingness indicators $\textbf{m}$ and the response vector $\textbf{y}$ is factorised as
\begin{equation}
\label{eq:selectionmodels_online}
p(\textbf{y}, \textbf{m} \mid \boldsymbol{\xi}, \boldsymbol{\psi})
= p(\textbf{m} \mid \textbf{y}, \boldsymbol{\psi})\, p(\textbf{y} \mid \boldsymbol{\xi}).
\end{equation}

Thus, the methodological framework developed in the main paper, namely, extending SEMs to non-Gaussian settings, jointly modelling the missingness indicators and the spatial process via the selection model factorisation, and performing inference using the HVB algorithm, is fully general and can be readily adapted to other SAR-type models with alternative spatial dependence structures.

\section{Additional derivations for the HVB algorithm}

This section provides additional proofs and derivations for the proposed HVB algorithm used to estimate non-Gaussian SEMs with missing data, introduced in Section~\ref{sec:HVB} of the main paper.

\subsection{Derivation of the reparameterisation gradient}
\label{sec:der.hvb}

Since $\mathcal{L}(\boldsymbol{\lambda})=\mathbb{E}_{q_{\boldsymbol{\lambda}}}\left[\text{log}~p(\textbf{y}_o, \textbf{m}\mid \boldsymbol{\xi},\boldsymbol{\psi})+\text{log}~p(\boldsymbol{\xi},\boldsymbol{\psi})-\text{log}~  q_{\boldsymbol{\lambda}}^0(\boldsymbol{\xi},\boldsymbol{\psi}) \right]=  \mathcal{L}^0(\boldsymbol{\lambda})$, as shown in Equation~\eqref{eq:ELBO.expectation_wrt_qaug} of the main paper, the reparameterisation gradient of $\mathcal{L}$ is the same as that of $\mathcal{L}^{0}$ and given by
\begin{equation}
\begin{aligned}
    \label{eq:ELBO.grad.lamda.aug_1}
                    \nabla_{\boldsymbol{\lambda}}\mathcal{L}(\boldsymbol{\lambda})&=\mathbb{E}_{f_{\boldsymbol{\delta}^0}}\Bigg[\frac{dt^0(\boldsymbol{\delta}^0,\boldsymbol{\lambda})^\top}{d\boldsymbol{\lambda}}\Big(\nabla_{(\boldsymbol{\xi}^\top,\boldsymbol{\psi}^\top)^\top}\text{log}~p(\boldsymbol{\xi},\boldsymbol{\psi})\\
                    &\qquad+\nabla_{(\boldsymbol{\xi}^\top,\boldsymbol{\psi}^\top)^\top}\text{log}~p(\textbf{y}_o,\textbf{m}\mid \boldsymbol{\xi},\boldsymbol{\psi})-\nabla_{(\boldsymbol{\xi}^\top,\boldsymbol{\psi}^\top)^\top}\text{log}~  q_{\boldsymbol{\lambda}}^0(\boldsymbol{\xi},\boldsymbol{\psi}) \Big)\Bigg],
                    \end{aligned}
\end{equation}

\noindent where, the random vector $\boldsymbol{\delta}^0$ has density $f_{\boldsymbol{\delta}^0}$, which follows a standard normal, and does not depend on $\boldsymbol{\lambda}$, and $t^0$ is the one-to-one vector-valued
transformation from ${\boldsymbol{\delta}^0=({\boldsymbol{\eta}^0}^\top,{\boldsymbol{\epsilon}^0}^\top)^\top}$ to the parameter vector, such that ${(\boldsymbol{\xi}^\top,\boldsymbol{\psi}^\top)^\top=t^0(\boldsymbol{\delta}^0,\boldsymbol{\lambda})=\boldsymbol{\mu}+\textbf{B}\boldsymbol{\eta}^0+\textbf{d}\circ \boldsymbol{\epsilon}^0}$. 

Then, the Fisher’s identity is given by 
\begin{equation}
    \label{eq:fishidentiy}
\begin{split}
    \nabla_{(\boldsymbol{\xi}^\top,\boldsymbol{\psi}^\top)^\top}\text{log}~p(\textbf{y}_o,\textbf{m}\mid \boldsymbol{\xi},\boldsymbol{\psi})&=\int  \nabla_{(\boldsymbol{\xi}^\top,\boldsymbol{\psi}^\top)^\top}\left[\text{log}~ (p(\textbf{y}_o,\textbf{m}\mid \textbf{y}_u, \boldsymbol{\xi},\boldsymbol{\psi})p(\textbf{y}_u \mid \boldsymbol{\xi}))\right] \\
    &p(\textbf{y}_u \mid \textbf{y}_o,\textbf{m}, \boldsymbol{\xi},\boldsymbol{\psi})d\textbf{y}_u,
\end{split}
\end{equation}
\noindent see~\citet{poyiadjis2011particle}. Substituting this expression into Equation~\eqref{eq:ELBO.grad.lamda.aug_1}, and writing $\mathbb{E}_{f_{\boldsymbol{\delta}}}[\cdot]$ for expectation with respect to $f_{\boldsymbol{\delta}}(\boldsymbol{\delta})=f_{\boldsymbol{\delta}^0}(\boldsymbol{\delta}^0)p(\textbf{y}_u \mid  \textbf{y}_o,\textbf{m},  \boldsymbol{\xi},\boldsymbol{\psi})$ and because $h(\boldsymbol{\xi},\boldsymbol{\psi},\textbf{y}_u)=p(\textbf{y}_o, \textbf{m} \mid \textbf{y}_u,\boldsymbol{\xi},\boldsymbol{\psi})p(\textbf{y}_u\mid \boldsymbol{\xi})p(\boldsymbol{\xi},\boldsymbol{\psi})$, we get
\begin{equation}
\label{eq:ELBO.grad.lamda.aug_2}
\begin{aligned}
\nabla_{\boldsymbol{\lambda}} \mathcal{L}(\boldsymbol{\lambda}) 
&= \mathbb{E}_{f_{\boldsymbol{\delta}}} \Bigg[ 
    \frac{d t^0(\boldsymbol{\delta}^0, \boldsymbol{\lambda})^\top}
         {d \boldsymbol{\lambda}} \Big(
         \nabla_{(\boldsymbol{\xi}^\top, \boldsymbol{\psi}^\top)^\top} \log p(\boldsymbol{\xi}, \boldsymbol{\psi}) \\
&\qquad + \nabla_{(\boldsymbol{\xi}^\top, \boldsymbol{\psi}^\top)^\top} \log p(\textbf{y}_u \mid \boldsymbol{\xi})
         + \nabla_{(\boldsymbol{\xi}^\top, \boldsymbol{\psi}^\top)^\top} \log p(\textbf{y}_o, \textbf{m} \mid \textbf{y}_u, \boldsymbol{\xi}, \boldsymbol{\psi}) \\
&\qquad - \nabla_{(\boldsymbol{\xi}^\top, \boldsymbol{\psi}^\top)^\top} \log q_{\boldsymbol{\lambda}}^0(\boldsymbol{\xi}, \boldsymbol{\psi}) \Big) \Bigg] \\
&= \mathbb{E}_{f_{\boldsymbol{\delta}}} \left[
    \frac{d t^0(\boldsymbol{\delta}^0, \boldsymbol{\lambda})^\top}
         {d \boldsymbol{\lambda}}  \left(
         \nabla_{(\boldsymbol{\xi}^\top, \boldsymbol{\psi}^\top)^\top} \log h(\boldsymbol{\xi}, \boldsymbol{\psi}, \textbf{y}_u)
       - \nabla_{(\boldsymbol{\xi}^\top, \boldsymbol{\psi}^\top)^\top} \log q_{\boldsymbol{\lambda}}^0(\boldsymbol{\xi}, \boldsymbol{\psi})
    \right) \right].
\end{aligned}
\end{equation}

\subsection{Derivation of $\frac{dt^0(\boldsymbol{\delta}^0, \boldsymbol{\lambda})}{d\boldsymbol{\lambda}}$, $\nabla_{(\boldsymbol{\xi}^\top, \boldsymbol{\psi}^\top)^\top} \log q_{\boldsymbol{\lambda}}^0(\boldsymbol{\xi}, \boldsymbol{\psi})$, and $ \nabla_{\boldsymbol{\lambda}}\mathcal{L}(\boldsymbol{\lambda})$ for HVB}

\label{nongousar_appendix:deri_t_log_q}

In the reparameterisation gradient of the ELBO given in Equation~\eqref{eq:ELBO.grad.lamda.aug_2}, $\frac{dt^0(\boldsymbol{\delta}^0,\boldsymbol{\lambda})}{d\boldsymbol{\lambda}}$ denotes the derivative of the transformation ${t^0(\boldsymbol{\delta}^0,\boldsymbol{\lambda})=\boldsymbol{\mu}+\textbf{B}\boldsymbol{\eta}^0+\textbf{d}\circ \boldsymbol{\epsilon}^0}$ with respect to the variational parameters ${\boldsymbol{\lambda}=(\boldsymbol{\mu}^\top,\text{vech}(\textbf{B})^\top,\textbf{d}^\top)^\top}$. We can express that ${t^0(\boldsymbol{\delta}^0,\boldsymbol{\lambda})=\boldsymbol{\mu}+(\boldsymbol{\eta}^0 \otimes \textbf{I}_s)\text{vech}(\textbf{B})+\textbf{d}\circ \boldsymbol{\epsilon}^0}$, where $\textbf{I}_s$ is the identity matrix of size $s$. It can be further shown that  $\nabla_{(\boldsymbol{\xi}^\top,\boldsymbol{\psi}^\top)^\top}\text{log}~  q^0_{\boldsymbol{\lambda}}(\boldsymbol{\xi},\boldsymbol{\psi})=-(\textbf{B}\textbf{B}^\top+\textbf{D}^2)^{-1}((\boldsymbol{\xi}^\top,\boldsymbol{\psi}^\top)^\top-\boldsymbol{\mu})$, 
\begin{equation}
    \label{eq:du_by_dmu_aug}
    \frac{\partial t^0(\boldsymbol{\delta}^0,\boldsymbol{\lambda})}{\partial\boldsymbol{\mu}}=\textbf{I}_s\\,~~~~~~~~~~
    \frac{\partial t^0(\boldsymbol{\delta}^0,\boldsymbol{\lambda})}{\partial\text{vech}(\textbf{B})}={\boldsymbol{\eta}^0}^\top\otimes\textbf{I}_s\\,~~~~~\text{and}~~~~~\frac{\partial t^0(\boldsymbol{\delta}^0,\boldsymbol{\lambda})}{\partial\textbf{d}}=\boldsymbol{\epsilon}^0.
\end{equation}

Then the derivatives of $\mathcal{L}(\boldsymbol{\lambda})$ with respect to $\boldsymbol{\mu}, \textrm{vech}(\textbf{B})$, and $\textbf{d}$ are:
    \begin{equation}
  \label{eq:grad_wrt_mu_vb2}
    \begin{split}
            \nabla_{\boldsymbol{\mu}}\mathcal{L}(\boldsymbol{\lambda})&=\mathbb{E}_{f_{\boldsymbol{\delta}}}[\nabla_{(\boldsymbol{\xi}^\top,\boldsymbol{\psi^\top)^\top}}~\text{log}~h(\boldsymbol{\mu}+\textbf{B}\boldsymbol{\eta}^0+\textbf{d}\circ\boldsymbol{\epsilon}^0,\textbf{y}_u)\\
            &+(\textbf{B}\textbf{B}^\top+\textbf{D}^2)^{-1}(\textbf{B}\boldsymbol{\eta}^0+\textbf{d}\circ\boldsymbol{\epsilon}^0)],
    \end{split}
\end{equation}
\begin{equation}
 \label{eq:grad_wrt_B_vb2}
    \begin{split}
            \nabla_{\textrm{vech}(\textbf{B})}\mathcal{L}(\boldsymbol{\lambda})&=\mathbb{E}_{f_{\boldsymbol{\delta}}}[\nabla_{(\boldsymbol{\xi}^\top,\boldsymbol{\psi^\top)^\top}}~\text{log}~h(\boldsymbol{\mu}+\textbf{B}\boldsymbol{\eta}^0+\textbf{d}\circ\boldsymbol{\epsilon}^0,\textbf{y}_u){\boldsymbol{\eta}^0}^\top\\
            &+(\textbf{B}\textbf{B}^\top+\textbf{D}^2)^{-1}(\textbf{B}\boldsymbol{\eta}^0+\textbf{d}\circ\boldsymbol{\epsilon}^0){\boldsymbol{\eta}^0}^\top],
    \end{split}
\end{equation}
\begin{equation}
 \label{eq:grad_wrt_d_vb2}
    \begin{split}
    \nabla_{\textbf{d}}\mathcal{L}(\boldsymbol{\lambda})&=\mathbb{E}_{f_{\boldsymbol{\delta}}}[\text{diag}(\nabla_{(\boldsymbol{\xi}^\top,\boldsymbol{\psi^\top)^\top}}\text{log}~h(\boldsymbol{\mu}+\textbf{B}\boldsymbol{\eta}+\textbf{d}\circ\boldsymbol{\epsilon}^0,\textbf{y}_u){\boldsymbol{\epsilon}^0}^\top\\
            &+(\textbf{B}\textbf{B}^\top+\textbf{D}^2)^{-1}(\textbf{B}\boldsymbol{\eta}^0+\textbf{d}\circ\boldsymbol{\epsilon}^0){\boldsymbol{\epsilon}^0}^\top)].
    \end{split}
\end{equation}

The expressions for  ${\nabla_{(\boldsymbol{\xi}^\top,\boldsymbol{\psi}^\top)^\top}\text{log}~h(\boldsymbol{\mu}+\textbf{B}\boldsymbol{\eta}+\textbf{d}\circ\boldsymbol{\epsilon}^0,\textbf{y}_u)=\nabla_{(\boldsymbol{\xi}^\top,\boldsymbol{\psi}^\top)^\top}\text{log}~h(\boldsymbol{\xi}, \boldsymbol{\psi}, \textbf{y}_u)}$ in Equations~\eqref{eq:grad_wrt_mu_vb2}-\eqref{eq:grad_wrt_d_vb2} can be found in Section~\ref{sec:grad_with_miss}. 
The expectations in these gradients are estimated using a single sample of $\boldsymbol{\delta}=({\boldsymbol{\delta}^0}^\top,\textbf{y}_u^\top)^\top$, which is drawn from $f_{\boldsymbol{\delta}^0}$, and $p(\textbf{y}_u \mid \boldsymbol{\xi}, \boldsymbol{\psi},\textbf{y}_o, \textbf{m})=p(\textbf{y}_u \mid t^0(\boldsymbol{\delta}^0,\boldsymbol{\lambda}),\textbf{y}_o, \textbf{m})$; see ~\citet{wijayawardhana2025variational} for further details.

\subsection{Expressions for $\nabla_{(\boldsymbol{\xi}^\top,\boldsymbol{\psi}^\top)^\top}\text{log}~h(\boldsymbol{\xi},\boldsymbol{\psi},\textbf{y}_u)$}
\label{sec:grad_with_miss}

This section presents the derivatives (gradients) $\nabla_{(\boldsymbol{\xi}^\top,\boldsymbol{\psi}^\top)^\top}\text{log}~h(\boldsymbol{\xi},\boldsymbol{\psi},\textbf{y}_u)$, which are required to compute the gradients of the reparameterised ELBO, defined in Equation~\eqref{eq:ELBO.grad.lamda.aug} of the main paper.

\label{sec:grad_with_miss}

\subsubsection{Gradients for SEM-Gau}
\label{sec:grad_SEM_Gau_miss}

For the SEM-Gau with missing data, ${h(\boldsymbol{\xi},\boldsymbol{\psi},\textbf{y}_u)=p(\textbf{y}\mid \boldsymbol{\phi})p(\textbf{m}\mid \textbf{y},\boldsymbol{\psi})p(\boldsymbol{\phi})p(\boldsymbol{\psi})}$, where ${\boldsymbol{\phi}=(\boldsymbol{\beta}^\top, \sigma^2_{\textbf{e}}, \rho)^\top.}$ 

The logarithm of $h(\boldsymbol{\xi},\boldsymbol{\psi},\textbf{y}_u)$ is ${\textrm{log}~p(\textbf{y}\mid \boldsymbol{\phi})+\textrm{log}~p(\textbf{m}\mid \textbf{y},\boldsymbol{\psi})+\textrm{log}~p(\boldsymbol{\phi})+\textrm{log}~p(\boldsymbol{\psi})}$. Note that, for $\sigma^2_{\textbf{e}}$, and $\rho$, we utilise the transformations described in Section~\ref{sec:grad_with_full}, and  place the priors on the transformed parameters as in Table~\ref{tab:priors}. This leads to 
\begin{equation} 
\label{eq:h_SEM_Gau_miss}
\begin{split}
    \text{log}~h(\boldsymbol{\xi}, \boldsymbol{\psi}, \textbf{y}_u) & \propto -\frac{n}{2}\omega^\prime+\frac{1}{2}\textrm{log}|\textbf{M}|-\frac{e^{-\omega{^\prime}}}{2}\textbf{r}^\top\textbf{M}\textbf{r}\\
    &+\sum_{i=1}^{n}m_i({\textbf{x}^*_i}^{\top}\boldsymbol{\psi}_{\textbf{x}}+{y}_i\boldsymbol{\psi}_{y})-\text{log}(1+e^{({\textbf{x}^*_i}^{\top}\boldsymbol{\psi}_{\textbf{x}}+{y}_i\boldsymbol{\psi}_{y})})\\
    &-\frac{\boldsymbol{\beta}^\top\boldsymbol{\beta}}{2 \sigma^2_{\boldsymbol{\beta}}}-\frac{{\omega^\prime}^2}{2\sigma^2_{\omega^\prime}}-\frac{{\rho^\prime}^2}{2\sigma^2_{\rho^\prime}}-\frac{\boldsymbol{\psi}^\top\boldsymbol{\psi}}{2 \sigma^2_{\boldsymbol{\psi}}},
\end{split}
\end{equation} 

\noindent where the values of $\sigma^2_{\boldsymbol{\beta}}$, $\sigma^2_{\omega^\prime}$, $\sigma^2_{\rho^\prime}$, and $\sigma^2_{\boldsymbol{\psi}}$ are each set to 100, as detailed in Table~\ref{tab:priors}, while $\textbf{r}$ and $\textbf{M}$ are provided in Table~\ref{tab:propertis_full_sem}.

The derivatives of $\text{log}~h(\boldsymbol{\xi}, \boldsymbol{\psi}, \textbf{y}_u)$ in Equation~\eqref{eq:h_SEM_Gau_miss} with respect to $\boldsymbol{\beta}$, ${\omega^\prime}$, and ${\rho^\prime}$ are similar to that of without missing data given in Equations~\eqref{eq:SEM_Gau_grad_beta}, ~\eqref{eq:SEM_Gau_grad_omegadash}, and~\eqref{eq:SEM_Gau_grad_rhodash}, respectively. The derivative of $\text{log}~h(\boldsymbol{\xi}, \boldsymbol{\psi}, \textbf{y}_u)$ with respect to $\boldsymbol{\psi}$ is

\begin{equation}
    \label{eq:SEM_Gau_grad_psi}
     \frac{\partial \text{log}~h(\boldsymbol{\xi}, \boldsymbol{\psi}, \textbf{y}_u)}{\partial \boldsymbol{\psi}}=\sum_{i=1}^n\left(m_i-\frac{e^{{\textbf{x}^*_i}^\top\boldsymbol{\psi}_{\textbf{x}}+{y}_i\boldsymbol{\psi}_{y}}}{1+e^{{\textbf{x}^*_i}^\top\boldsymbol{\psi}_{\textbf{x}}+{y}_i\boldsymbol{\psi}_{y}}}\right)\textbf{z}_i-\frac{\boldsymbol{\psi}^\top}{\sigma^2_{\boldsymbol{\psi}}},
\end{equation}

\noindent where $\textbf{z}_i=({\textbf{x}^*_i}^\top,y_i)$, ${\textbf{x}^*_i}^\top$ is the $i^{\textrm{th}}$ row of matrix $\textbf{X}^*$, and $y_i$ is the $i^{\textrm{th}}$ element of the vector $\textbf{y}$; see Section~\ref{sec:SEMmissing} of the main paper.

\subsubsection{Gradients for SEM-t}
\label{sec:grad_SEM_t_miss}

For the SEM-t with missing data, ${h(\boldsymbol{\xi},\boldsymbol{\psi},\textbf{y}_u)=p(\textbf{y}\mid \boldsymbol{\tau}, \boldsymbol{\phi})p(\textbf{m}\mid \textbf{y},\boldsymbol{\psi})p(\boldsymbol{\tau}\mid \boldsymbol{\phi})p(\boldsymbol{\phi})p(\boldsymbol{\psi})}$, where $\boldsymbol{\phi}=(\boldsymbol{\beta}^\top, \sigma^2_{\textbf{e}}, \nu, \rho)^\top$. 

The logarithm of $h(\boldsymbol{\xi},\boldsymbol{\psi},\textbf{y}_u)$ is $\textrm{log}~p(\textbf{y}\mid \boldsymbol{\tau}, \boldsymbol{\phi})+\textrm{log}~p(\textbf{m}\mid \textbf{y},\boldsymbol{\psi})+\textrm{log}~p(\boldsymbol{\tau}\mid\boldsymbol{\phi})+\textrm{log}~p(\boldsymbol{\phi})+\textrm{log}~p(\boldsymbol{\psi})$. Note that, for $\sigma^2_{\textbf{e}}$, $\rho$, and $\nu$, we utilise the transformations described in Section~\ref{sec:grad_with_full}, and place the priors on the transformed parameters as in Table~\ref{tab:priors}. This leads to 
\begin{equation} 
\label{eq:h_SEM_t_miss}
\begin{split}
    \text{log}~h(\boldsymbol{\xi}, \boldsymbol{\psi}, \textbf{y}_u) & \propto -\frac{n}{2}\omega^\prime+\frac{1}{2}\textrm{log}|\textbf{M}|-\frac{e^{-\omega{^\prime}}}{2}\textbf{r}^\top\textbf{M}\textbf{r}\\
    &+\sum_{i=1}^{n}m_i({\textbf{x}^*_i}^{\top}\boldsymbol{\psi}_{\textbf{x}}+{y}_i\boldsymbol{\psi}_{y})-\text{log}(1+e^{({\textbf{x}^*_i}^{\top}\boldsymbol{\psi}_{\textbf{x}}+{y}_i\boldsymbol{\psi}_{y})})\\
    &+\sum_{i=1}^n\textrm{log}~p(\tau_i \mid \nu)+\textrm{log}\left|\frac{\partial\tau_i}{\partial \tau_i^\prime}\right|-\frac{\boldsymbol{\beta}^\top\boldsymbol{\beta}}{2 \sigma^2_{\boldsymbol{\beta}}}-\frac{{\omega^\prime}^2}{2\sigma^2_{\omega^\prime}}-\frac{{\rho^\prime}^2}{2\sigma^2_{\rho^\prime}}-\frac{{\nu^\prime}^2}{2\sigma^2_{\nu^\prime}},
\end{split}
\end{equation} 

\noindent where the values of $\sigma^2_{\boldsymbol{\beta}}$, $\sigma^2_{\omega^\prime}$, $\sigma^2_{\rho^\prime}$, $\sigma^2_{\nu^\prime}$, and $\sigma^2_{\boldsymbol{\psi}}$ are each set to 100, as detailed in Table~\ref{tab:priors}, while $\textbf{r}$ and $\textbf{M}$ are provided in Table~\ref{tab:propertis_full_sem}.

The derivatives of $\text{log}~h(\boldsymbol{\xi}, \boldsymbol{\psi}, \textbf{y}_u)$ in Equation~\eqref{eq:h_SEM_t_miss} with respect to $\boldsymbol{\beta}$, ${\omega^\prime}$, ${\rho^\prime}$, ${\nu^\prime}$ and $\tau_i^{\prime}$ are similar to that of without missing data given in Equations~\eqref{eq:SEM_t_grad_beta}, ~\eqref{eq:SEM_t_grad_omegadash},~\eqref{eq:SEM_t_grad_rhodash}, \eqref{eq:SEM_t_grad_nudash}, and~\eqref{eq:SEM_t_grad_taudash}, respectively. The derivative of $\text{log}~h(\boldsymbol{\xi}, \boldsymbol{\psi}, \textbf{y}_u)$ with respect to $\boldsymbol{\psi}$ is similar to that of SEM-Gau given in Equation~\eqref{eq:SEM_Gau_grad_psi}.

\subsubsection{Gradients for YJ-SEM-Gau}
\label{sec:grad_YJ_SEM_Gau_miss}

For the YJ-SEM-Gau with missing data, ${h(\boldsymbol{\xi},\boldsymbol{\psi},\textbf{y}_u)=p(\textbf{y}\mid \boldsymbol{\phi})p(\textbf{m}\mid \textbf{y},\boldsymbol{\psi})p(\boldsymbol{\phi})p(\boldsymbol{\psi})}$, where $\boldsymbol{\phi}=(\boldsymbol{\beta}^\top, \sigma^2_{\textbf{e}}, \rho, \gamma)^\top$.

The logarithm of $h(\boldsymbol{\xi},\boldsymbol{\psi},\textbf{y}_u)$ is ${\textrm{log}~p(\textbf{y}\mid \boldsymbol{\phi})+\textrm{log}~p(\textbf{m}\mid \textbf{y},\boldsymbol{\psi})+\textrm{log}~p(\boldsymbol{\phi})+\textrm{log}~p(\boldsymbol{\psi})}$. Note that, for $\sigma^2_{\textbf{e}}$, $\rho$, and $\gamma$, we utilise the transformations described in in Section~\ref{sec:grad_with_full}, and place the priors on the transformed parameters as in Table~\ref{tab:priors}. This leads to 
\begin{equation} 
\label{eq:h_YJ_SEM_Gau_miss}
\begin{split}
    \text{log}~h(\boldsymbol{\xi}, \boldsymbol{\psi}, \textbf{y}_u) & \propto -\frac{n}{2}\omega^\prime+\frac{1}{2}\textrm{log}|\textbf{M}|-\frac{e^{-\omega{^\prime}}}{2}\textbf{r}^\top\textbf{M}\textbf{r}+ \sum_{i=1}^n \text{log}\left( \frac{dt_{\gamma}(y_i)}{dy_i} \right)\\
    &+\sum_{i=1}^{n}m_i({\textbf{x}^*_i}^{\top}\boldsymbol{\psi}_{\textbf{x}}+{y}_i\boldsymbol{\psi}_{y})-\text{log}(1+e^{({\textbf{x}^*_i}^{\top}\boldsymbol{\psi}_{\textbf{x}}+{y}_i\boldsymbol{\psi}_{y})})\\
    &-\frac{\boldsymbol{\beta}^\top\boldsymbol{\beta}}{2 \sigma^2_{\boldsymbol{\beta}}}-\frac{{\omega^\prime}^2}{2\sigma^2_{\omega^\prime}}-\frac{{\rho^\prime}^2}{2\sigma^2_{\rho^\prime}}-\frac{{\gamma^\prime}^2}{2\sigma^2_{\gamma^\prime}}-\frac{\boldsymbol{\psi}^\top\boldsymbol{\psi}}{2 \sigma^2_{\boldsymbol{\psi}}},
\end{split}
\end{equation} 

\noindent where the values of $\sigma^2_{\boldsymbol{\beta}}$, $\sigma^2_{\omega^\prime}$, $\sigma^2_{\rho^\prime}$, $\sigma^2_{\nu^\prime}$, $\sigma^2_{\gamma^\prime}$, and $\sigma^2_{\boldsymbol{\psi}}$ are each set to 100, as detailed in Table~\ref{tab:priors}, while $\textbf{r}$ and $\textbf{M}$ are provided in Table~\ref{tab:propertis_full_sem}.

The derivatives of $\text{log}~h(\boldsymbol{\xi}, \boldsymbol{\psi}, \textbf{y}_u)$ in Equation~\eqref{eq:h_YJ_SEM_Gau_miss} with respect to $\boldsymbol{\beta}$, ${\omega^\prime}$, ${\rho^\prime}$, and ${\gamma^\prime}$ are similar to that of without missing data given in Equations~\eqref{eq:YJSEMGau_grad_beta}, ~\eqref{eq:YJSEMGau_grad_omegadash},~\eqref{eq:YJSEMGau_grad_rhodash}, and~\eqref{eq:YJSEMGau_grad_gammadash}, respectively. The derivative of $\text{log}~h(\boldsymbol{\xi}, \boldsymbol{\psi}, \textbf{y}_u)$ with respect to $\boldsymbol{\psi}$ is similar to that of SEM-Gau given in Equation~\eqref{eq:SEM_Gau_grad_psi}.

\subsubsection{Gradients for YJ-SEM-t}
\label{sec:grad_YJ_SEM_t_miss}

For the YJ-SEM-t with missing data, ${h(\boldsymbol{\xi},\boldsymbol{\psi},\textbf{y}_u)=p(\textbf{y}\mid \boldsymbol{\tau}, \boldsymbol{\phi})p(\textbf{m}\mid \textbf{y},\boldsymbol{\psi})p(\boldsymbol{\tau}\mid \boldsymbol{\phi})p(\boldsymbol{\phi})p(\boldsymbol{\psi})}$, where $\boldsymbol{\phi}=(\boldsymbol{\beta}^\top, \sigma^2_{\textbf{e}}, \rho, \nu,\gamma)^\top$.

The logarithm of $h(\boldsymbol{\xi},\boldsymbol{\psi},\textbf{y}_u)$ is $\textrm{log}~p(\textbf{y}\mid \boldsymbol{\tau}, \boldsymbol{\phi})+\textrm{log}~p(\textbf{m}\mid \textbf{y},\boldsymbol{\psi})+\textrm{log}~p(\boldsymbol{\tau}\mid \boldsymbol{\phi})+\textrm{log}~p(\boldsymbol{\phi})+\textrm{log}~p(\boldsymbol{\psi})$. Note that, for $\sigma^2_{\textbf{e}}$, $\rho$, $\nu$, and $\gamma$, we utilise the transformations described in Section~\ref{sec:grad_with_full}, and place the priors on the transformed parameters as in Table~\ref{tab:priors}. This leads to  
\begin{equation} 
\label{eq:h_YJ_SEM_t_miss}
\begin{split}
    \text{log}~h(\boldsymbol{\xi}, \boldsymbol{\psi}, \textbf{y}_u) & \propto -\frac{n}{2}\omega^\prime+\frac{1}{2}\textrm{log}|\textbf{M}|-\frac{e^{-\omega{^\prime}}}{2}\textbf{r}^\top\textbf{M}\textbf{r}+\sum_{i=1}^n \text{log}\left( \frac{dt_{\gamma}(y_i)}{dy_i} \right)\\
    &+\sum_{i=1}^{n}m_i({\textbf{x}^*_i}^{\top}\boldsymbol{\psi}_{\textbf{x}}+{y}_i\boldsymbol{\psi}_{y})-\text{log}(1+e^{({\textbf{x}^*_i}^{\top}\boldsymbol{\psi}_{\textbf{x}}+{y}_i\boldsymbol{\psi}_{y})})\\
    &+\sum_{i=1}^n\textrm{log}~p(\tau_i \mid \nu)+\textrm{log}\left|\frac{\partial\tau_i}{\partial \tau_i^\prime}\right|-\frac{\boldsymbol{\beta}^\top\boldsymbol{\beta}}{2 \sigma^2_{\boldsymbol{\beta}}}-\frac{{\omega^\prime}^2}{2\sigma^2_{\omega^\prime}}-\frac{{\rho^\prime}^2}{2\sigma^2_{\rho^\prime}}-\frac{{\nu^\prime}^2}{2\sigma^2_{\nu^\prime}}-\frac{{\gamma^\prime}^2}{2\sigma^2_{\gamma^\prime}},
\end{split}
\end{equation} 

\noindent where the values of $\sigma^2_{\boldsymbol{\beta}}$, $\sigma^2_{\omega^\prime}$, $\sigma^2_{\rho^\prime}$, $\sigma^2_{\nu^\prime}$, $\sigma^2_{\gamma^\prime}$, and $\sigma^2_{\boldsymbol{\psi}}$ are each set to 100, as detailed in Table~\ref{tab:priors}, while $\textbf{r}$ and $\textbf{M}$ are provided in Table~\ref{tab:propertis_full_sem}.

The derivatives of $\text{log}~h(\boldsymbol{\xi}, \boldsymbol{\psi}, \textbf{y}_u)$ in Equation~\ref{eq:h_YJ_SEM_t_miss} with respect to $\boldsymbol{\beta}$, ${\omega^\prime}$, ${\rho^\prime}$, ${\nu^\prime}$, ${\gamma^\prime}$  and $\tau_i^{\prime}$ are similar to that of without missing data given in Equations~\eqref{eq:YJSEMt_grad_beta}, ~\eqref{eq:YJSEMt_grad_omegadash},~\eqref{eq:YJSEMt_grad_rhodash}, \eqref{eq:YJSEMt_grad_nudash},~\eqref{eq:YJSEMt_grad_gammadash}, and~\eqref{eq:YJSEM_t_grad_taudash}, respectively. The derivative of $\text{log}~h(\boldsymbol{\xi}, \boldsymbol{\psi}, \textbf{y}_u)$ with respect to $\boldsymbol{\psi}$ is similar to that of SEM-Gau given in Equation~\eqref{eq:SEM_Gau_grad_psi}.




\subsection{HVB-AllB algorithm}
\label{sec:HVB_All}



This section presents the MCMC steps used in the HVB-AllB algorithm for estimating SEMs with missing values, which is particularly suited for cases where both $n$ and $n_u$ are large.

The MCMC steps in Algorithm~\ref{alg:AugvbMCMCstep} of the main paper generates samples from the conditional distribution $p(\textbf{y}_u\mid \textbf{y}_o,\textbf{m},  \boldsymbol{\xi},\boldsymbol{\psi})$. However, as $n$ and $n_u$ increase, the HVB algorithm implemented using these MCMC steps does not estimate the parameters accurately because of the low acceptance percentage. 
To improve the acceptance percentage, we partition $\textbf{y}_{u}$ into $k$ blocks and update one block at a time.

We start with partitioning the unobserved responses vector into $k$ blocks, such that $\textbf{y}_u=(\textbf{y}_{u_1}^\top,\dots,\textbf{y}_{u_k}^\top)^\top$. Using proposals from $p(\textbf{y}_{u_j}\mid \boldsymbol{\xi}^{(t)},\textbf{y}_o, \textbf{y}_u^{(-j)})$,  Algorithm~\ref{alg:AugvbMCMCstepB} outlines the MCMC steps for sampling the missing values one block at a time, for $j=1,..,k$, where $\textbf{y}_{u_j}$ is the updated block and $\textbf{y}_u^{(-j)}=(\textbf{y}_{u_{1}}^\top, \dots, \textbf{y}_{u_{j-1}}^\top,\textbf{y}_{u_{j+1}}^\top, \dots, \textbf{y}_{u_k}^\top)^\top$ is the remaining blocks. The complete response vector $\textbf{y}$ can now be written as $\textbf{y}=({\textbf{y}^\top_{s_j}}, \textbf{y}_{u_j}^{\top})^\top$, where ${\textbf{y}_{s_j}}=(\textbf{y}_o^{\top},{\textbf{y}_u^{(-j)}}^\top)^\top$. 
Based on this partitioning of $\textbf{y}$, the following partitioned vector and matrices are defined:
\begin{equation}
\label{mat:portions_of_xwMB}
\textbf{r}=
\begin{pmatrix}
    \textbf{r}_{s_j}\\
   \textbf{r}_{u_j}
\end{pmatrix},
~\textbf{X}=
\begin{pmatrix}
    \textbf{X}_{s_j}\\
   \textbf{X}_{u_j}
\end{pmatrix},
~\textbf{M}=
\begin{pmatrix}
    \textbf{M}_{{s_j}{s_j}}&  \textbf{M}_{{s_j}{u_j}}\\
    \textbf{M}_{{u_j}{s_j}} & \textbf{M}_{{u_j}{u_j}}
\end{pmatrix},
\end{equation}

\noindent where $\textbf{r}_{s_j}$ is the corresponding sub-vector of $\textbf{r}$ for the observed responses ($\textbf{y}_o$), and the unobserved responses that are not in the $j^{th}$ block (${\textbf{y}_u^{(-j)}}$). i.e. $\textbf{r}_{s_j}=( {{\textbf{r}}}^{\top}_o,{{\textbf{r}}^\top_{u^{(-j)}}})^\top$ and ${\textbf{r}}_{u_j}$ is the corresponding sub vector of $\textbf{r}$ for $j^{th}$ block of unobserved responses, $\textbf{y}_{u_j}$. Similarly, $\textbf{X}_{s_j}$ and $\textbf{X}_{u_j}$ are represent the sub-matrices of $\textbf{X}$, and  $\textbf{M}_{{s_j}{s_j}}$, $\textbf{M}_{{s_j}{u_j}}$, $\textbf{M}_{{u_j}{s_j}}$, and $\textbf{M}_{{u_j}{u_j}}$ are sub-matrices of $\textbf{M}$. Further, $\textbf{r}$ and $\textbf{M}$ for different SEMs are provided in Table~\ref{tab:propertis_full_sem}.

\begin{algorithm}
  \caption{MCMC steps within the $t^{th}$ iteration of the HVB algorithm under MNAR. The missing values are updated one block at a time.}
  \begin{algorithmic}[1]
  \label{alg:AugvbMCMCstepB}
  \STATE Initialise missing values $\textbf{y}_{u,0}=(\textbf{y}^\top_{u_1,0},\dots,\textbf{y}^\top_{u_k,0})^\top~\sim p(\textbf{y}_u\mid \boldsymbol{\xi}^{(t)}, \textbf{y}_o)$
  \FOR{$i=1, \dots, N_1$} 
  \FOR{$j=1, \dots, k$}
      \STATE Sample $\widetilde{\textbf{y}}_{u_j}$ from the proposal distribution $p(\widetilde{\textbf{y}}_{u_j}\mid \boldsymbol{\xi}^{(t)},\textbf{y}_o,\textbf{y}_{u,{i-1}}^{(-j)})$, where ${\textbf{y}_{u,i-1}^{(-j)}=(\textbf{y}^\top_{u_1,i}, \textbf{y}^\top_{u_2,i}, \hdots ,\textbf{y}^\top_{u_{j-1},i},\textbf{y}^\top_{u_{j+1},i-1},\hdots, ,\textbf{y}^\top_{u_{k},i-1})^\top}$.  
      \STATE Sample $u$ from uniform distribution, $u~\sim \mathcal{U}(0,1)$
      \STATE Calculate $a=\text{min}\left (1,\frac{p(\textbf{m} \mid \widetilde{\textbf{y}},\boldsymbol{\psi}^{(t)})}{p(\textbf{m} \mid \textbf{y}_{i-1},\boldsymbol{\psi}^{(t)})}\right)$, where $\widetilde{\textbf{y}}=(\textbf{y}_o^{\top},{\textbf{y}_{u,i-1}^{(-j)}}^\top,{\widetilde{\textbf{y}}_{u_j}}^\top)^\top$ and ${\textbf{y}_{i-1}=(\textbf{y}_o^{\top},{\textbf{y}_{u,i-1}^{(-j)}}^\top,{\textbf{y}_{u_j,{i-1}}}^\top)^\top}$
      \IF{$a>u$}
        \STATE $\textbf{y}_{u_j,i}=\widetilde{\textbf{y}}_{u_j}$
      \ELSE
        \STATE $\textbf{y}_{u_j,i}=\textbf{y}_{{u_j,i-1}}$
      \ENDIF
  \ENDFOR
   \STATE $\textbf{y}_{u,i}=(\textbf{y}_{u_1,i}^\top,\hdots ,\textbf{y}_{u_k,i}^\top)^\top$
  \ENDFOR
\STATE  Output $\textbf{y}^{(t)}_{u}=\textbf{y}_{u,N_1}$
  \end{algorithmic}
\end{algorithm}

In step 4 of Algorithm~\ref{alg:AugvbMCMCstepB}, proposals from the conditional distributions, ${p(\widetilde{\textbf{y}}_{u_j}\mid \boldsymbol{\xi}^{(t)},\textbf{y}_o,\textbf{y}_{u,{i-1}}^{(-j)})={p(\widetilde{\textbf{y}}_{u_j}\mid \boldsymbol{\xi}^{(t)},\textbf{y}_{s_j})}}$ needs to be generated. For the SEM-Gau, and SEM-t, this distribution follows a multivariate Gaussian distribution with the mean $\textbf{X}_{u_j}\boldsymbol{\beta}-\textbf{M}_{{u_j}{u_j}}^{-1}\textbf{M}_{{u_j}{s_j}}\textbf{r}_{s_j}$ and the covariance matrix $\sigma^2_{\textbf{e}}\textbf{M}_{{u_j}{u_j}}^{-1}$. Expressions of $\textbf{r}$ and $\textbf{M}$ for SEM-Gau and SEM-t are provided in Table~\ref{tab:propertis_full_sem}, and details on the partitioning of $\textbf{r}$, $\textbf{X}$, and $\textbf{M}$ are given in Equation~\eqref{mat:portions_of_xwMB}.

For YJ-SEM-Gau and YJ-SEM-t, the proposals are generated in two steps. First, we sample $\widetilde{\textbf{y}}_{u_j}^{\ast}$ from the full conditional distribution $p({\widetilde{\textbf{y}}^{\ast}}_{u_j}\mid \boldsymbol{\xi}^{(t)},\textbf{y}_{s_j})$, which follows a multivariate Gaussian with the mean vector given by $\textbf{X}_{u_j}\boldsymbol{\beta}-\textbf{M}_{{u_j}{u_j}}^{-1}\textbf{M}_{{u_j}{s_j}}(\textbf{y}^\ast_{s_j}-\textbf{X}_{s_j}\boldsymbol{\beta})$ and the covariance matrix $\sigma^2_{\textbf{e}}\textbf{M}_{{u_j}{u_j}}^{-1}$, where $\textbf{y}^\ast_{s_j}=t_{\gamma}(\textbf{y}_{s_j})$. Then, we apply the inverse
 Yeo-Johnson (YJ) transformation to obtain the final proposal: \( \widetilde{\textbf{y}}_{u_j} = t_{\gamma}^{-1}(\widetilde{\textbf{y}}_{u_j}^*) \).




\section{Bayesian model comparison for SEMs}
\label{sec:online_model_comp}

This section presents expressions for the term \textrm{log}~$p(\textbf{y} \mid \boldsymbol{\phi})$ for different SEMs. These terms are used to compute the deviance information criterion (DIC)~\citep{10.1214/06-BA122} values discussed in Section~\ref{NonGauSAR_sec:DIC} of the main paper. Note that for the SEM-Gau, and YJ-SEM-Gau, $\boldsymbol{\xi}=\boldsymbol{\phi}$, meaning for these two models $p(\textbf{y} \mid \boldsymbol{\phi})=p(\textbf{y} \mid \boldsymbol{\xi})$. For the SEM-Gau, $p(\textbf{y} \mid \boldsymbol{\xi})$ is multivariate Gaussian; see Section~\ref{sec:models_withoutyj} of the main paper. Corresponding log-likelihood is given by:
\begin{equation}
\label{eq:log.like.SEM.Gau_online}
    \text{log}~p(\textbf{y} \mid \boldsymbol{\phi} )=-\frac{n}{2}\textrm{log}(2\pi)-\frac{n}{2}\textrm{log}(\sigma^2_{\boldsymbol{e}})+\frac{1}{2}\textrm{log}|\textbf{M}|-\frac{1}{2\sigma^2_{\boldsymbol{e}}}\textbf{r}^\top\textbf{M}\textbf{r},
\end{equation}

\noindent where $\textbf{M}$ and $\textbf{r}$ are given in Table~\ref{tab:propertis_full_sem}.

For the YJ-SEM-Gau, $p(\textbf{y} \mid \boldsymbol{\phi})=p(\textbf{y} \mid \boldsymbol{\xi})$ is derived in Section~\ref{sec:SEM_YJ} of the main paper. Corresponding log-likelihood is given by:
\begin{equation}
\label{eq:log.like.YJ.SEM.Gau}
    \begin{split}
           \text{log}~p(\textbf{y} \mid \boldsymbol{\phi} )&=-\frac{n}{2}\textrm{log}(2\pi)-\frac{n}{2}\textrm{log}(\sigma^2_{\boldsymbol{e}})+\frac{1}{2}\textrm{log}|\textbf{M}|-\frac{1}{2\sigma^2_{\boldsymbol{e}}}\textbf{r}^\top\textbf{M}\textbf{r} \\ &+
           \sum_{i=1}^n \text{log}\left(  \frac{dt_{\gamma}(y_i)}{dy_i} \right),
    \end{split}
\end{equation}

\noindent where $\textbf{M}$, and $\textbf{r}$ are given in Table~\ref{tab:propertis_full_sem}, while $\frac{dt_{\gamma}(y_i)}{dy_i} $ is given in Equation~\eqref{eq:der_yj_by_yi}.

For the SEM-t, 
\begin{equation}
    p(\textbf{y} \mid \boldsymbol{\phi})=\int_{0}^{\infty}\hdots\int_{0}^{\infty}p(\textbf{y} \mid \boldsymbol{\tau},\boldsymbol{\phi})p({\tau}_1\mid \boldsymbol{\phi})\hdots p({\tau}_n\mid \boldsymbol{\phi})d{\tau}_1\hdots d{\tau}_n
\end{equation}

\noindent where $p(\textbf{y} \mid \boldsymbol{\tau},\boldsymbol{\phi})=p(\textbf{y} \mid \boldsymbol{\xi})$ is multivariate Gaussian, with its mean vector and covariance matrix specified in Table~\ref{tab:propertis_full_sem}. Each  $p({\tau}_i\mid \boldsymbol{\phi})$ (for $i = 1, \ldots, n$) follows an inverse gamma distribution where both the shape parameter and the rate parameter are set to $\nu/2$; see Section~\ref{sec:models_withoutyj} of the main paper. This hierarchical representation implies that the marginal distribution 
$p(\mathbf{y} \mid \boldsymbol{\phi})$ is a multivariate Student’s $t$-distribution with mean vector $\textbf{X}\boldsymbol{\beta}$, scale matrix $\sigma^2_{\boldsymbol{e}}(\mathbf{A}^\top \mathbf{A})^{-1}$, and degrees of freedom $\nu$. Then, the corresponding log-likelihood is given by:
\begin{equation}
\label{eq:log.like.SEM.t_online}
\begin{split}
    \text{log}~p(\textbf{y} \mid \boldsymbol{\phi} )&=\textrm{log}~\Gamma\left(\frac{\nu+2}{2}\right)-\textrm{log}~\Gamma\left(\frac{\nu}{2}\right)+\frac{1}{2}\textrm{log}|\textbf{A}^\top \textbf{A}|\\
    &-\frac{n}{2}\textrm{log}(\pi\nu\sigma^2_{\textbf{e}})-\frac{(n+\nu)}{2}\textrm{log}\left(1+\frac{1}{\nu\sigma^2_{\boldsymbol{e}}}\textbf{r}^\top\textbf{A}^\top \textbf{A}\textbf{r}\right),
    \end{split}
\end{equation}

\noindent where $\textbf{r}=\textbf{y}-\textbf{X}\boldsymbol{\beta}$, and $\Gamma (\cdot)$ is the gamma function.

Similarly, the logarithm of $p(\textbf{y} \mid \boldsymbol{\phi})$ for YJ-SEM-t can be derived as
\begin{equation}
\label{eq:log.like.YJSEM.t_online}
\begin{split}
    \text{log}~p(\textbf{y} \mid \boldsymbol{\phi} )&=\textrm{log}~\Gamma\left(\frac{\nu+2}{2}\right)-\textrm{log}~\Gamma\left(\frac{\nu}{2}\right)+\frac{1}{2}\textrm{log}|\textbf{A}^\top \textbf{A}|\\
    &-\frac{n}{2}\textrm{log}(\pi\nu\sigma^2_{\textbf{e}})-\frac{(n+\nu)}{2}\textrm{log}\left(1+\frac{1}{\nu\sigma^2_{\boldsymbol{e}}}\textbf{r}^\top\textbf{A}^\top \textbf{A}\textbf{r}\right)\\ &+
           \sum_{i=1}^n \text{log}\left(  \frac{dt_{\gamma}(y_i)}{dy_i} \right),
    \end{split}
\end{equation}
\noindent where $\textbf{r}=t_\gamma(\textbf{y})-\textbf{X}\boldsymbol{\beta}$.

\section{Additional figures for simulation study: Assessing the accuracy of VB methods in Section~\ref{sec:simulationstudy-1}}
\label{sec:sim_comp_online}

\begin{figure}[H]
    \centering
    \includegraphics[width=0.5\linewidth]{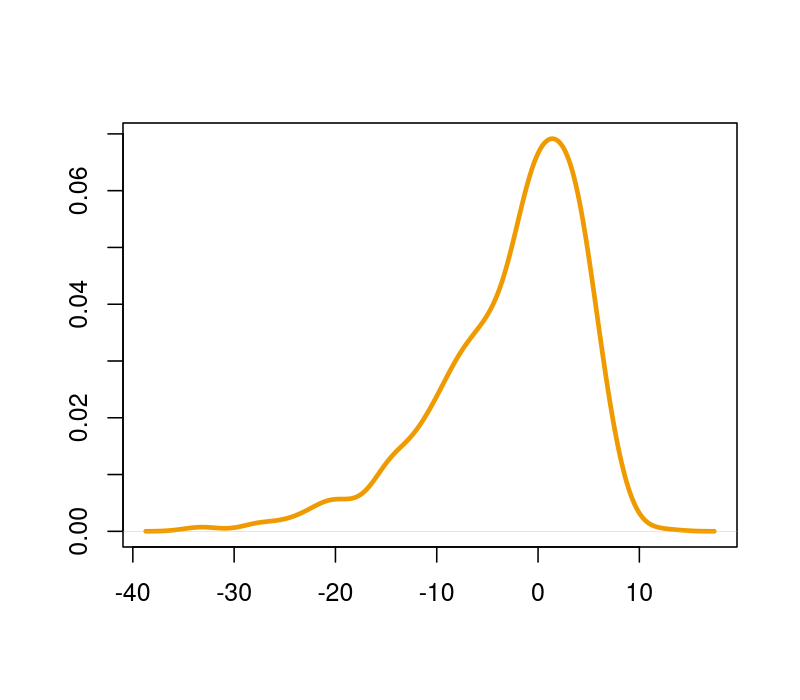}  
    \caption{
      Kernel density plot of the simulated response variable generated using the YJ-SEM-Gau with $\gamma = 1.25$ in the simulation study described in Section~\ref{sec:simulationstudy-1} of the main paper.  
    }
    \label{fig:HMCvsVB_density}
\end{figure}

Section~\ref{sec:simulationstudy-1} of the main paper compares the posterior density estimates of the model parameters for YJ-SEM-Gau obtained using VB (and HVB when missing values are present) and HMC~\citep{neal2011HMCchapter}, based on a simulated dataset analysed both with and without missing values. To generate the data, the model parameters are set as follows: the fixed effects ($\boldsymbol{\beta}$) are randomly drawn from discrete uniform values between \(-3\) and \(3\) (excluding \(0\)). The error variance parameter is set to $\sigma^2_{\mathbf{e}} = 1$ and the spatial autoregressive parameter is $\rho = 0.8$. The Yeo-Johnson transformation parameter is set to $\gamma = 1.25$, which introduces moderate left skewness in the simulated response variable. The kernel density plot of the simulated response variable is provided in Figure~\ref{fig:HMCvsVB_density}. As expected, the density is moderately skewed to the left. 

\section{Simulation study: Assessing the accuracy and robustness of proposed SEMs (full data case)}
\label{sec:sim_full}

This section fits SEM-Gau, SEM-t, YJ-SEM-Gau, and YJ-SEM-t to simulated datasets 1 and 2 (without missing data) described in Section~\ref{sec:simulationstudy-2} of the main paper, and evaluates how effectively these models, along with their VB estimation methods, capture the characteristics of these datasets.

The initial values for the VB algorithms are set as follows:  the elements of the variational mean vector ($\boldsymbol{\mu}$) corresponding to $\boldsymbol{\beta}$, $\sigma^2_{\textbf{e}}$, and $\rho$ are initialised using estimates obtained by fitting the SEM-Gau via maximum likelihood (ML) estimation. The variational means corresponding to the parameters $\nu$ and $\gamma$ are initialised to 4 and 1, respectively. For the models with Student's $t$ errors, the set of latent variables $\boldsymbol{\tau} = (\tau_1, \dots, \tau_n)^\top$ must also be initialised. The variational means corresponding to these latent variables are initialised by sampling from the prior distribution ${p(\tau_i \mid \nu^{(0)}) \sim \text{IG}(\frac{\nu^{(0)}}{2}, \frac{\nu^{(0)}}{2})}$ for $i = 1, \dots, n$, where $\nu^{(0)} = 4$ denotes the initial value of $\nu$. For the variational covariance matrix parameters, the elements of $\textbf{B}$ and the diagonal elements of $\textbf{D}$ are all initialised to 0.01. We use $p=4$ factors.

After some experimentation, we found that the models with Student’s $t$ errors required a relatively large number of iterations to converge. This higher computational cost is due to the added complexity of estimating the latent variables $\boldsymbol{\tau}$ and the parameter $\nu$. For simulated dataset 1, the VB algorithms are run for 10,000 iterations for both SEM-Gau and YJ-SEM-Gau, and 20,000 iterations for SEM-t and YJ-SEM-t. For simulated dataset 2, convergence is achieved after 10,000 iterations for SEM-Gau and YJ-SEM-Gau, and 40,000 iterations for SEM-t and YJ-SEM-t. Convergence is assessed visually by inspecting the variational mean plots over iterations, which are presented in \textcolor{black}{Section~\ref{sec:sim_conv_analysis_7}}.  Once convergence is achieved, 10,000 samples of model parameters are drawn from the variational distribution $q_{\boldsymbol{\lambda}}(\boldsymbol{\xi})$ to form the posterior sample.

\subsection{Results for the dataset 1}

Table~\ref{tab:sim_full_dset1} shows the posterior means and the 95\% credible intervals for some of the model parameters of the SEM-Gau, SEM-t, YJ-SEM-Gau, and YJ-SEM-t obtained using the VB algorithm applied to the simulated dataset 1. The table also includes the computation times (in seconds) for one VB iteration, as well as the $\text{DIC}_1$ and $\text{DIC}_2$ values for each model, calculated using Equations~\eqref{eq:DIC1} and~\eqref{eq:DIC2}, respectively, in Section~\ref{NonGauSAR_sec:DIC} of the main paper. 

\begin{table}[ht]
\centering
\small
\caption{Posterior means and 95\% credible intervals for selected parameters of different SEMs based on the simulated dataset 1 (full data), along with the computation time (CT) in seconds for one VB iteration. The table also presents the DIC values for each model. Parameters labelled as 'NA' indicate that they are not applicable to the corresponding model. 
}
\begin{tabular}{ccccc}
\hline
 & SEM-Gau & SEM-t & YJ-SEM-Gau   & YJ-SEM-t \\
\hline
${\beta}_0=-1$ & \makecell{3.1962 \\ (2.9423, 3.4453)} & \makecell{ 0.9017\\ (0.8048, 0.9978)} & \makecell{-1.0319\\ (-1.0709, -0.9931)}   &  \makecell{-1.0065 \\ (-1.0391, -0.9742)} \\
${\beta}_1=1$ & \makecell{1.5352\\ (1.3089, 1.7603)} & \makecell{0.9097\\ (0.8079, 1.0136)} & \makecell{0.9941\\ (0.9669, 1.0212)}   &  \makecell{0.9880\\ (0.9568, 1.0192)} \\

$\sigma^2_{\textbf{e}}=0.5$ & \makecell{58.4686\\ (56.0301, 60.9582)} & \makecell{8.2516\\ (7.7781, 8.7454)} & \makecell{1.0362\\ (0.9909, 1.0823)}   &  \makecell{0.6318\\ (0.6090, 0.6554)}   \\

$\rho=0.8$ &\makecell{0.2274\\ (0.1969, 0.2577)} & \makecell{0.1626\\ (0.1416, 0.1835)} & \makecell{0.7945\\ (0.7832, 0.8053)}   &  \makecell{0.7925\\ (0.7798, 0.8048)}   \\

$\nu=4$ & NA & \makecell{3.0049\\ (3.0000, 3.0063)}   & NA & \makecell{5.9993\\ (5.7126, 6.3156)}  \\

$\gamma=0.5$ & NA & NA  & \makecell{0.4964\\ (0.4918, 0.5010)}  & \makecell{0.4985\\ (0.4954, 0.5016)}  \\
\hline

$\text{DIC}_1$ & 30328.85  & 31535.42  & 14734.73       & {14732.85} \\
$\text{DIC}_2$ & 30328.34 & 31655.98& 14737.32   &  {14732.32}\\
CT & 0.1322& 0.1206& 0.6715&  0.7252\\
\hline
\end{tabular}
\label{tab:sim_full_dset1}
\end{table}

We begin by comparing the estimated values of the fixed effects (specifically $\beta_0$ and $\beta_1$), the variance parameter $ \sigma_{\textbf{e}}^2 $, and the spatial autocorrelation parameter $ \rho $, as these parameters are present in all four models. 
The posterior means for $\beta_0$, $\beta_1$ and $ \rho $ are nearly identical for YJ-SEM-Gau and YJ-SEM-t, closely matching the true values of $-1$ for $ \beta_0 $, $1$ for $\beta_1$, and $0.8$ for $ \rho $. In contrast, the estimates from SEM-Gau and SEM-t show significant deviations from the true values. The estimated posterior means of $ \sigma_{\textbf{e}}^2 $ from SEM-Gau, SEM-t, and YJ-SEM-Gau deviate considerably from the true value, while the posterior mean obtained from YJ-SEM-t is much closer to the true value. The models incorporating the YJ transformation, YJ-SEM-Gau and YJ-SEM-t, successfully recover the parameter $\gamma $. The degrees of freedom ($\nu$) estimated by the SEM-t and YJ-SEM-t models show slight deviations from the true value.

For the simulated dataset 1, the model fit assessment indicates that YJ-SEM-t is the best-performing model, as it achieves the lowest $\text{DIC}_1$ and $\text{DIC}_2$ values among all four models. This outcome is expected, given that the dataset is simulated from YJ-SEM-t. The second-best model is YJ-SEM-Gau, which has the next lowest DIC values. Notably, the difference in DIC between YJ-SEM-t and YJ-SEM-Gau is significantly small, suggesting that YJ-SEM-Gau provides a model fit nearly as good as YJ-SEM-t for the dataset 1.

\begin{figure}[H]
    \centering
    \includegraphics[width=0.8\linewidth]{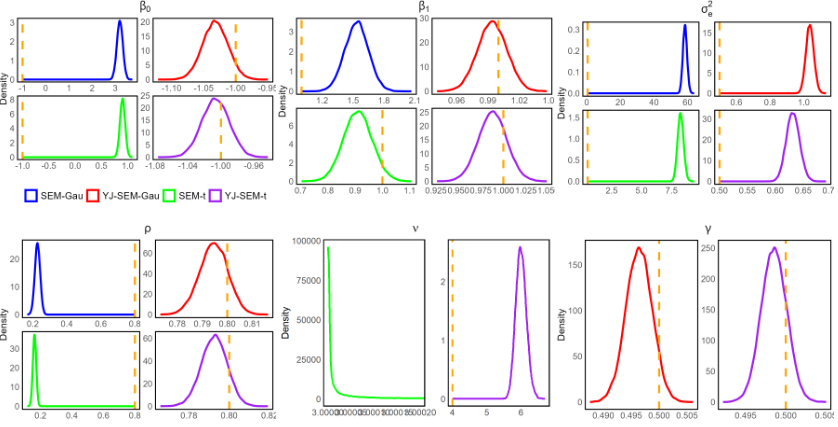}  
    \caption{
      Posterior densities of some of the parameters for the SEM-Gau, YJ-SEM-Gau, SEM-t, and YJ-SEM-t, fitted to the simulated dataset 1 (without missing data), using the VB method. The vertical dotted lines indicate the true parameter values.
    }
    \label{fig:densities_sk_full}
\end{figure}

Figure~\ref{fig:densities_sk_full} compares the posterior densities of selected model parameters obtained using the proposed VB method, applied to different SEMs for the simulated dataset 1. As expected, the YJ-SEM-t recovers the true parameter values more accurately than the other three models, since dataset 1 is simulated using the YJ-SEM-t. 

\subsection{Results for the dataset 2}

Table~\ref{tab:sim_full_dset2} presents the posterior means and 95\% credible intervals for selected model parameters obtained from the SEM-Gau, SEM-t, YJ-SEM-Gau, and YJ-SEM-t applied to the simulated dataset 2, along with the computation cost for one VB iteration. The table also includes the $\text{DIC}_1$ and $\text{DIC}_2$ values for each model. Unlike the simulated dataset 1, which is skewed and has heavy tails, the simulated dataset 2 is more symmetrical with lighter tails. The posterior mean estimates for parameters $\beta_0$, $\beta_1$, $\sigma^2_{\textbf{e}}$, and $\rho$ across all four models closely match their true values. Likewise, the posterior mean estimate of $\nu$ for SEM-t and YJ-SEM-t is close to the true value of $30$, while the posterior mean estimate of $\gamma$ for YJ-SEM-Gau and YJ-SEM-t is close to the true value of $1$. The overall accuracy of parameter estimation across all models is further supported by the nearly identical $\text{DIC}_1$ and $\text{DIC}_2$ values, with SEM-Gau having the lowest values, indicating the best fit for the simulated dataset 2.

\begin{table}[ht]
\centering
\small
\caption{Posterior means and 95\% credible intervals for selected parameters of different SEMs based on the simulated dataset 2 (full data), along with the computation time (CT) in seconds for one VB iteration. The table also presents the DIC values for each model. Parameters labelled as 'NA' indicate that they are not applicable to the corresponding model.}
\begin{tabular}{ccccc}
\hline
 & SEM-Gau & SEM-t & YJ-SEM-Gau   & YJ-SEM-t \\
\hline
${\beta}_0=-3$ & \makecell{ -2.9976\\(-3.0258, -2.9692)} & \makecell{-3.0008\\(-3.0305, -2.9707)}   & \makecell{-2.9781\\(-3.0108, -2.9449)}   &  \makecell{-2.9808\\(-3.0033, -2.9584)}  \\
${\beta}_1=2$ & \makecell{1.9968\\(1.9773, 2.0161)}  & \makecell{1.9947\\(1.9774, 2.0121)}  & \makecell{1.9920\\(1.9739, 2.0104)}  & \makecell{1.9950\\(1.9716, 2.0183)}\\

$\sigma^2_{\textbf{e}}=0.5$ & \makecell{0.5414\\(0.5175, 0.5661)}  & \makecell{0.4921\\(0.4734, 0.5113)}  & \makecell{0.5378\\(0.5167, 0.5598)}  & \makecell{0.4981\\(0.4775, 0.5192)} \\

$\rho=0.8$ & \makecell{0.7987\\(0.7880, 0.8091)}  & \makecell{0.7970\\(0.7856, 0.8080)}  & \makecell{0.7984\\(0.7873, 0.8091)}  & \makecell{0.7977\\(0.7857, 0.8092)}  \\

$\nu=30$ & NA & \makecell{22.7544\\(21.8282, 23.7190)}   & NA & \makecell{26.1797 \\(24.8903, 27.5045)}   \\

$\gamma=1$ & NA & NA  & \makecell{1.0028\\(0.9998, 1.0057)}  & \makecell{1.0035\\( 0.9987, 1.0083)}   \\
\hline

$\text{DIC}_1$ &  {11315.17} &  11320.44   &   11475.1     &  11324.83\\
$\text{DIC}_2$  &     {11314.81} & 11321.4 &   11535.12 &  11324.85 \\
CT & 0.1280 & 0.6718 & 0.1197 & 0.7241  \\
\hline
\end{tabular}
\label{tab:sim_full_dset2}
\end{table}

\begin{figure}[H]
    \centering
    \includegraphics[width=0.8\linewidth]{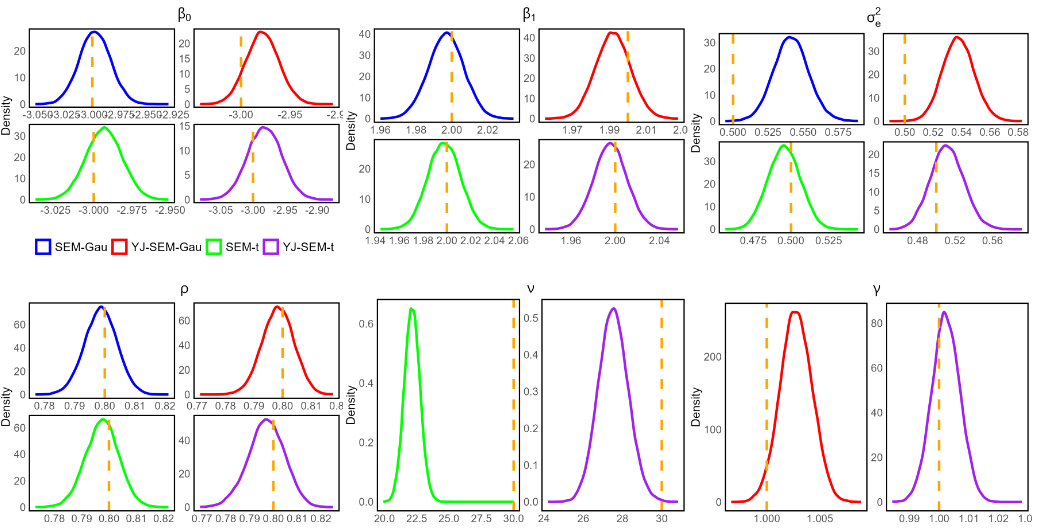}  
    \caption{
      Posterior densities of some of the parameters for the SEM-Gau, YJ-SEM-Gau, SEM-t, and YJ-SEM-t, fitted to the simulated dataset 2 (without missing data), using the VB method. The vertical dotted lines indicate the true parameter values.
    }
    \label{fig:densities_norm_full}
\end{figure}

Figure~\ref{fig:densities_norm_full} compares the posterior densities of selected model parameters obtained using the proposed VB method, applied to different SEMs for simulated dataset 2. It can be seen that,  all four models successfully recover the true values of the parameters.

\subsection{Summary and computation time comparison}
\label{online_sec:summary_sim}

According to the computation times presented in Tables~\ref{tab:sim_full_dset1} and~\ref{tab:sim_full_dset2}, VB algorithms for both SEM-t and YJ-SEM-t, which incorporate Student's $t$ errors, require significantly more computation time per iteration compared to SEM-Gau and YJ-SEM-Gau. This is because, in models with Student's $t$ errors, the VB optimisation is substantially more complex than in models with Gaussian errors. Specifically, for models with Student's $t$ errors, the set of model parameters is $\boldsymbol{\xi} = (\boldsymbol{\phi}^\top, \boldsymbol{\tau}^\top)^\top$, whereas for models with Gaussian errors, it simplifies to $\boldsymbol{\xi} = \boldsymbol{\phi}$. Here, $\boldsymbol{\tau}$ is a latent variable vector of length $n$; see Section~\ref{sec:models_withoutyj} of the main paper. In addition to the higher per-iteration cost, SEM-t and YJ-SEM-t also require more VB iterations to achieve convergence. Together, these factors lead to substantially higher computational costs for models with Student's $t$ errors than for their Gaussian-error counterparts.

In terms of estimation performance, the results indicate that both YJ-SEM-Gau and YJ-SEM-t are particularly effective at capturing the key characteristics of non-Gaussian data. For simulated dataset 1, which exhibits skewness and heavy tails, YJ-SEM-t achieves the lowest DIC values, indicating the best fit, while YJ-SEM-Gau attains the second-lowest values. The differences between the DIC values of YJ-SEM-t and YJ-SEM-Gau are small, suggesting that YJ-SEM-Gau provides a model fit comparable to that of YJ-SEM-t. For simulated dataset 2, which is closer to Gaussian, all four models recover the true parameter values accurately, and the DIC values are nearly identical, with SEM-Gau achieving the lowest values and therefore providing the best fit.



\section{Additional figures for simulation study: Assessing the accuracy and robustness of proposed SEMs (missing data case in Section~\ref{sec:sim_miss_results})}
\label{sec:addi_fig_sec7}

This section provides additional figures supporting the analysis in Section~\ref{sec:sim_miss_results} of the main paper, which evaluates the accuracy and robustness of the proposed SEMs and HVB (HVB-AllB) methods using two simulated datasets with missing values.




\subsection{Comparison of posterior densities of model parameters}
\label{sec:comp_rob_para_den}

\begin{figure}[H]
    \centering
    \includegraphics[width=0.8\linewidth]{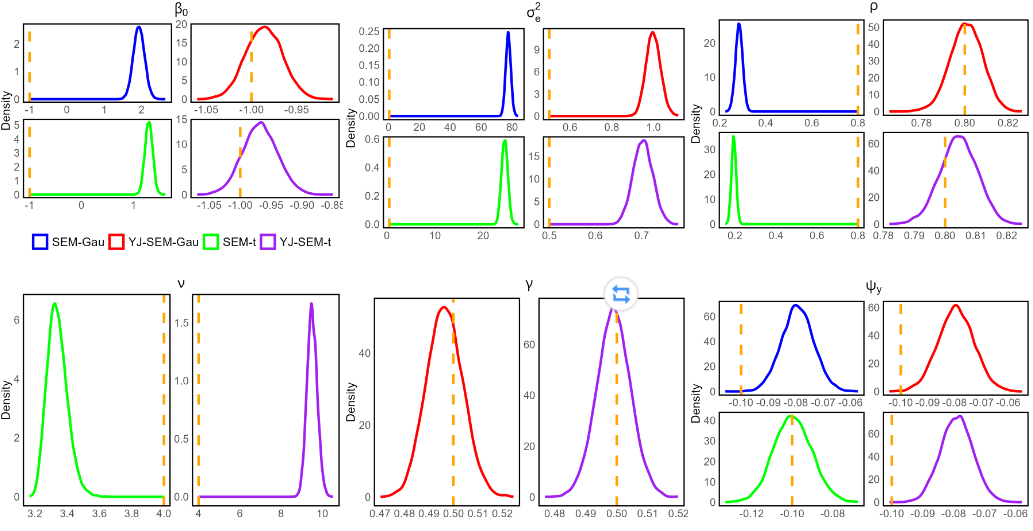}  
    \caption{
  Posterior densities of some of the parameters for the SEM-Gau, YJ-SEM-Gau, SEM-t, and YJ-SEM-t, fitted to the simulated dataset 1 with missing data, using the HVB-AllB method. The vertical dotted lines indicate the true parameter values.
    }
    \label{fig:densities_sk_miss}
\end{figure}

\begin{figure}[H]
    \centering
    \includegraphics[width=0.8\linewidth]{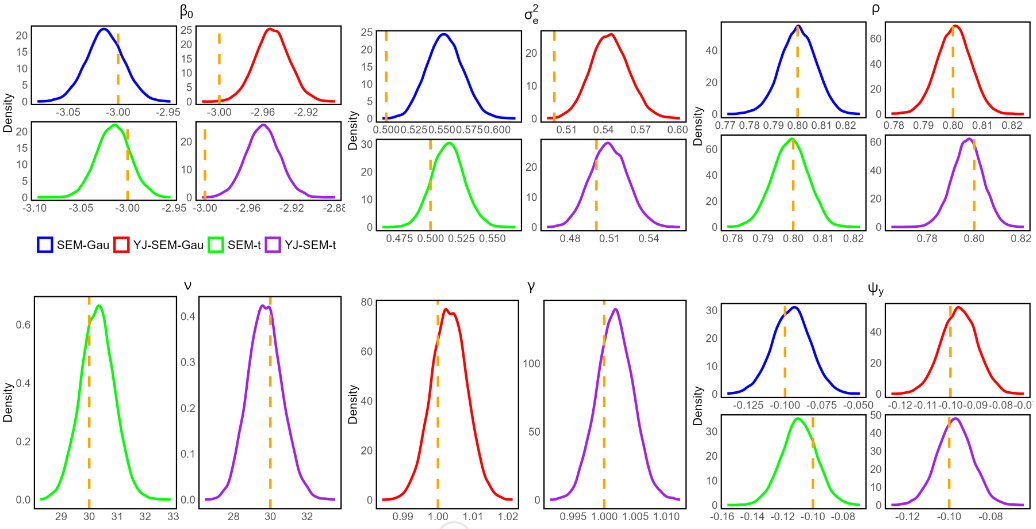}  
    \caption{
  Posterior densities of some of the parameters for the SEM-Gau, YJ-SEM-Gau, SEM-t, and YJ-SEM-t, fitted to the simulated dataset 2 with missing data, using the HVB-AllB method. The vertical dotted lines indicate the true parameter values.
    }
    \label{fig:densities_norm_miss}
\end{figure}

Figures~\ref{fig:densities_sk_miss} and~\ref{fig:densities_norm_miss} present posterior densities of selected model parameters from different SEMs, estimated using the proposed HVB-AllB method on simulated datasets~1 and~2 with missing values. Figure~\ref{fig:densities_sk_miss} shows that YJ-SEM-t recovers the 
true parameter values most accurately for simulated dataset~1, as expected, 
since this dataset was generated from the YJ-SEM-t. In contrast, 
Figure~\ref{fig:densities_norm_miss} demonstrates that, for simulated 
dataset~2, all four models recover the true parameter values well.

\subsection{Comparison of posterior densities of missing values}
\label{sec:comp_rob_missing_den}

\begin{figure}[H]
    \centering
    \begin{minipage}{0.5\textwidth}
        \centering
        \includegraphics[width=\linewidth, height=0.2\textheight]{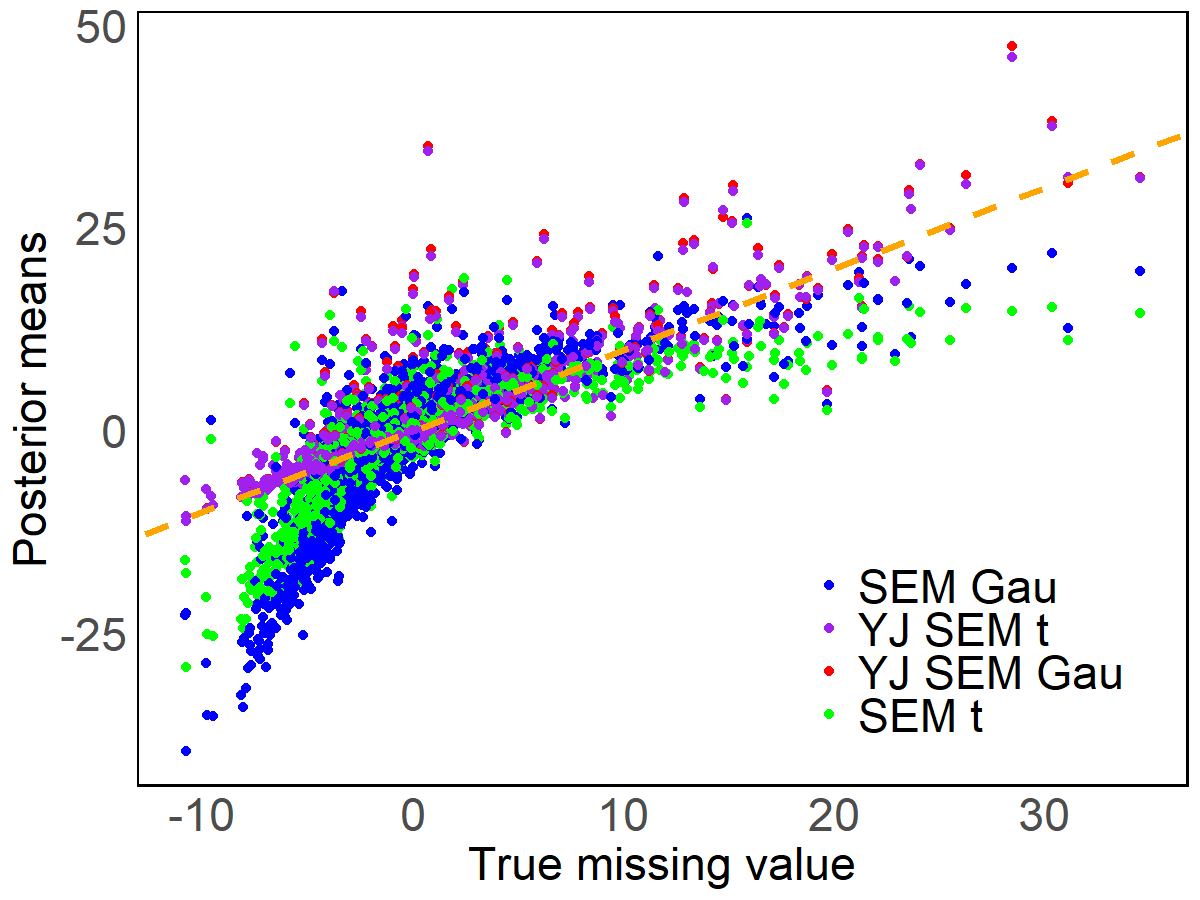} 
    \end{minipage}%
    \begin{minipage}{0.5\textwidth}
        \centering
        \includegraphics[width=\linewidth, height=0.2\textheight]{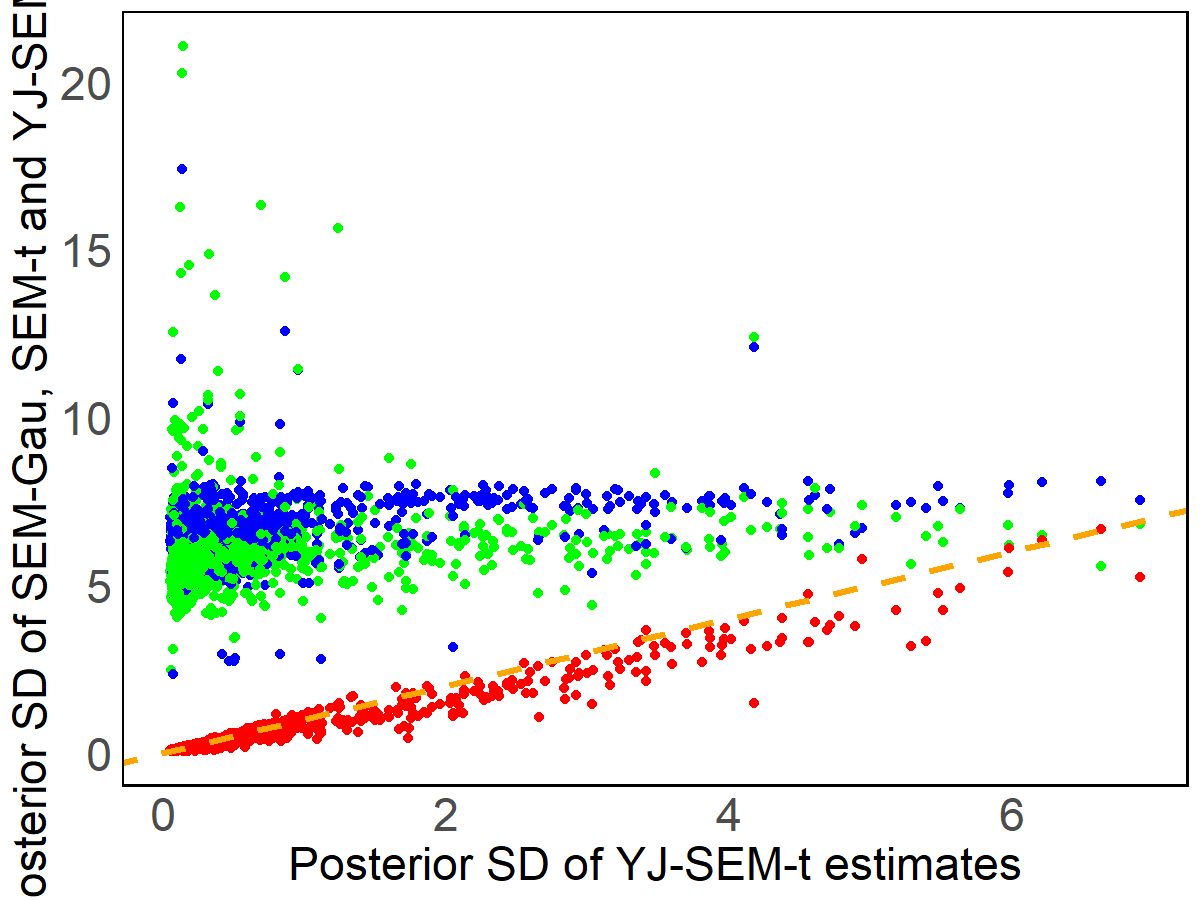} 
    \end{minipage}%
    \caption{Comparison of posterior means and standard deviations of missing values for the simulated dataset 1: The left panel displays the comparison of the posterior means of predicted missing values from the SEM-Gau, SEM-t, YJ-SEM-Gau, and YJ-SEM-t with true missing values. The right panel compares the posterior standard deviations of the missing values from SEM-Gau, SEM-t, and YJ-SEM-Gau against those obtained from YJ-SEM-t.}
    \label{fig:sem_hsk_mean_sd_yu}
\end{figure}

Figure~\ref{fig:sem_hsk_mean_sd_yu} compares the posterior means and standard deviations of the missing values $\textbf{y}_u={(y_{u_1}, \ldots, y_{u_{n_u}})}^\top$ estimated by the HVB-AllB algorithms under different SEMs for the simulated dataset 1 with missing values. The left panel compares the posterior means of the predicted missing values with their true values. The right panel compares the posterior standard deviations of the missing values estimated under the YJ-SEM-t (x-axis) with those obtained from the SEM-Gau, SEM-t, and YJ-SEM-Gau (y-axis). The posterior means of the missing values estimated under the YJ-SEM-Gau and YJ-SEM-t are closer to the true missing values compared to those estimated under the SEM-Gau and SEM-t. The posterior standard deviations of the missing values from the YJ-SEM-Gau and YJ-SEM-t are nearly identical, whereas those from the SEM-Gau and SEM-t deviate from the standard deviations obtained by the models with YJ transformation.



\begin{figure}[H]
    \centering
    \begin{minipage}{0.5\textwidth}
        \centering
        \includegraphics[width=\linewidth, height=0.2\textheight]{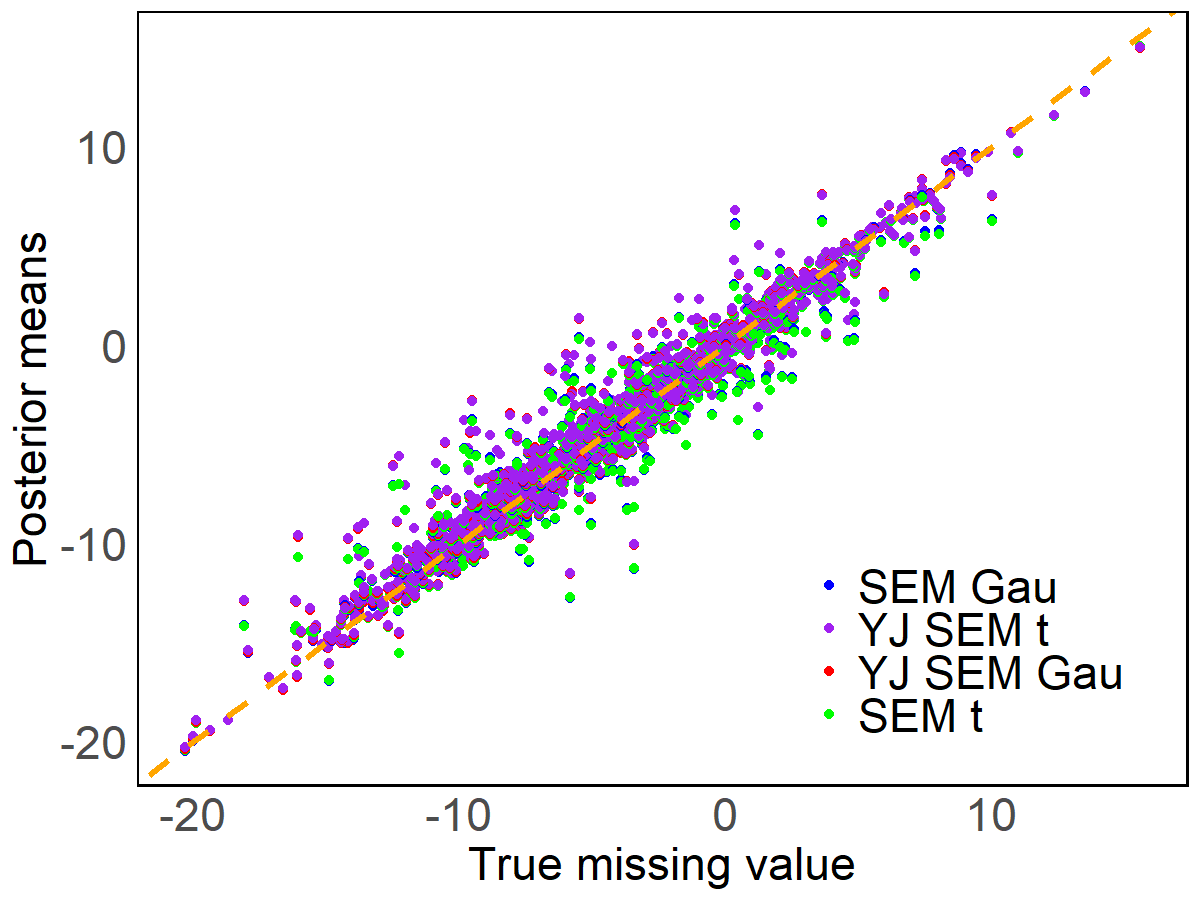} 
    \end{minipage}%
    \begin{minipage}{0.5\textwidth}
        \centering
        \includegraphics[width=\linewidth, height=0.2\textheight]{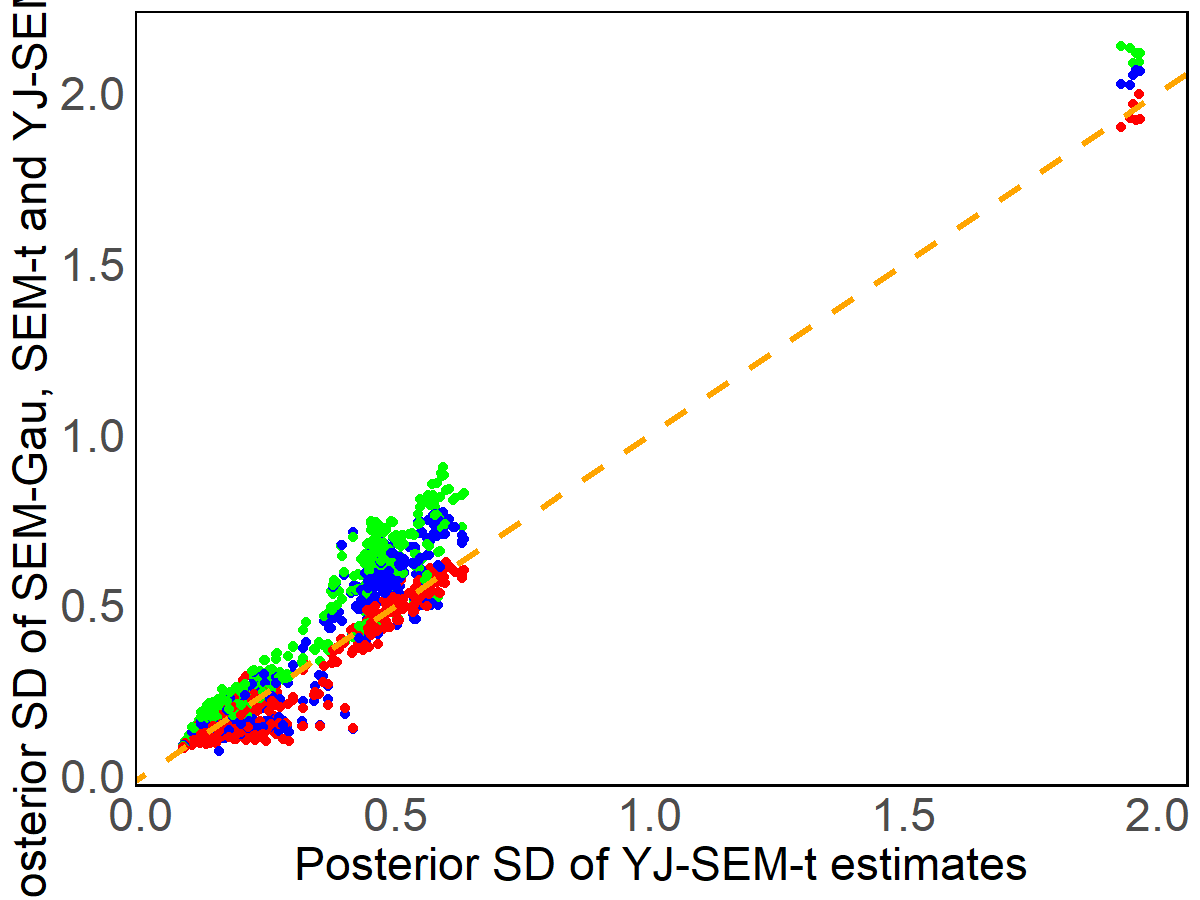} 
    \end{minipage}%
    \caption{Comparison of posterior means and standard deviations of predicted missing values for the simulated dataset 2: The left panel displays the comparison of the posterior means of missing values from the SEM-Gau, SEM-t, YJ-SEM-Gau, and YJ-SEM-t with true missing values. The right panel compares the posterior standard deviations of the missing values from SEM-Gau, SEM-t, and YJ-SEM-Gau against those obtained from YJ-SEM-t.}
    \label{fig:sem_norm_mean_sd_yu}
\end{figure}

Figure~\ref{fig:sem_norm_mean_sd_yu} compares the posterior means and standard deviations of the missing values $\textbf{y}_u={(y_{u_1}, \ldots, y_{u_{n_u}})}^\top$ estimated by the HVB-AllB algorithms under different SEMs for the simulated dataset 2 with missing values. The left panel compares the posterior means of the predicted missing values with their true values. The right panel compares the posterior standard deviations of the missing values estimated under the YJ-SEM-t (x-axis) with those obtained from the SEM-Gau, SEM-t, and YJ-SEM-Gau (y-axis). The posterior means of the missing values estimated under all four models are close to the true values. Additionally, the posterior standard deviations from the SEM-Gau, SEM-t, and YJ-SEM-Gau closely align with those from the YJ-SEM-t, with YJ-SEM-Gau showing the closest match.

\begin{figure}[H]
    \centering

    \begin{minipage}{0.49\textwidth}
        \centering
        \begin{minipage}{0.49\textwidth}
            \centering
            \includegraphics[width=\linewidth]{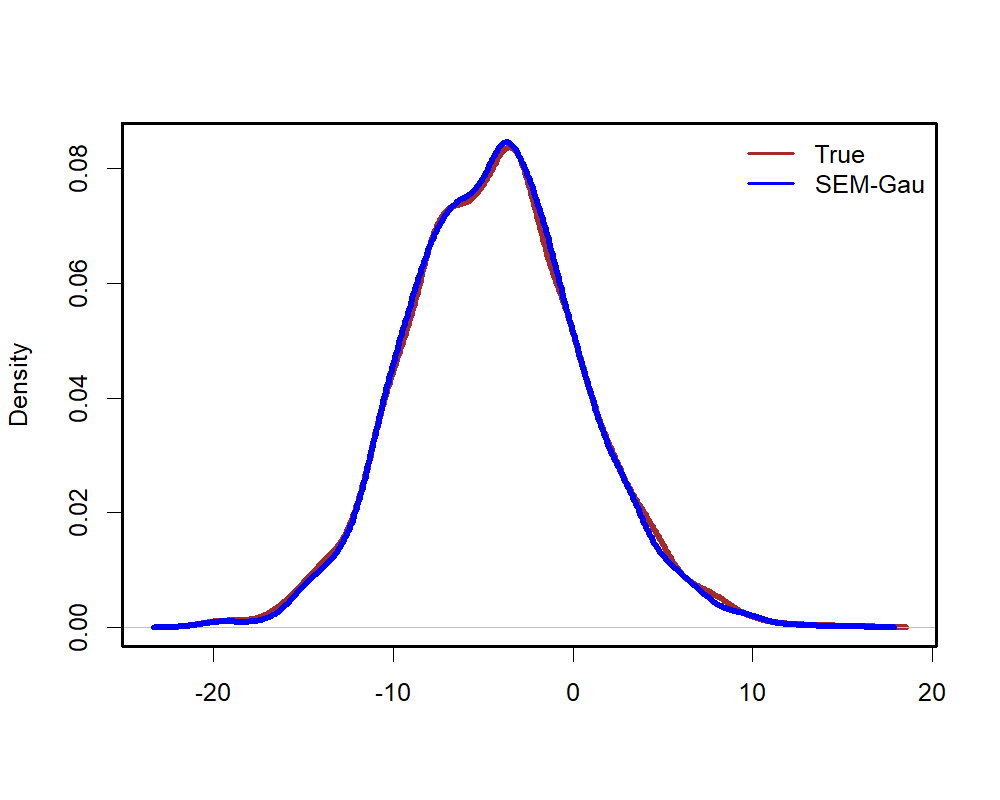}
        \end{minipage}%
        \begin{minipage}{0.49\textwidth}
            \centering
            \includegraphics[width=\linewidth]{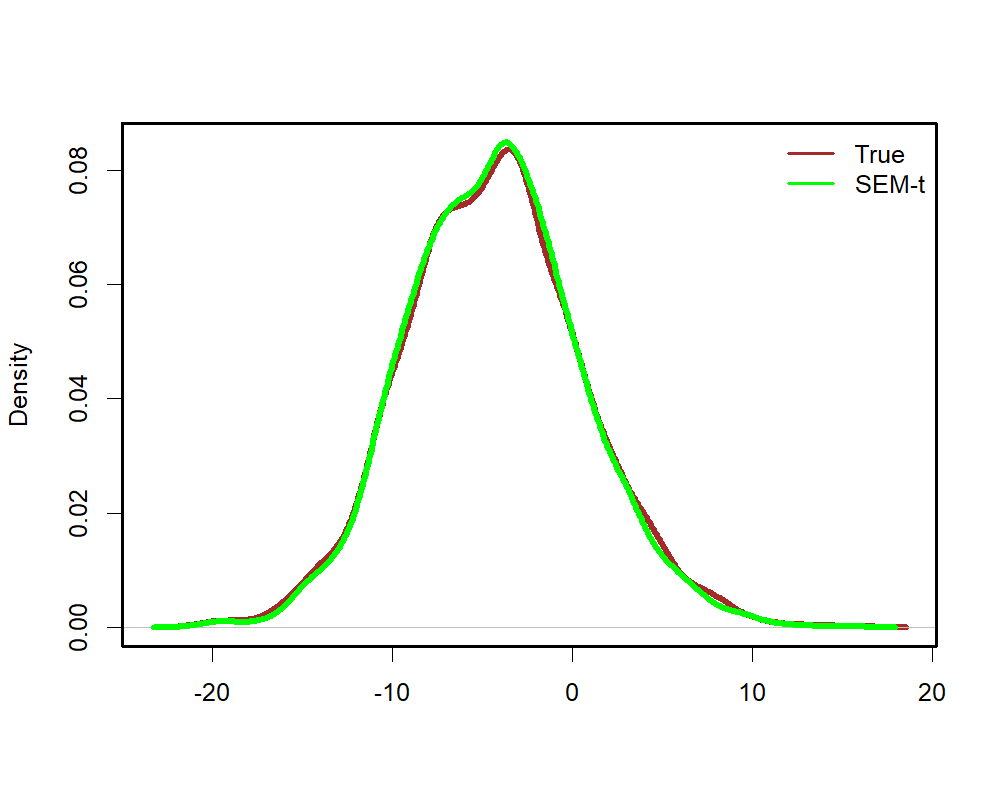}
        \end{minipage}

        \vspace{0.25cm}

        \begin{minipage}{0.49\textwidth}
            \centering
            \includegraphics[width=\linewidth]{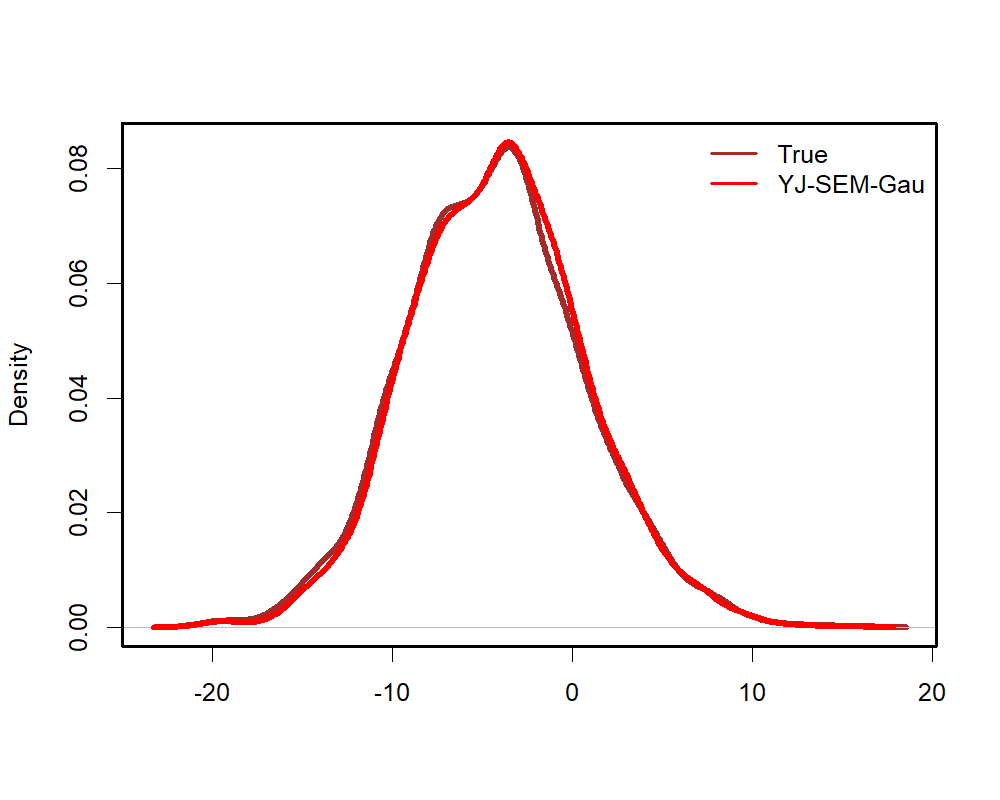}
        \end{minipage}%
        \begin{minipage}{0.49\textwidth}
            \centering
            \includegraphics[width=\linewidth]{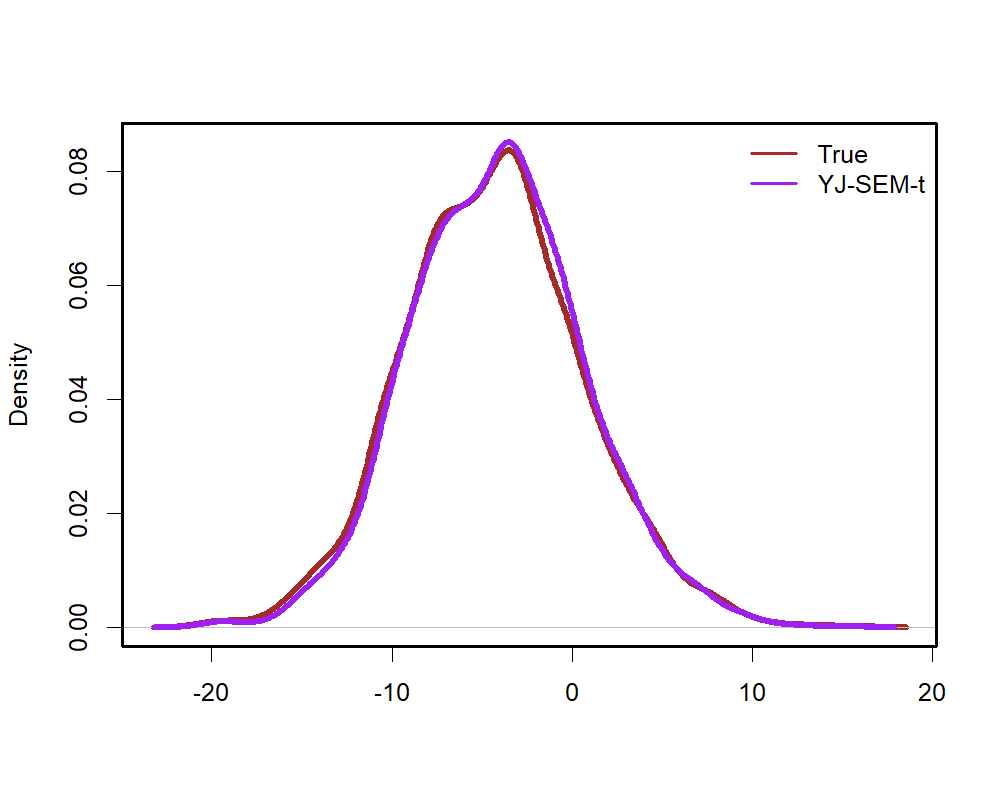}
        \end{minipage}
    \end{minipage}%
    \hfill
    \begin{minipage}{0.49\textwidth}
        \centering
        \includegraphics[width=\linewidth]{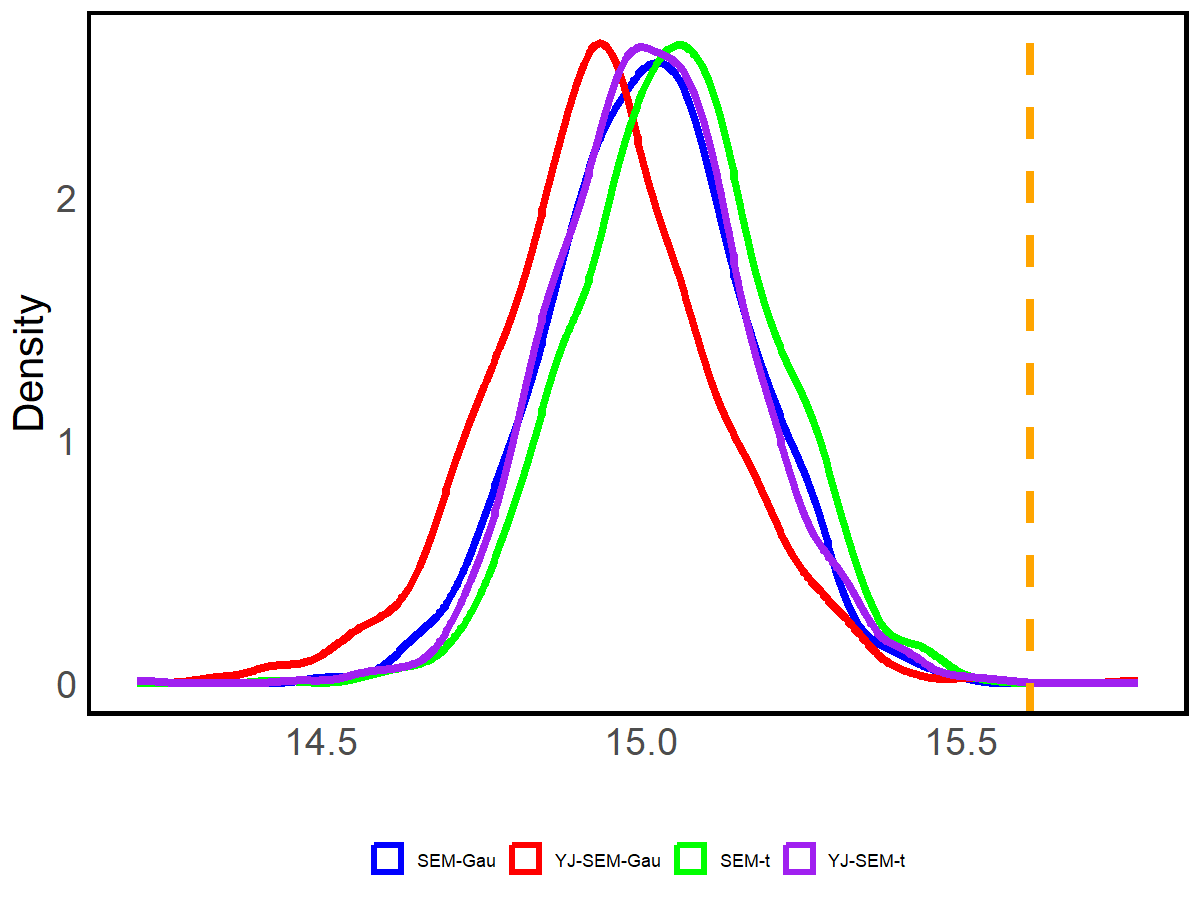}
    \end{minipage}

    \caption{
        Left panel: The kernel densities of the true missing values ($\textbf{y}_u$) and the posterior means of the missing values ($\textbf{y}_u$) obtained using the HVB-AllB method for different SEMs for the simulated dataset 2 with missing values. Right panel: The posterior density of the maximum missing value, $\textrm{max}(\textbf{y}_u)$, predicted by different SEMs for the simulated dataset 2 with missing values. The true value of the maximum missing value is indicated by the vertical line. 
    }
    \label{fig:sim_norm_miss_densities}
\end{figure}

The left four panels of Figure~\ref{fig:sim_norm_miss_densities} compare the kernel density of the true missing values ($\textbf{y}_u$) with the kernel densities of the posterior means of the missing values estimated under different SEMs for the simulated dataset 2. All four models produce nearly identical densities of posterior means of missing values that closely match the density of the true missing values. The right panel of Figure~\ref{fig:sim_norm_miss_densities} presents the posterior density of the maximum missing value, $\textrm{max}(\textbf{y}_u)$, estimated under each SEM. The true value of $\textrm{max}(\textbf{y}_u)$ lies within the posterior distribution estimated by all models. These results are consistent with expectations, as all four models produce similar parameter estimates and $\text{DIC}_5$ values (see Table~\ref{tab:sim_miss_dset2} in Section~\ref{sec:sim_miss_results} of the main paper). A corresponding figure for simulated dataset 1 is presented in Figure~\ref{fig:sim_sk_miss_densities} in Section~\ref{sec:sim_miss_results} of the main paper.


\section{Additional figures for real data application in Section~\ref{sec:real}}
\label{sec:online_real}

This section provides additional figures related to the real data application discussed in Section~\ref{sec:real} of the main paper.

\begin{figure}[H]
    \centering
    \begin{minipage}{0.5\textwidth}
        \centering
        \includegraphics[width=\linewidth, height=0.3\textheight]{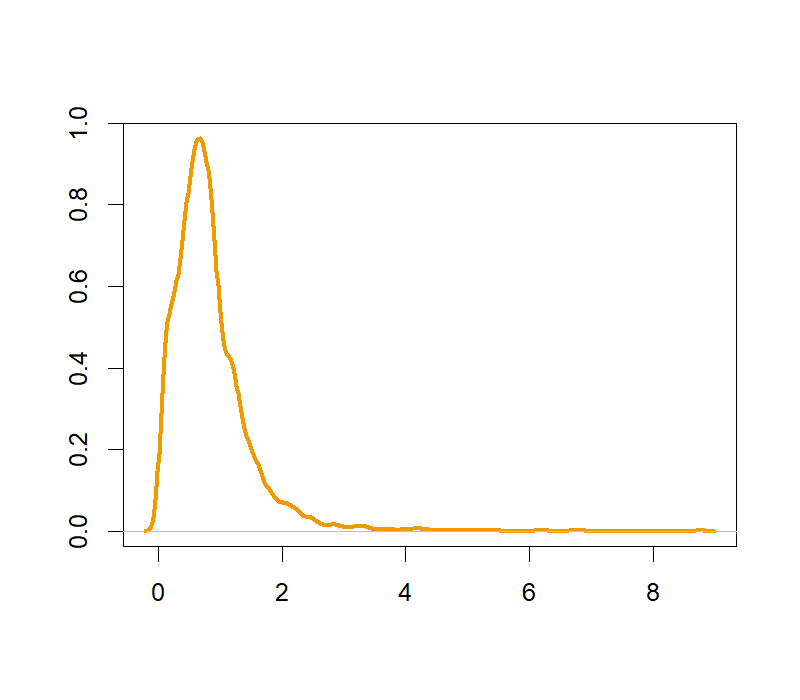} 
    \end{minipage}%
    \begin{minipage}{0.5\textwidth}
        \centering
        \includegraphics[width=\linewidth, height=0.3\textheight]{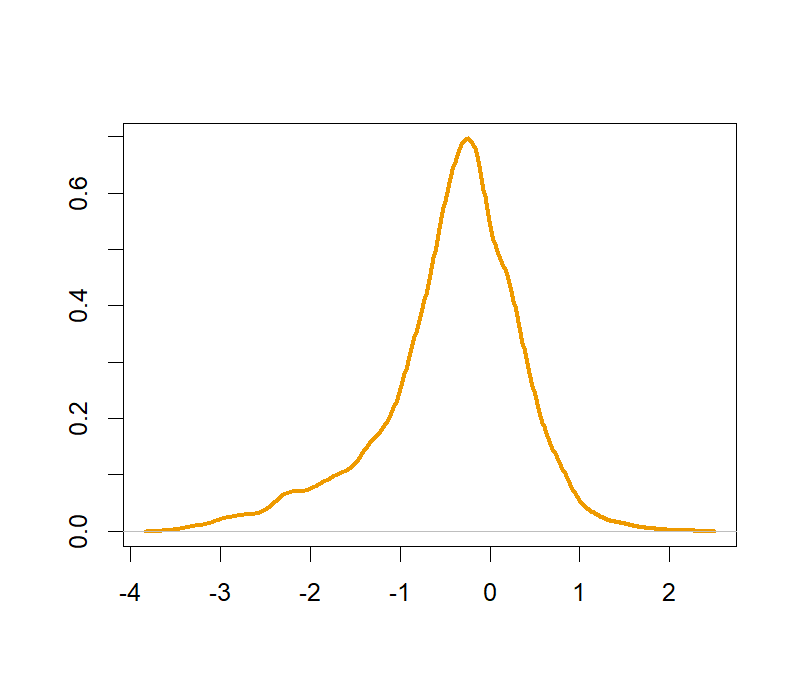} 
    \end{minipage}%
    \caption{Kernel density plots of house prices from the Lucas-1998-HP data: The left panel displays the kernel density of house prices (in hundreds of thousands). The right panel shows the kernel density of the logarithm of house prices (in hundreds of thousands).}
    \label{fig:densities_realdata}
\end{figure}

Figure~\ref{fig:densities_realdata} shows kernel density plots of house prices from the Lucas-1998-HP dataset. The left panel displays the distribution of house prices (in hundreds of thousands), while the right panel shows the distribution of their natural logarithms. For a detailed description of the Lucas-1998-HP dataset, see Section~\ref{sec:real} of the main paper. As expected, house prices being strictly positive, exhibit right skewness, whereas their logarithms appear to be left-skewed.

\label{sec:sim_miss}

\begin{figure}[H]
    \centering
    \includegraphics[width=0.8\linewidth]{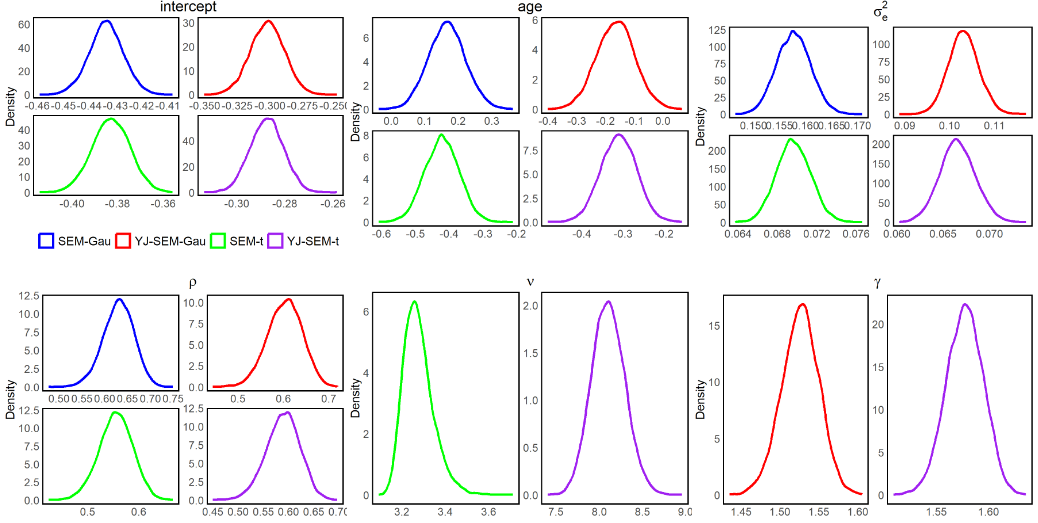}  
    \caption{
  Posterior densities of some of the parameters of the SEM-Gau, YJ-SEM-Gau, SEM-t, and YJ-SEM-t, fitted to the Lucas-1998-HP dataset without missing values, using the VB method.
    }
    \label{fig:para_densities_real_full}
\end{figure}

\begin{figure}[H]
    \centering
    \includegraphics[width=0.8\linewidth]{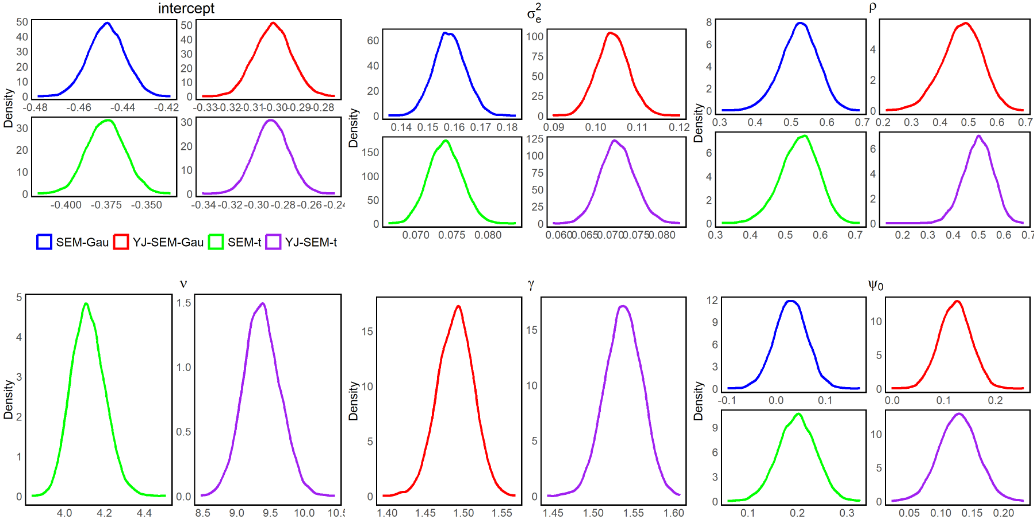}  
    \caption{
Posterior densities of some of the parameters of the SEM-Gau, YJ-SEM-Gau, SEM-t, and YJ-SEM-t, fitted to the Lucas-1998-HP dataset with missing values, using the HVB-AllB method.
    }
    \label{fig:para_densities_real_miss}
\end{figure}

Figures~\ref{fig:para_densities_real_full} and~\ref{fig:para_densities_real_miss} compare the posterior densities of selected model parameters, estimated using 
the proposed VB and HVB-AllB methods, across the different SEMs applied to the 
Lucas-1998-HP dataset, without and with missing values, respectively.

\section{Convergence analysis}

For the variational Bayes (VB) and hybrid VB (HVB) algorithms, convergence is assessed by examining the trajectories of the variational means of model parameters across iterations. For the Hamiltonian Monte Carlo (HMC) algorithm~\citep{neal2011mcmc}, convergence is evaluated by inspecting the trace plots of the posterior draws of model parameters.

Section~\ref{sec:sim_conv_analysis_6} presents convergence analysis plots for the simulation study described in Section~\ref{sec:simulationstudy-1} of the main paper. Section~\ref{sec:sim_conv_analysis_7} provides the corresponding convergence analysis plots for the simulation study in Section~\ref{sec:simulationstudy-2} of the main paper (with missing data), as well as for the full-data case in Section~\ref{sec:sim_full}. Finally, Section~\ref{sec:real_conv_analysis} reports the convergence analysis plots for the real-data example discussed in Section~\ref{sec:real} of the main paper.

\subsection{Convergence analysis for simulation study: Assessing the accuracy of VB methods}
\label{sec:sim_conv_analysis_6}


This section presents convergence plots from the simulation study designed to compare the performance of the VB approximation with the HMC algorithm, as described in Section~\ref{sec:simulationstudy-1} of the main paper. In this study, data are simulated from the YJ-SEM-Gau, and the model is estimated using both VB and HMC methods.

\begin{figure}[H]  
    \centering
    \includegraphics[width=12cm]{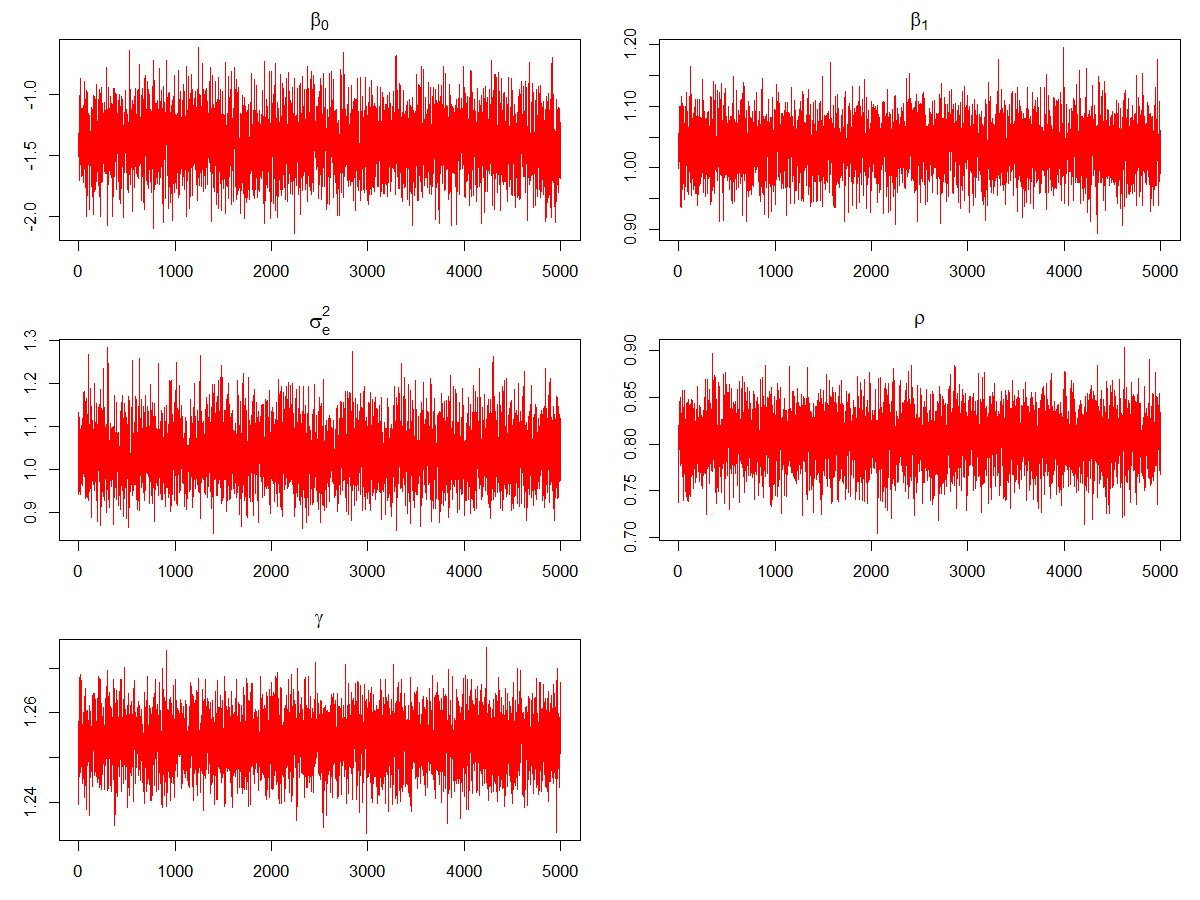}  
    \caption{Trace plots of posterior samples obtained using HMC for selected parameters of the YJ-SEM-Gau  (without missing values), after excluding burn-in samples.}
    \label{fig:con_VB_vs_HMC_full}
\end{figure}

\begin{figure}[H]  
    \centering
    \includegraphics[width=12cm]{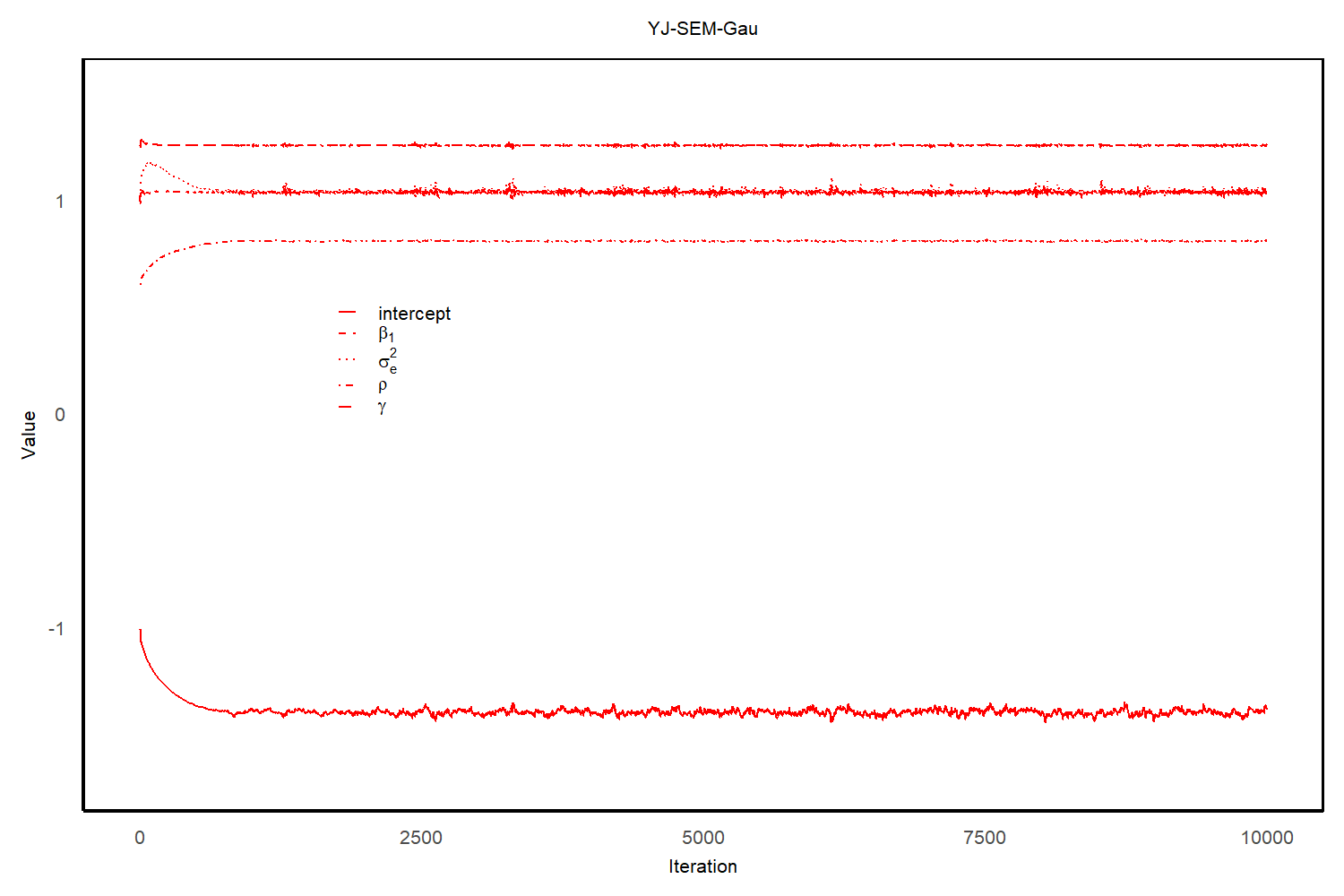}  
    \caption{Trajectories of the variational means of selected parameters of the YJ-SEM-Gau (without missing values) across VB iterations.}
    \label{fig:con_VB_vs_HMC_VB_full}
\end{figure}

Figure~\ref{fig:con_VB_vs_HMC_full} presents trace plots of posterior samples obtained using the HMC algorithm for selected parameters of the YJ-SEM-Gau with full data, after excluding burn-in samples. Figure~\ref{fig:con_VB_vs_HMC_VB_full} displays the trajectories of the variational means for selected parameters across VB iterations. Both plots indicate that the algorithms have converged.

Next, missing values are introduced into the simulated dataset, and the YJ-SEM-Gau is fitted using both the HVB-NoB and HMC algorithms.

\begin{figure}[H]  
    \centering
    \includegraphics[width=12cm]{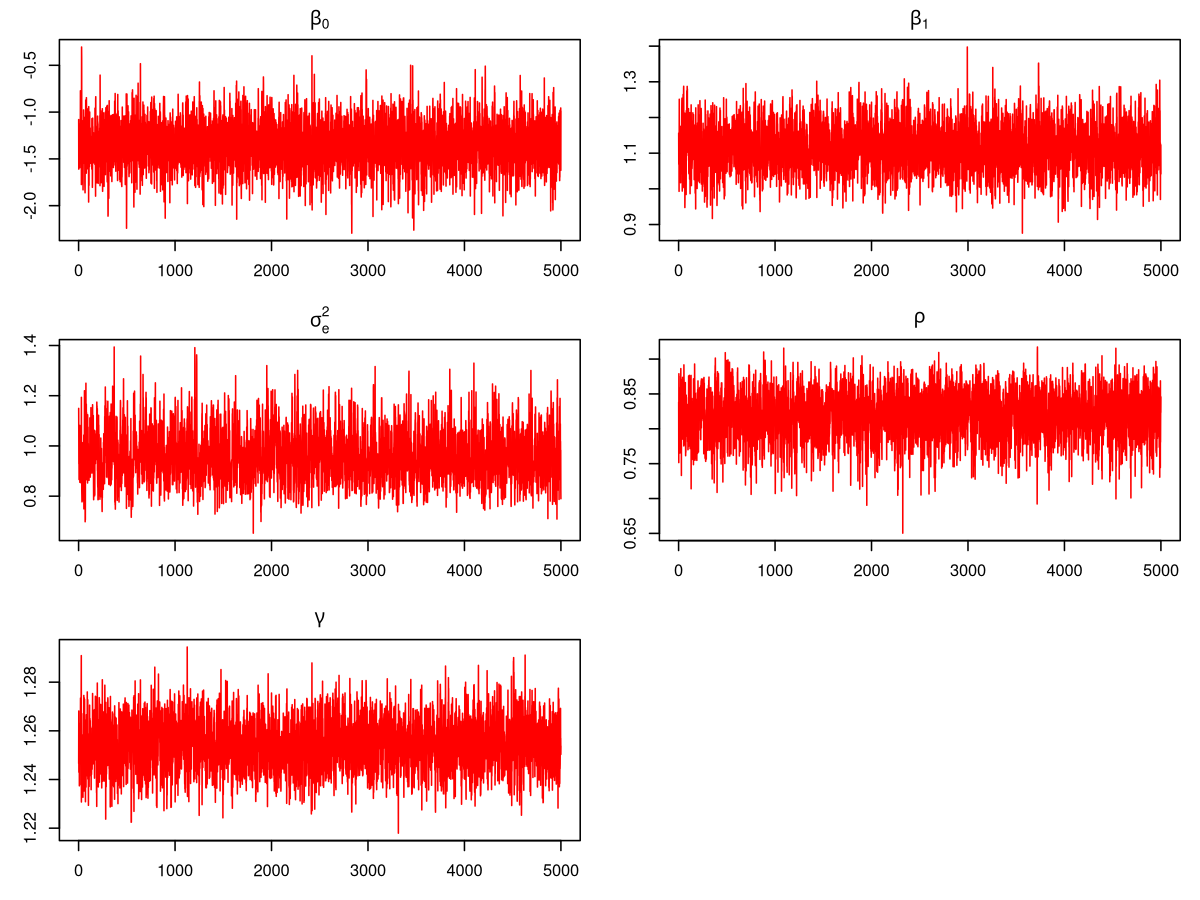}  
    \caption{Trace plots of posterior samples obtained using HMC for selected parameters of the YJ-SEM-Gau  (with missing values), after excluding burn-in samples.}
    \label{fig:con_VB_vs_HMC_miss}
\end{figure}

\begin{figure}[H]  
    \centering
    \includegraphics[width=12cm]{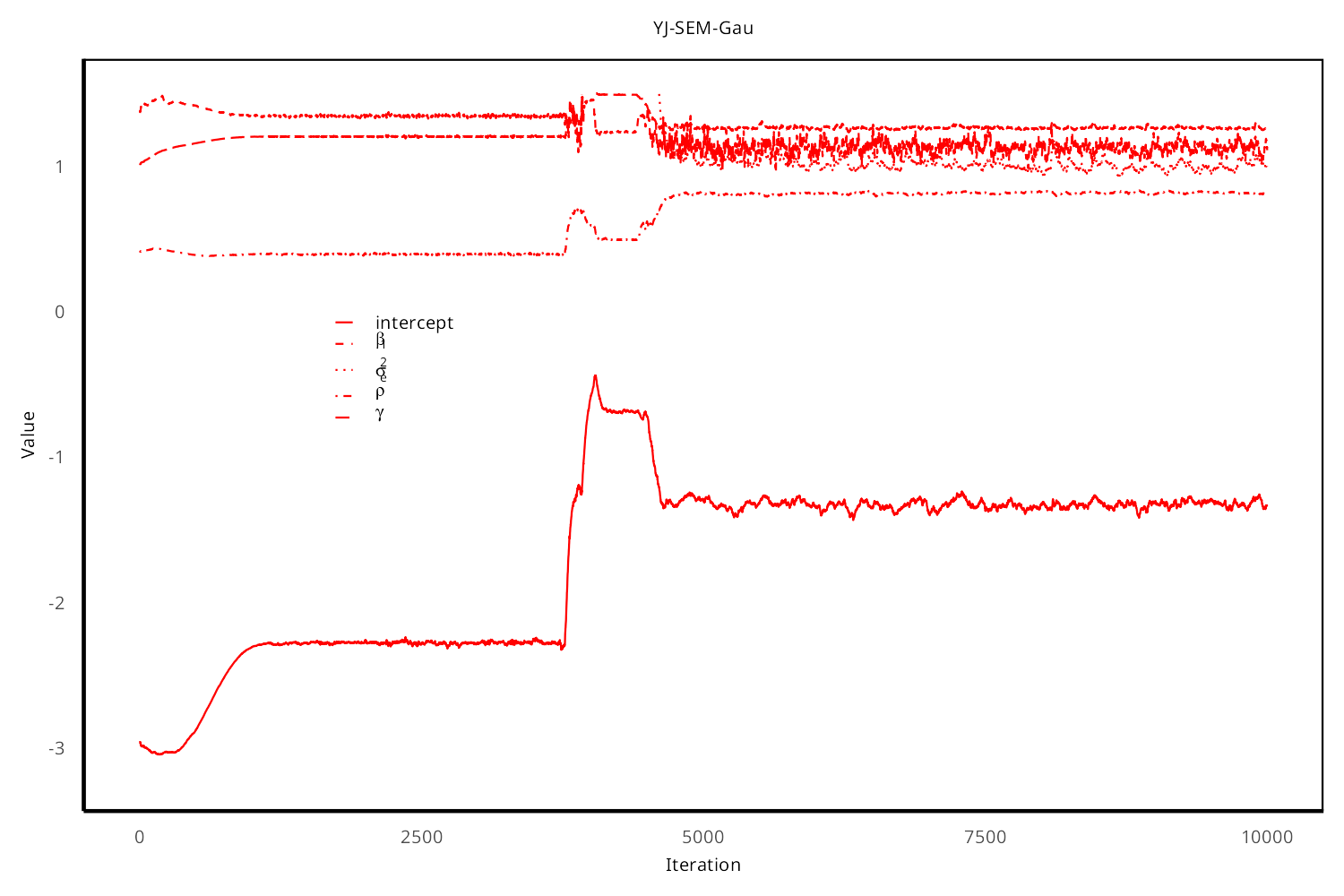}  
    \caption{Trajectories of the variational means of selected parameters of the YJ-SEM-Gau (with missing values) across HVB-NoB iterations.}
    \label{fig:con_VB_vs_HMC_VB_miss}
\end{figure}

Figure~\ref{fig:con_VB_vs_HMC_miss} presents trace plots of posterior samples obtained using the HMC algorithm for selected parameters of the YJ-SEM-Gau with missing values, after excluding burn-in samples. Figure~\ref{fig:con_VB_vs_HMC_VB_miss} displays the trajectories of the variational means for selected parameters across HVB-NoB iterations. Both plots indicate that the algorithms have converged.

\subsection{Convergence analysis for 
simulation study: Assessing the accuracy and robustness of proposed SEMs}
\label{sec:sim_conv_analysis_7}

This section presents the convergence plots for the simulation study in Section~\ref{sec:simulationstudy-2} of the main paper (with missing data), as well as for the full-data case in Section~\ref{sec:sim_full}.

\begin{figure}[H]  
    \centering
    \includegraphics[width=15cm]{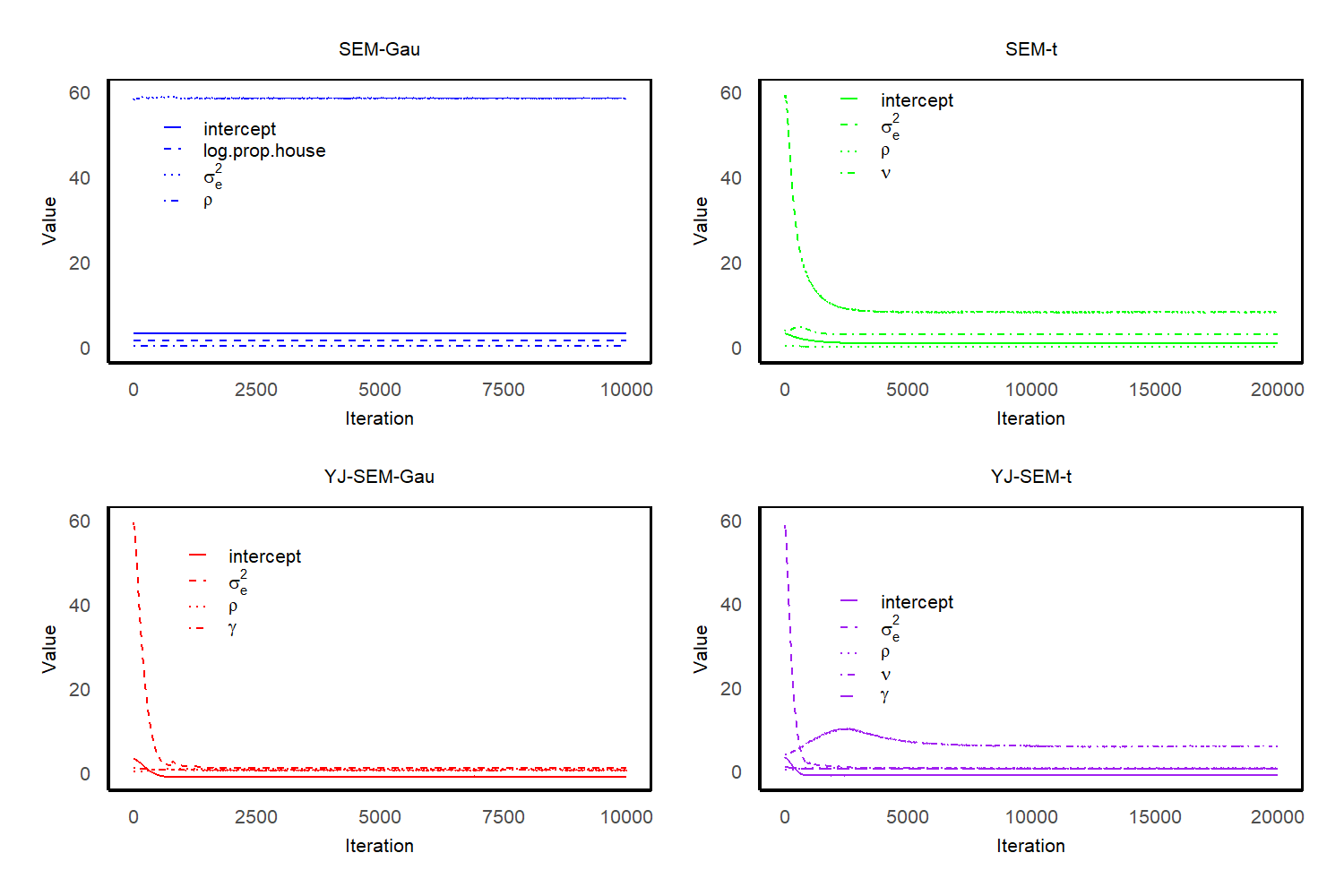}  %
    \caption{Trajectories of the variational means of selected parameters across VB iterations for different SEMs fitted on the simulated dataset 1.}
    \label{fig:con_sim_hs_ev_full}
\end{figure}

Figure~\ref{fig:con_sim_hs_ev_full} shows trajectories of the variational means for selected model parameters of different SEMs fitted to the simulated dataset 1, without missing values. It can be observed that for each parameter in each model, the variational means converge by the end of the final iterations,  indicating that the VB algorithms for all models have successfully converged.

\begin{figure}[H]  
    \centering
    \includegraphics[width=15cm]{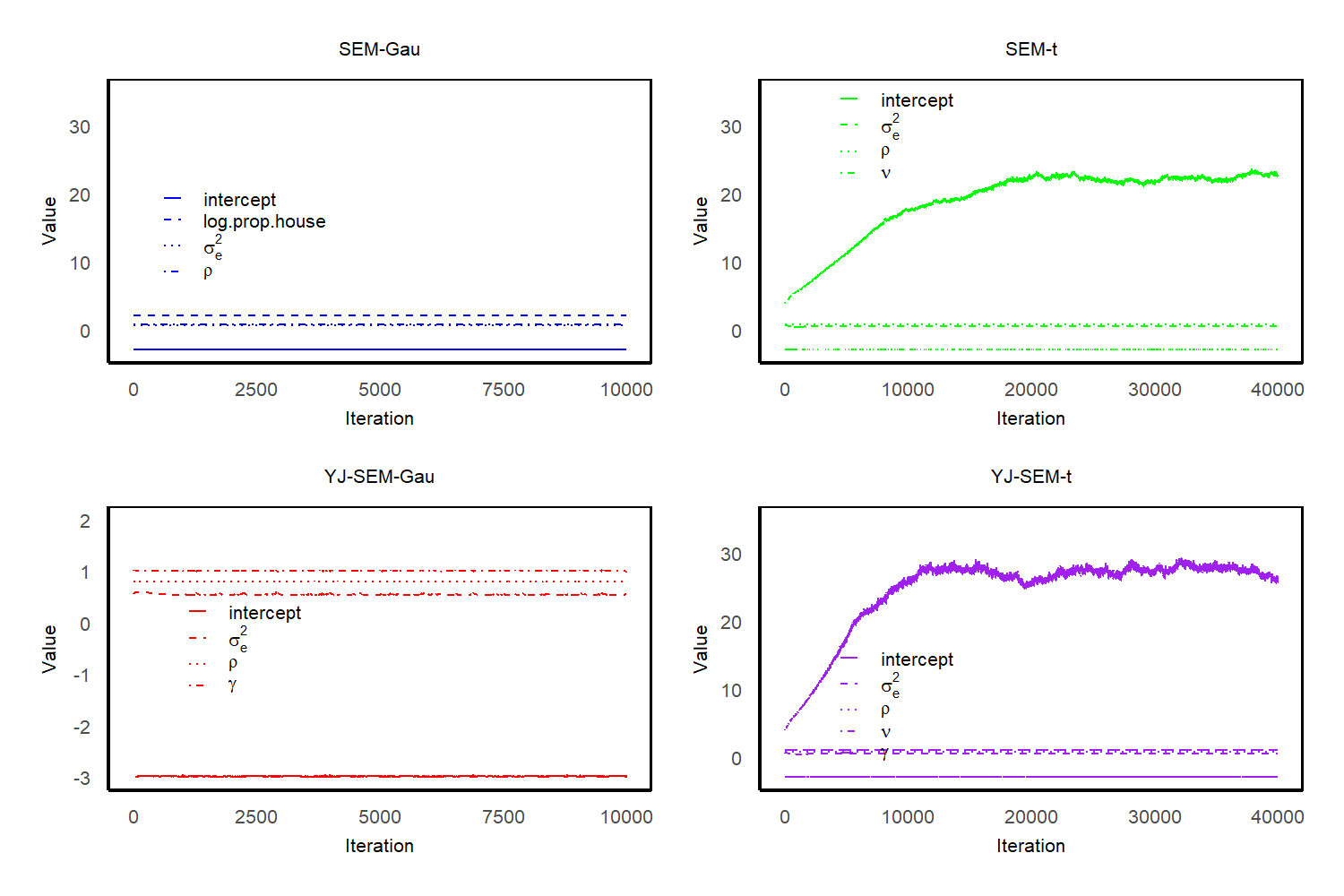}  
    \caption{Trajectories of the variational means of selected parameters across VB iterations for different SEMs fitted on the simulated dataset 2.}
    \label{fig:con_sim_norm_ev_full}
\end{figure}

Figure~\ref{fig:con_sim_norm_ev_full} shows trajectories of the variational means for selected model parameters of different SEMs fitted to the simulated dataset 2, without missing values. It can be observed that for each parameter in each model, the variational means converge by the end of the final iterations,  indicating that the VB algorithms for all models have successfully converged.

\begin{figure}[H]  
    \centering
    \includegraphics[width=15cm]{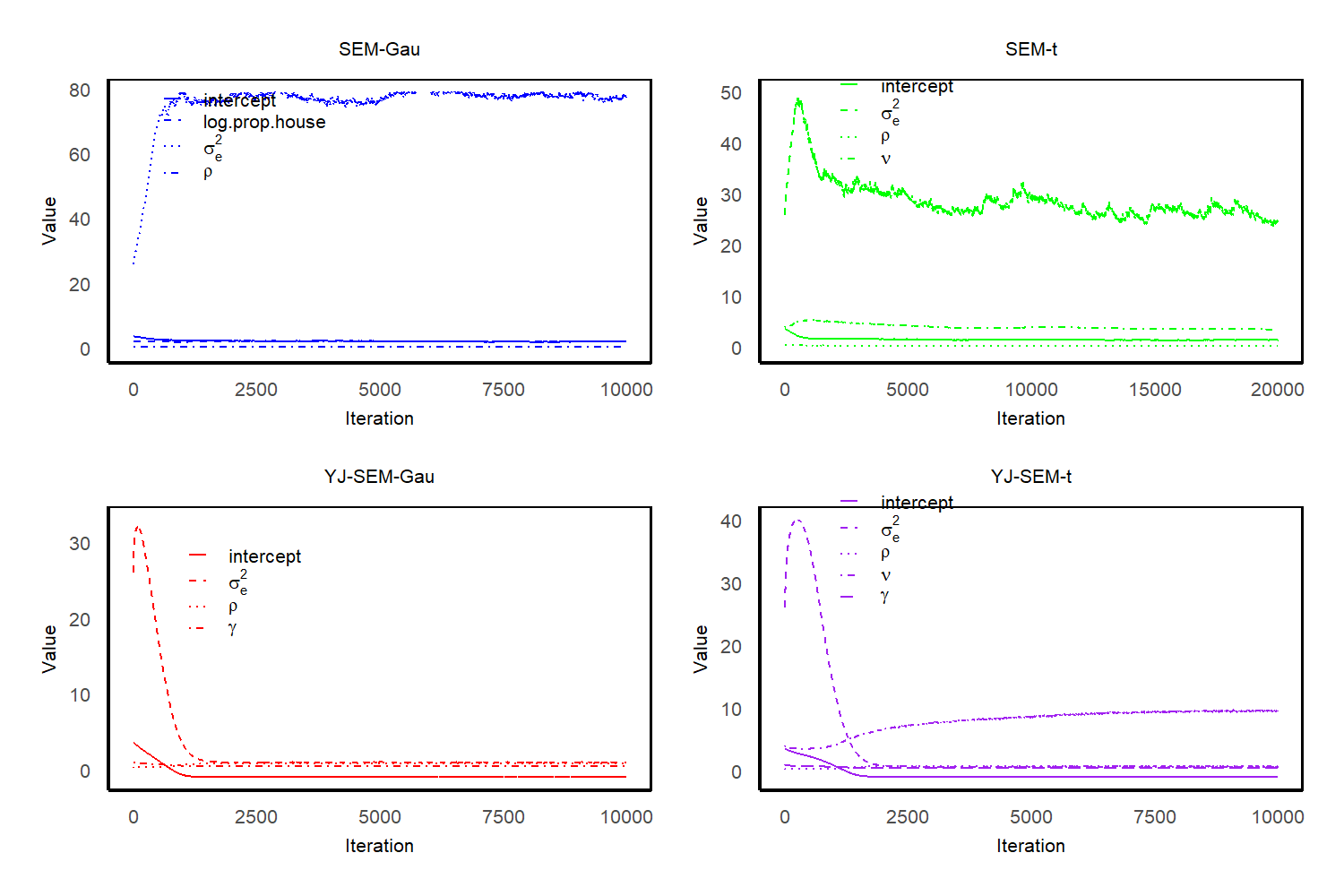}  
    \caption{Trajectories of the variational means of selected parameters across HVB-AllB iterations for different SEMs fitted on the simulated dataset 1 with missing values.}
    \label{fig:con_sim_hs_miss}
\end{figure}

Figure~\ref{fig:con_sim_hs_miss} shows trajectories of the variational means for selected model parameters of different SEMs fitted to the simulated dataset 1, with missing values. It can be observed that for each parameter in each model, the variational means converge by the end of the final iterations, indicating that the HVB-AllB algorithms for all models have successfully converged.

\begin{figure}[H]  
    \centering
    \includegraphics[width=15cm]{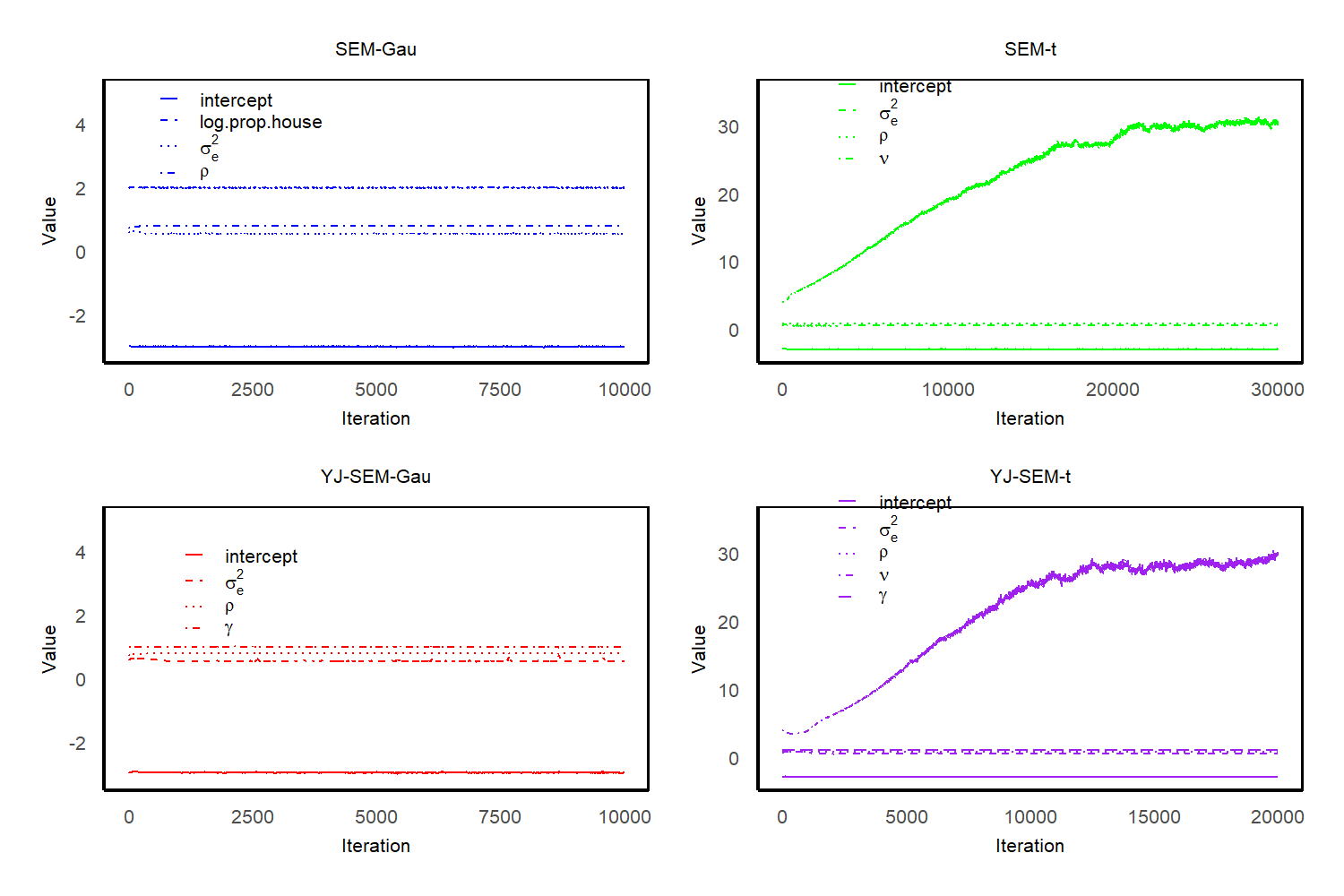}  
    \caption{Trajectories of the variational means of selected parameters across HVB-AllB iterations for different SEMs fitted on the simulated dataset 2 with missing values.}
    \label{fig:con_sim_norm_miss}
\end{figure}

Figure~\ref{fig:con_sim_norm_miss} shows trajectories of the variational means for selected model parameters of different SEMs fitted to the simulated dataset 2, with missing values. It can be observed that for each parameter in each model, the variational means converge by the end of the final iterations,  indicating that the HVB-AllB algorithms for all models have successfully converged.

\subsection{Convergence analysis of real data application}
\label{sec:real_conv_analysis}

This section presents the convergence analysis plots for the real data application discussed in Section~\ref{sec:real} of the main paper.

\begin{figure}[H]  
    \centering
    \includegraphics[width=15cm]{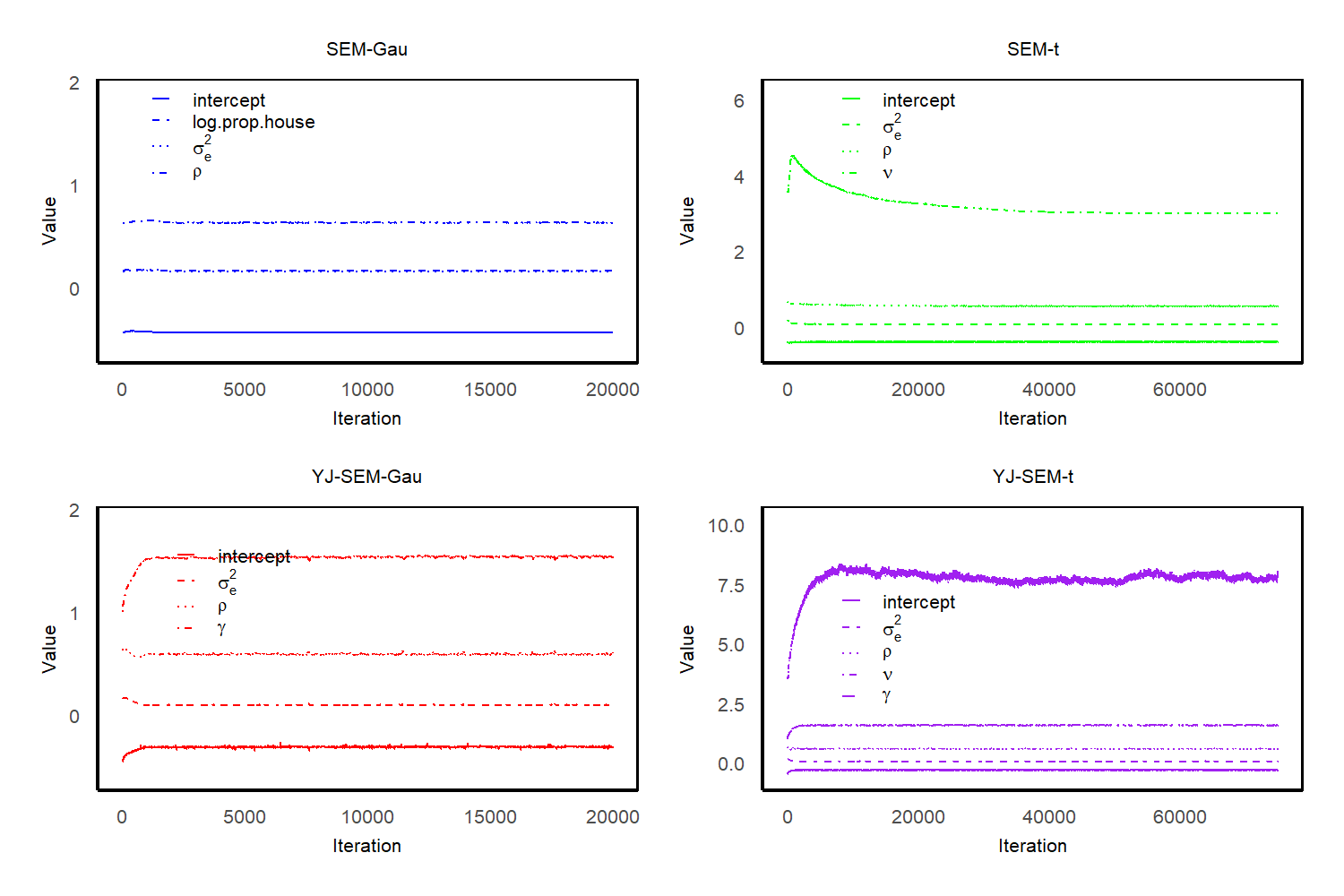}  
    \caption{Trajectories of the variational means of selected parameters across VB iterations for different SEMs fitted on the Lucas-1998-HP dataset without missing values.}
    \label{fig:con_real_full}
\end{figure}

Figure~\ref{fig:con_real_full} shows trajectories of the variational means for selected model parameters of different SEMs fitted to the Lucas-1998-HP dataset (full data). It can be observed that for each parameter in each model, the variational means converge by the end of the final iterations,  indicating that the VB algorithms for all models have successfully converged.

\begin{figure}[H]  
    \centering
    \includegraphics[width=15cm]{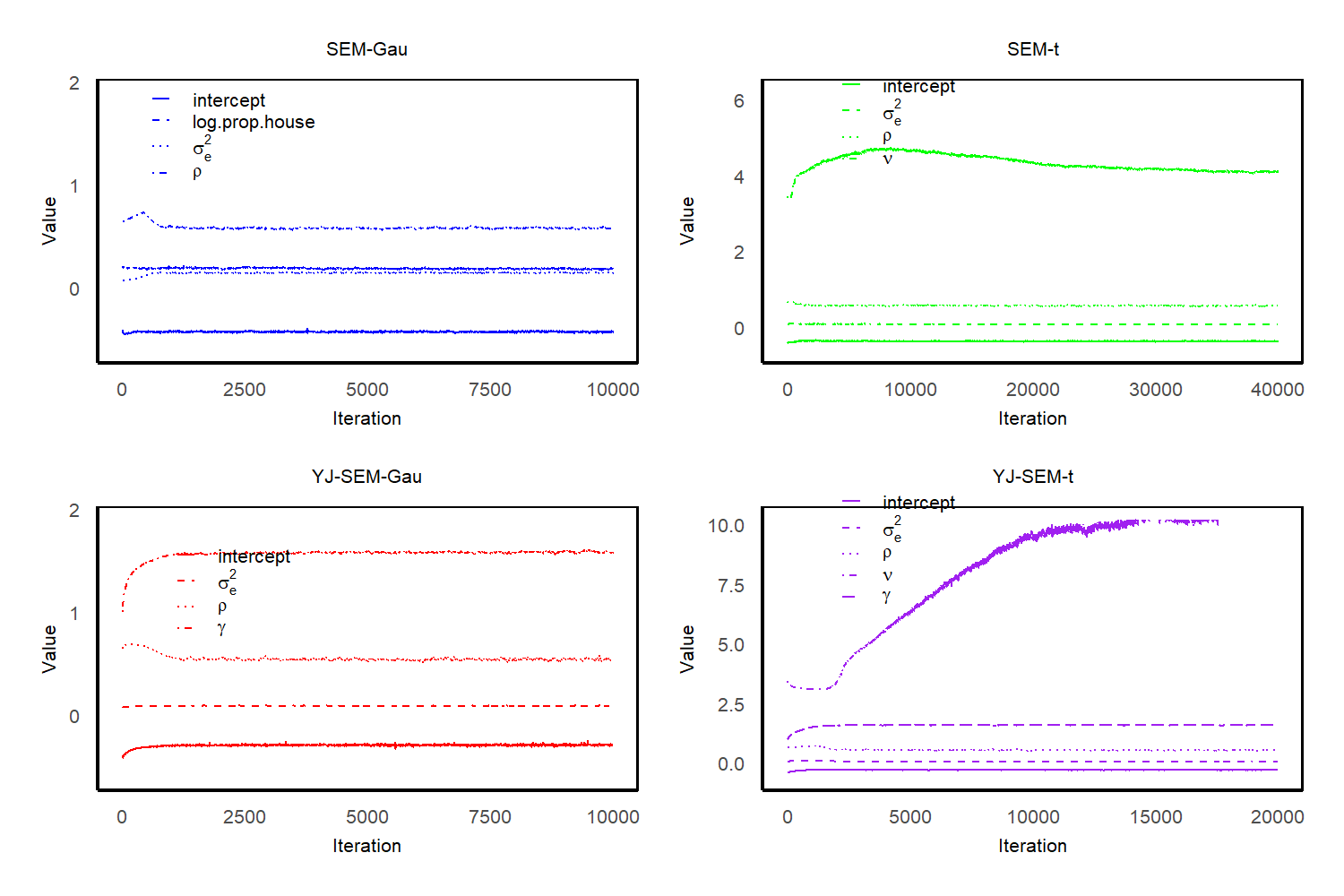}  
    \caption{Trajectories of the variational means of selected parameters across HVB-AllB iterations for different SEMs fitted on the Lucas-1998-HP dataset with missing values.}
    \label{fig:con_real_miss}
\end{figure}

Figure~\ref{fig:con_real_miss} shows trajectories of the variational means for selected model parameters of different SEMs fitted to the Lucas-1998-HP dataset, with missing values. It can be observed that for each parameter in each model, the variational means converge by the end of the final iterations,  indicating that the HVB-AllB algorithms for all models have successfully converged.

\end{document}